%% file: GR-review-1_3.tex
\documentclass[ppnp,twoside,12pt,graphicx,psfig]{article}
\usepackage{epsfig}
\usepackage{graphicx}
\usepackage{dcolumn}
\usepackage{amssymb}
\usepackage{booktabs}
\usepackage{longtable}
\usepackage{color}

\newcommand{\be}{\begin{equation}}
\newcommand{\ee}{\end{equation}}
\newcommand{\bea}{\begin{eqnarray}}
\newcommand{\eea}{\end{eqnarray}}

\begin{document}

\title{ \vspace{1cm} 
The Compression-Mode Giant Resonances and Nuclear Incompressibility
}

\author{Umesh Garg,$^{1}$ Gianluca Col\`o,$^{2,3}$
\\
$^1$Physics Department, University of Notre Dame, \\ 
Notre Dame, IN 46556, USA\\
$^2$Dipartimento di Fisica, Universit\`a
degli Studi di Milano,\\ via Celoria 16, 20133 Milano, Italy\\
$^3$INFN, Sezione di Milano, via Celoria 16, 20133 Milano, Italy}

\date{\today}

\maketitle

\begin{abstract} 
The compression-mode giant resonances, namely the isoscalar giant
monopole and isoscalar giant dipole modes, are examples of collective
nuclear motion. Their main interest stems from the fact that 
one hopes to extrapolate from their properties the incompressibility
of uniform nuclear matter, which is a key parameter of the nuclear Equation
of State (EoS). Our understanding of these issues has undergone two
major jumps, one in the late 1970s when the Isoscalar Giant Monopole Resonance 
(ISGMR) was experimentally 
identified, and another around the turn of the millennium since when
theory has been able to start giving reliable error bars to the
incompressibility. However, mainly magic nuclei have been involved
in the deduction of the incompressibility from the vibrations of finite
nuclei.

The present review deals with the developments beyond all this.
Experimental techniques have been improved, and new open-shell, and deformed, 
nuclei have been investigated. The associated changes in our understanding of
the problem of the nuclear incompressibility are discussed. 
New theoretical models, decay measurements, and the search for
the evolution of compressional modes in exotic nuclei are also
discussed.
\end{abstract}


\input{introduction_2}

\input{general_def}
\input{experiment_techniques_3}

\input{theory}

\input{experiment_results_2}

\input{incompressibility_2}

\input{deformation_plus_other_3}

\input{ktau_2}

\input{decay_2}

\input{unstable_nuclei_2}

\input{conclusions_2}

\input{acknowledgments_2}

\bibliography{review2}

\end{document}

%% file: introduction_2.tex
\section{Introduction}

Giant resonances (GRs) are the clearest manifestation of nuclear collective 
motion. 
The domain of GRs is a mature field, in which the basic issues have been
established quite some time ago. Monographs exist that review the 
classification of these collective modes and illustrate the findings
obtained up to the turn of the century \cite{BBB_book,Harakeh_book}. 
We do not intend to restart from scratch but, in a way, this review
paper aims to be a supplement to the book by M. Harakeh and A. Van 
Der Woude \cite{Harakeh_book}; consequently, the results presented here 
have all been obtained after the years 2000-2001, with only a few exceptions.

There are essentially three new and interesting lines of research in the
current physics of GRs: use well-established experimental data as a benchmark for
new theories; identify GRs in neutron-rich, eventually weakly-bound isotopes;
and, search for elusive modes that have not been seen so far. These issues are
quite general, and will be touched upon in this review. However, our main
concern in this review is the compression-mode resonances. These are interesting
for one specific reason, namely one hopes to extract from them valuable
information about the incompressibility of nuclear matter, $K_\infty$. This quantity, 
as we shall discuss in detail below, is one of the key parameters of
the nuclear Equation of State (EoS). We strive to determine it for
our general knowledge, and also for some applications related to
astrophysics. The mechanism of core-collapse supernova explosion is
still under current intensive study. Although the weak-interaction
processes and the neutrino transport are the most crucial ingredients
for the understanding of the explosion and of the evolution towards the
(proto-)neutron star, the EoS is also an important input for the simulations.

Therefore, extending the study of compression modes to more systems
than it has been done so far (including open-shell, or deformed isotopes) is not
done merely to complete some systematics but rather with the goal to better
understand how nuclear matter behaves when it is compressed. There is
an interplay here with the general issue of ``elusive'' modes, 
that we mentioned previously. The nucleus is a complex system, 
and the question, in some cases, is
first whether a mode is compression or not --- and after that, whether 
its properties shed some new light on the value of the incompressibility. While the
isoscalar giant monopole resonance (ISGMR) in several medium-heavy nuclei
is definitely a well-defined compressional mode, it gets more fragmented 
in lighter systems and/or in deformed systems and associating it 
with a macroscopic compressional ``single mode'' becomes more problematic. 
As we move to the isoscalar dipole case, the issue is to understand 
the nature of the low-lying strength. Whether this strength has 
simple one-particle character or is hiding some new ``mode'' (like the elusive
toroidal mode), is still not completely clear and the answer 
may be different according to the region of the isotope chart. In short, 
while in the past only the ISGMR in $^{208}$Pb has been used to
extract the value of the nuclear incompressibility, our present 
understanding needs to be checked against other types of nuclei.
This may bring to a broader view, or to a change in perspective, 
concerning the nuclear incompressibility.

An ultimate goal is to further broaden our understanding 
by going towards exotic nuclei far from the stability valley. 
There is obvious interest in being able to pinpoint how 
the incompressibility, or the other quantities that characterise 
the nuclear EoS, change when the neutron-proton asymmetry increases. 
Obviously, we may expect qualitative changes when nuclei become 
weakly bound, resulting in a dilute form of nuclear matter, in 
keeping with the formation of a skin or a halo.
In fact, if isotope chains that are long enough can be investigated, 
one finds that when the neutron excess increases the protons become 
more bound due to the strong neutron-proton interaction, while the 
neutrons occupy higher levels that lie close to the continuum. A 
large difference between the Fermi energies of protons and neutrons may
produce a decoupling between the well-bound
nucleons in the “core” and the less bound neutrons in the skin or halo. 
These two components might behave like two fluids with different 
incompressibilities. If this happens, by studying the compression
modes in long chains extending far from the stability valley, we can have access to the
incompressibility of neutron matter at sub-saturation density. 
This is also of paramount importance for the
physics of the core-collapse supernova and the neutron stars.
The inner crust of neutron star is composed by matter at
densities between 10$^{-3}\rho_0$ and $\rho_0$ where $\rho_0$ is the
ordinary saturation density. Hence, the importance of knowing 
the parameters of the EoS below saturation.

In this paper, after some brief reminders about the definition of monopole
and dipole operators and strengths in Sec. 2, we discuss the experimental and
theoretical tools for the study of compression modes in Secs. 3 and 4, 
respectively. The experimental results obtained after the turn of the 
millennium are reviewed in Sec. 5. Sec. 6 is devoted to the extraction of
the nuclear incompressibility from the compression modes with the help
of theory. We will stress the differences that seem to emerge between 
magic and open-shell nuclei, as far as the deduced value of the 
incompressibility is concerned. Secs. 7 and 8 are devoted to other
possible effects (deformation and shell effects) that may play a role 
in making the systematic extraction of the incompressibility quite
difficult. Then, Sec. 9 
concerns the asymmetry term of the nuclear incompressibility, while Sec. 10
reports on recent particle-decay measurements. The main prespective for
the future, i.e. compression modes in unstable nuclei, is the subject of
Sec. 11. Finally, our conclusions are presented in Sec. 12.

%% file: general_def.tex
\section{General definitions}\label{general}

Giant resonances are states in which most of the strength associated
with a given external operator acting on the nucleus is concentrated. 
The external operator $F$ can be of various kinds, but in this 
paper we shall focus on the isoscalar monopole operator,
\begin{equation}
F_{\rm IS\ monopole} = \sum_i r_i^2,
\end{equation}
and on the isoscalar dipole operator,
\begin{equation}\label{eq:ISdipole}
F_{\rm IS\ dipole} = \sum_i r_i^3 Y_{1M}\left( \hat r_i \right),
\end{equation}
where $i$ labels the $i$-th nucleon while $r$ and $\hat r$ are
the radial coordinate and a shorthand notation for the polar angles.
The strength function, or simply strength, associated with an 
operator $F$ is a function of the excitation energy $E$ that reads
\begin{equation}\label{eq:strength}
S(E) = \sum_n \vert \langle n \vert F \vert 0 \rangle \vert^2
\delta(E-E_n),
\end{equation}
where $n$ labels a complete set of excited states of the nucleus, $\vert n \rangle$, 
having energy $E_n$ with respect to the ground state $\vert 0 \rangle$.
The latter equation (\ref{eq:strength}) is written in terms of discrete excited states 
for the sake of simplicity, but above the particle threshold 
$S(E)$ is actually a continuous function.  
The sum rules $m_k$ are defined by
\begin{equation}
m_k \equiv \int dE\ E^kS(E), 
\end{equation}
and have the obvious meaning of being the moments of the strength
function. 
Of special importance are the non-energy weighted sum rule (NEWSR) that
is simply the moment $m_0$, and the energy-weighted sum rule (EWSR) 
$m_1$. The centroid of the strength function can be written as
\begin{equation}
\bar E = \frac{m_1}{m_0},
\end{equation}
and this quantity is often referred to in what follows. The
inverse-energy weighted sum rule is also used in the context of
the constrained model (cf. Sec. \ref{sec:ka} below). 

We end this 
Section by providing the expected values of the EWSR for
the modes under study. In the case of the ISGMR one obtains 
\begin{equation}\label{eq:monopoleEWSR}
m_1(ISGMR) = \frac{2\hbar^2}{m}A\langle r^2\rangle.
\end{equation}
The corresponding value of EWSR for the ISGDR reads
\begin{equation}
m_1(ISGDR) = \frac{\hbar^2 A}{8\pi m} 11\langle r^4 \rangle.
\end{equation}
However, it is well known that part of the ISGDR sum rule is taken by the spurious
center-of-mass state that corresponds to the translation of the nucleus as a whole.
An effective way to get rid of the spurious state is to modify the operator 
(\ref{eq:ISdipole}) and write
\begin{equation}
F_{\rm IS\ dipole}^\prime = \sum_i \left( r_i^3 - \eta r_i \right) Y_{1M}\left( \hat r_i \right),
\end{equation}
where $\eta = \frac{5}{3}\langle r^2 \rangle$. Then, the EWSR becomes
\begin{equation}\label{eq:EWSRdipole_RPA}
m_1^\prime(ISGDR) = \frac{\hbar^2 A}{8\pi m} \left( 11\langle r^4 \rangle - 5 \eta \langle r^2 \rangle \right).
\end{equation}
If one starts from the small momentum limit of the dipole term in $\sum_i e^{i\vec q \vec r_i}$, 
namely if one considers the operator
\begin{equation}
F_{\rm IS\ dipole}^{(q)} = \sum_i j_1\left( qr_i \right) Y_{1M}\left( \hat r_i \right),
\end{equation}
and performs the limit $q \rightarrow 0$, then the ISGDR sum rule becomes
\begin{equation}\label{eq:EWSRdipole_final}
m_1^{\prime\prime}(ISGDR) = \frac{\hbar^2 A}{8\pi m} \left( 11\langle r^4 \rangle - 
5 \eta \langle r^2 \rangle - 10 \epsilon \langle r^2 \rangle \right),
\end{equation}
where
\begin{equation}\label{eq:epsilon}
\epsilon = \left( \frac{4}{E_{\rm ISGQR}} + \frac{5}{E_{\rm ISGMR}} \right) \frac{\hbar^2}{3mA}
\end{equation}
under the assumption that the ISGMR strength, as well as the isoscalar quadrupole resonance (ISGQR) one, 
are exhausted in a single peak\footnote{Note an extra factor 1/4 in Eq. (3) of Ref. \cite{Harakeh1981}, due to a factor 1/2 in front of the
operator [Eq. (A6b) of the same paper].}. The term proportional to $\epsilon$ in Eq. (\ref{eq:EWSRdipole_final}) can
safely be neglected in medium-heavy nuclei. In such a case, Eqs. (\ref{eq:EWSRdipole_RPA}) and 
(\ref{eq:EWSRdipole_final}) coincide. 

In all the expressions for the ISGDR sum rule we have taken into account the excitation of only one of the three $M$-components. In a spherical
system the sum rule associated with the sum over $M$ will have an extra factor three [as we write below, cf. Eq. (\ref{eq:Sdipole})].

%% file: experiment_techniques_3.tex
\section{Experimental and Data Analysis Techniques}\label{experiment_techniques}
Experimental determination of the ISGMR strength has been accomplished generally via inelastic scattering of isoscalar particles --- typically $\alpha$ particles or, in some cases, deuterons --- at energies of 35-100 MeV/nucleon. The cross sections for excitation of ISGMR rise sharply over this energy range, and rather slowly thereafter.
Because of the highly absorptive nature of the $\alpha$-nucleus and deuteron-nucleus interactions, the scattering may be treated as off a ``black disk'', with the cross sections, to the first order, given by the squares of the corresponding Bessel functions, $J_\lambda$ \cite{Bessel}:

\begin{equation}
\left(\frac{d\sigma}{d\Omega}\right)_{0^+\rightarrow 0^+} \propto  |J_0(qR_D)|^2,
\end{equation}

\begin{equation}
\left(\frac{d\sigma}{d\Omega}\right)_{0^+\rightarrow 1^-} \propto |J_1(qR_D)|^2,
\end{equation}

\begin{equation}
\left(\frac{d\sigma}{d\Omega}\right)_{0^+\rightarrow 2^+} \propto \left[\frac{1}{4}J_0(qR_D)^2 + \frac{3}{4}J_2(qR_D)^2\right].  
\end{equation}

\begin{figure}
\centering\includegraphics [height=0.27\textheight]{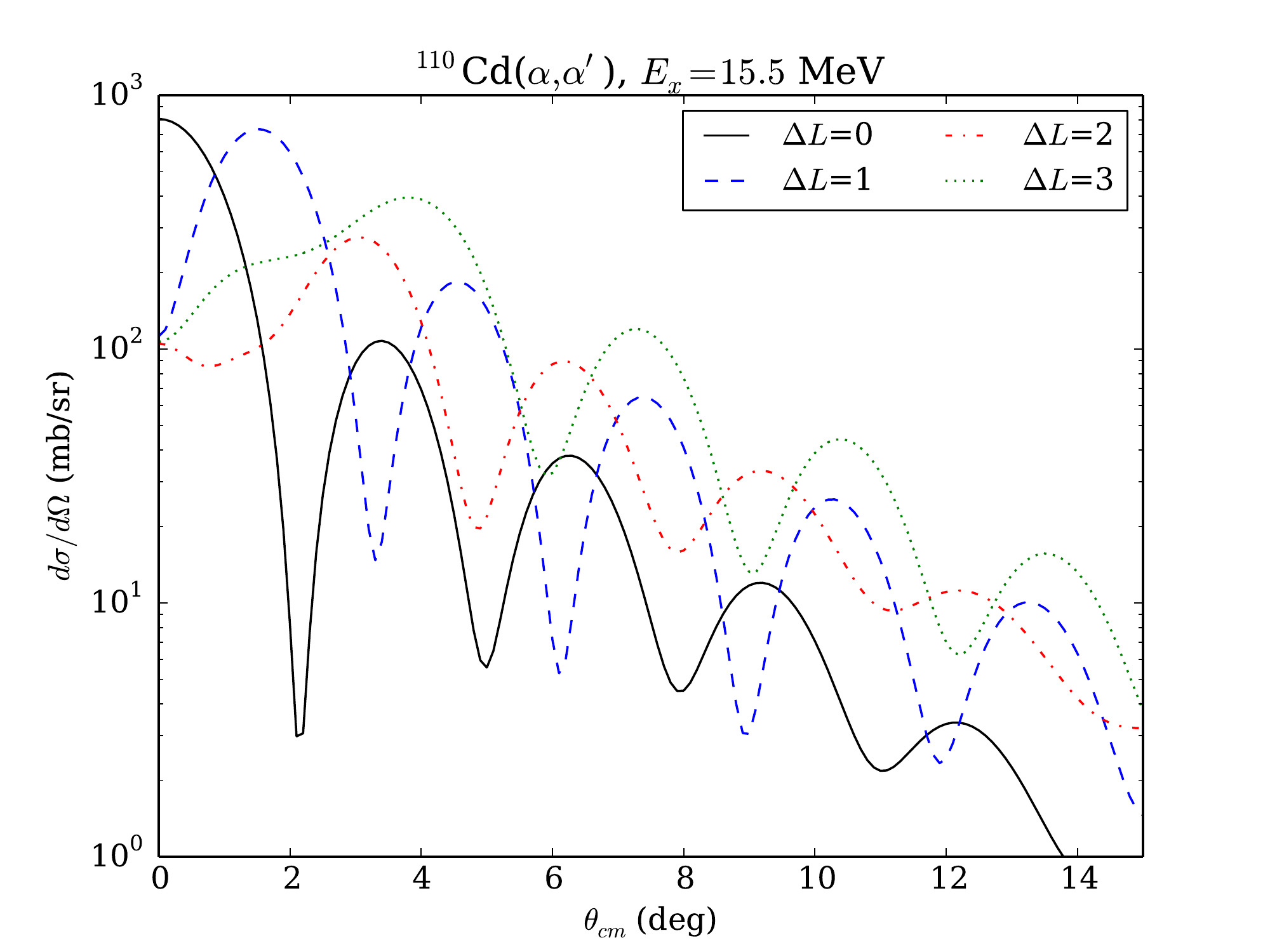}
\caption{
DWBA calculations of angular distributions of differential cross sections for excitation of isoscalar states of various multipoles ($\Delta L$=0--3) at an excitation energy of 15.5 MeV in inelastic scattering of 386-MeV $\alpha$ particles off $^{110}$Cd. Each calculation corresponds to exhaustion of 100\% of the EWSR in a single state at that energy.}
\label{angdist}
\end{figure}

\noindent
Here, $q$ is the momentum transfer, and $R_D$, the diffraction radius, is adjusted to fit the phase of the elastic scattering angular distribution. 
This leads to rather distinctive angular distributions of the inelastic scattering cross sections for various multipoles. Fig. \ref{angdist} shows distorted-wave Born approximation (DWBA) calculations for angular distributions for inelastic scattering of 386 MeV $\alpha$ particles off $^{110}$Cd for excitation of a state at an excitation energy of 15.5 MeV corresponding to  angular momentum transfers $\Delta${\em L} = 0--3.

A complication in this procedure arises because of the overlap between various giant resonances. For example, the ISGMR ($\Delta${\em L} = 0) overlaps significantly with the ISGQR ($\Delta${\em L} = 2); and the ISGDR ($\Delta${\em L} = 1) with the so-called high-energy octupole resonance (ISHEOR, $\Delta${\em L} = 3). 
As is evident from Fig. \ref{angdist}, the angular distributions corresponding to these overlapping $\Delta${\em L} values are clearly distinct only at very forward angles ($\leq$ 5$^{\circ}$). At higher beam energies, the angular distributions would be further ``compressed'', making the angular range for distinctive multipolar characteristics smaller, and the measurements even more difficult to carry out successfully. 
Thus, one needs to measure inelastic scattering at center-of-mass energies of 35-100 MeV/nucleon and at very small angles to clearly identify the strengths corresponding to various multipoles. The practical requirements for such measurements, therefore, are:
\begin {itemize}
\item
An accelerator--typically, a cyclotron--capable of providing beams at energies of 35-100 MeV/nucleon. At these energies, the cross sections for excitation of the ISGMR are sufficient to carry out the measurements in a reasonable time. At lower beam energies, the cross sections are rather low; at higher energies, the increase in cross sections is rather small and, as mentioned above, there are practical difficulties with determination of multipolarities based on angular distributions. It is also extremely important that the beam be free of any ``halos'' or ``wings'' at the target position; indeed, the process of obtaining a ``clean'' beam requires the right equipment and expertise, and can sometimes take a large fraction of the beam time typically allotted for such measurements.
\item
A high-quality magnetic spectrometer to allow for measurements at extremely forward angles, including 0$^{\circ}$, since only at small angles is it possible to clearly distinguish the various multipoles based on their distinct angular distributions. The 0$^{\circ}$ measurement is crucial because the ISGMR cross section is maximal there (see Fig. \ref{angdist}). 
[An exception to this requirement occurs when using inverse-kinematics reactions (with radioactive ion beams, for example); those measurements are discussed later.]
\end {itemize}
These requirements, in principle, are met at several laboratories around the world. However, almost all recent measurements on the compression-mode giant resonances with stable beams have been carried out at the Research Center for Nuclear Physics (RCNP), Osaka University, and at the Texas A \& M University Cyclotron Institute (TAMU) (with some decay measurements carried out at KVI, Groningen; these,  as also measurements at RCNP with deuterons, are discussed separately later in this article). In the RCNP measurements, a 386-MeV $\alpha$-particle beam is employed and inelastic scattering spectra are measured using the magnetic spectrometer Grand Raiden \cite{fujiwara99} for detection of the scattered particles. The TAMU group uses 240-MeV $\alpha$ beams and the MDM spectrometer \cite{pringle_mdm}. The scattered particles are momentum analyzed by the spectrometers and focused onto the focal-plane detector systems comprising a combination of multi-wire drift chambers, proportional counters, ionization chambers, and scintillation detectors, to allow for particle identification and determination of position (both x- and y- in case of RCNP work) and the angle of incidence of the scattered particle via the ray-tracing technique; typically, x-position resolutions of $<$1.0 mm and angular resolution of $<$0.15$^{\circ}$ have been achieved. The 0$^{\circ}$ measurements present an especial challenge because the beam itself also has to go through the system. In the TAMU arrangement, the beam passes beside the detector and is stopped on a carbon block inside a Faraday cup behind the detector \cite{dhy1998}, whereas in the RCNP system, it passes through a hole in the detectors and is stopped in a Faraday cup (FC) placed several meters downstream from the detectors \cite{Itoh_prc2003}; Fig. 1 of Ref. \cite{Itoh_prc2003}, for example, shows the arrangement of the three Faraday cups utilized in the RCNP measurements for different angular ranges.
In both cases, inelastic scattering spectra are measured typically over the angular range  0$^{\circ}$--10$^{\circ}$, with elastic scattering spectra also measured (over a much wider angular range), in many cases, in order to obtain appropriate optical model parameters used in the distorted-wave Born approximation (DWBA) calculations, as described hereinafter.
Calibration of the excitation energy spectra is carried out by measuring elastic and low-energy excitation peaks in the nuclei $^{12}$C and $^{24}$Mg.
 
\begin{figure}
\centering\includegraphics [height=0.45\textheight]{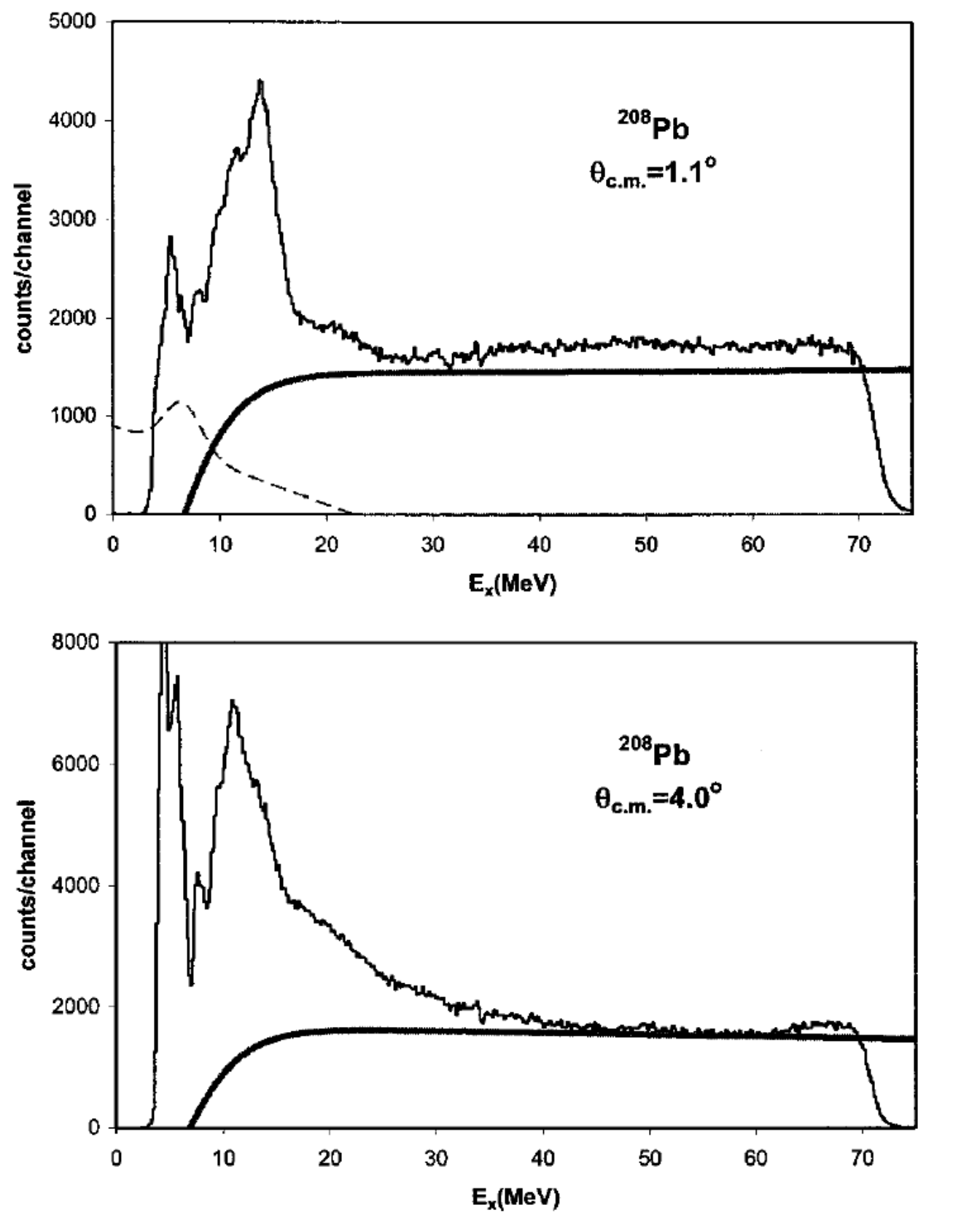}`
\caption{Excitation energy spectra for $^{208}$Pb from the TAMU ($\alpha, \alpha$') work at the scattering angles indicated. The thick solid lines show the continuum chosen for the analysis. The dashed line below 22 MeV represents a contaminant present at some angles in the spectra taken with the MDM spectrometer at 0$^{\circ}$. Figure from Ref. \cite{dhybg}. }
\label{tamu_bg}
\end{figure}

The inelastic scattering spectra for the medium-mass and heavier nuclei (A$\geq$90) typically consist of one or two broad giant resonance ``bumps'' on top of a ``background'' (see Fig. \ref{tamu_bg}). This ``background'' comprises excitation of nuclear continuum and contributions from three-body channels, such as knock-out reactions \cite{bran1}. At lower beam energies ($<$ 40 MeV/nucleon), there are also the contributions from the so-called ``pick-up and decay'' channels whereby the incoming $\alpha$ particle picks up a proton or a neutron from the target, forming $^{5}$Li and $^{5}$He, respectively. These unstable nuclei decay almost immediately, with the final $\alpha$ particle leading to a spurious ``bump'' in the inelastic scattering spectra, not dissimilar to a giant resonance. This is purely a kinematical effect and the position of  this ``bump'' depends on the scattering angle and beam energy. Still, this effect may lead to claims of identification of new resonances, one case being that of the ISHEOR in $^{208}$Pb using $\sim$20 MeV/nucleon $^{16}$O particles \cite{doll_prl}; later measurements, using
a $^{14}$N beam at 19 MeV/nucleon, clearly established \cite{garg_N14} that there was no evidence for excitation of resonances purported to have been observed in Ref. \cite{doll_prl}. 

The largest contributions to this ``background'' are, generally, instrumental, originating from rescattering of elastically scattered particles from the opening slits and other parts of the spectrograph. This problem is quite severe at small angles where the elastic scattering cross sections are very large. This large overall ``background'' had been a bane of all giant resonance measurements for the longest time because there is no direct way to calculate, or even estimate, its shape and magnitude. What one did was to subtract out from the spectra a background of  ``reasonable'' shape before further analysis. The ``reasonable shape'' could be a matter of debate, of course, and the process always led to questions about the correctness of the extracted results.

Of the two aforementioned laboratories from where most of the recent giant resonance measurements have come, TAMU uses this background subtraction process still. After extensive investigations of the results of different background shapes, they now employ an empirical background 
assuming that it has the shape of a straight line at high excitation, joining onto a Fermi shape at low excitation to model particle threshold effects \cite{dhybg,dhybg2}; examples of such empirical background are shown in Fig. \ref{tamu_bg}. 

\begin{figure}  [!h]
\centering\includegraphics [height=0.25\textheight]{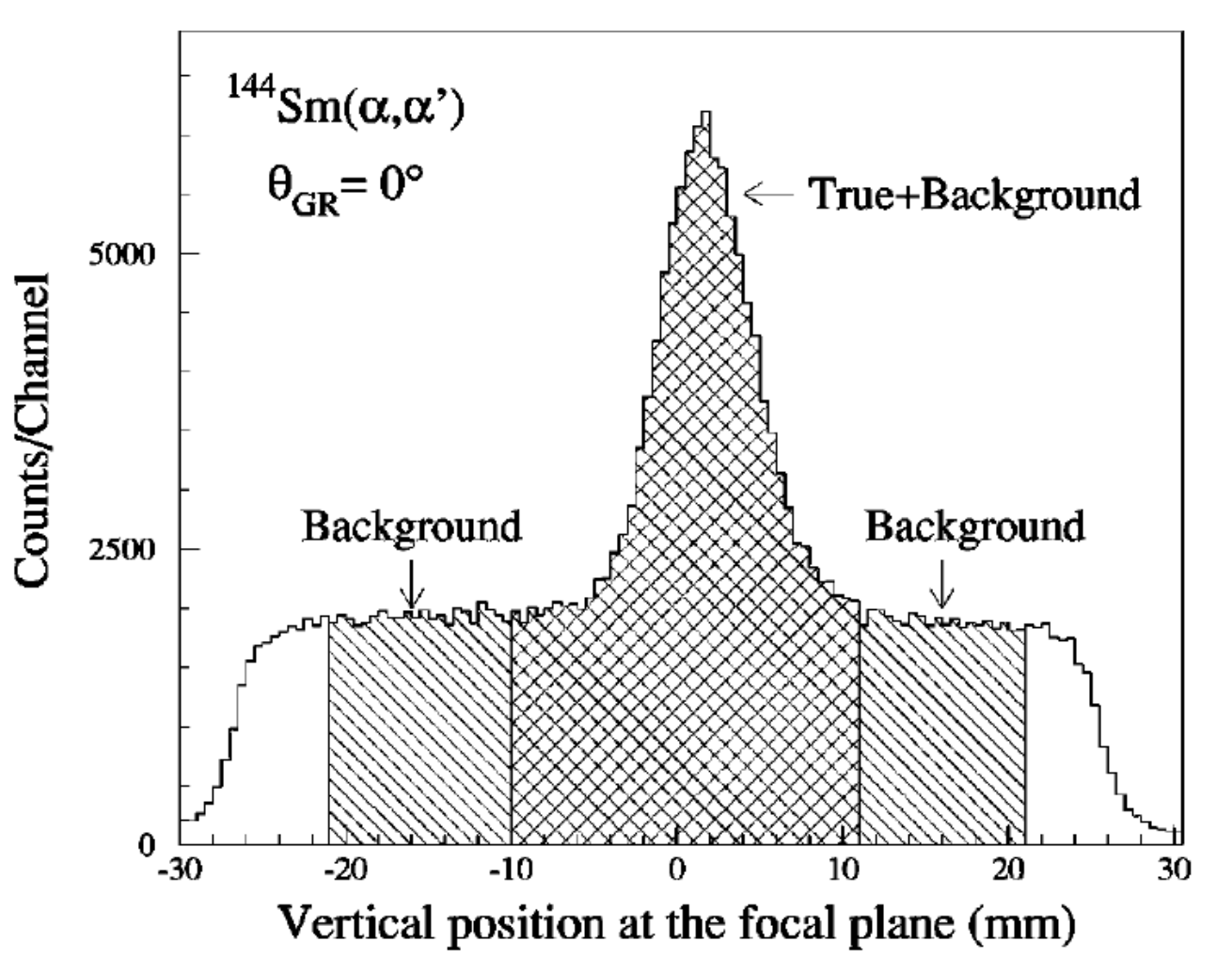}
\vspace{-0.3cm}
\caption{Vertical position spectra from the RCNP $^{144}$Sm ($\alpha, \alpha$') work, with the Grand Raiden spectrometer set at 0$^{\circ}$. The central, hatched region represents true+background events. The off-center, slanted lined regions represent only background events. The true events were obtained by subtracting background events from the true+background events; ``True'' stands for ``target-scattered''. Figure from Ref. \cite{Itoh_prc2003}.}
\label{bgsubtraction}
\end{figure}

The spectra in the RCNP measurements, on the other hand, are essentially free of all instrumental background. 
The ion-optics of Grand Raiden enables particles scattered from the target position to be focused vertically at the focal plane. On the other hand, background events due to the rescattering of $\alpha$ particles from the wall and pole surfaces of the spectrometer show a flat distribution in the vertical position spectra at the focal plane, as shown in Fig. \ref{bgsubtraction} for $^{144}$Sm($\alpha,\alpha$') with the Grand Raiden spectrometer set at 0$^{\circ}$. The vertical center (cross-hatched) region contains a combination of true (target-scattered) events plus those from the background; the off-center (slant-hatched) regions have comprise only the background. 

\begin{figure} [!h]
\centering\includegraphics [height=0.40\textheight]{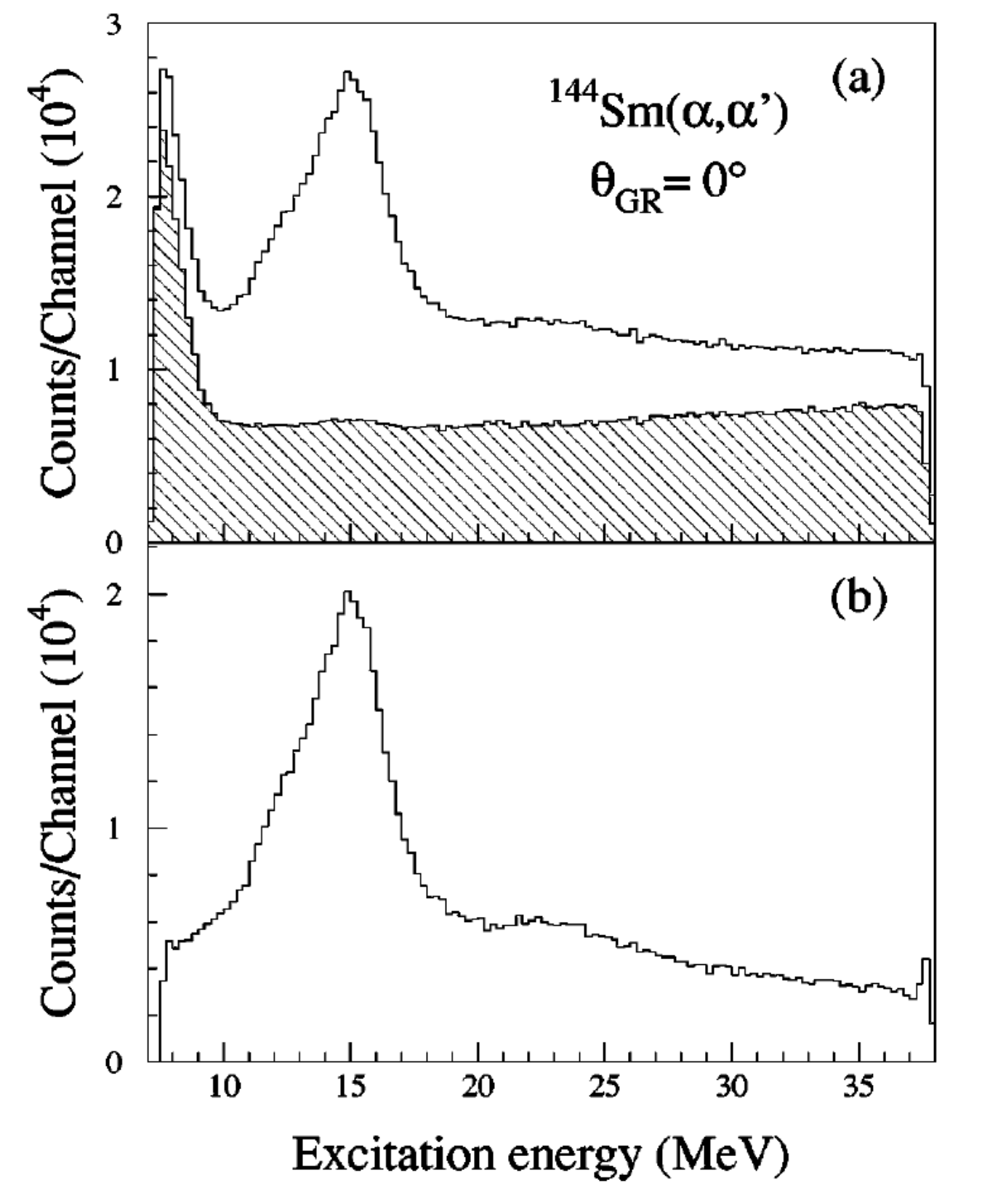}
\vspace{-0.5cm}
\caption{(a) Excitation-energy spectrum for $^{144}$Sm from the ``0$^{\circ}$'' RCNP ($\alpha, \alpha$') work obtained by gating on the central region of the vertical position in the focal plane. The hatched part corresponds to the off-center parts in Fig.~\ref{bgsubtraction}. (b) True excitation-energy spectrum after background subtraction. Figure from Ref. \cite{Itoh_prc2003}.}
\label{subtracted-spectrum}
\end{figure}

\begin{figure} 
\centering\includegraphics [height=0.40\textheight]{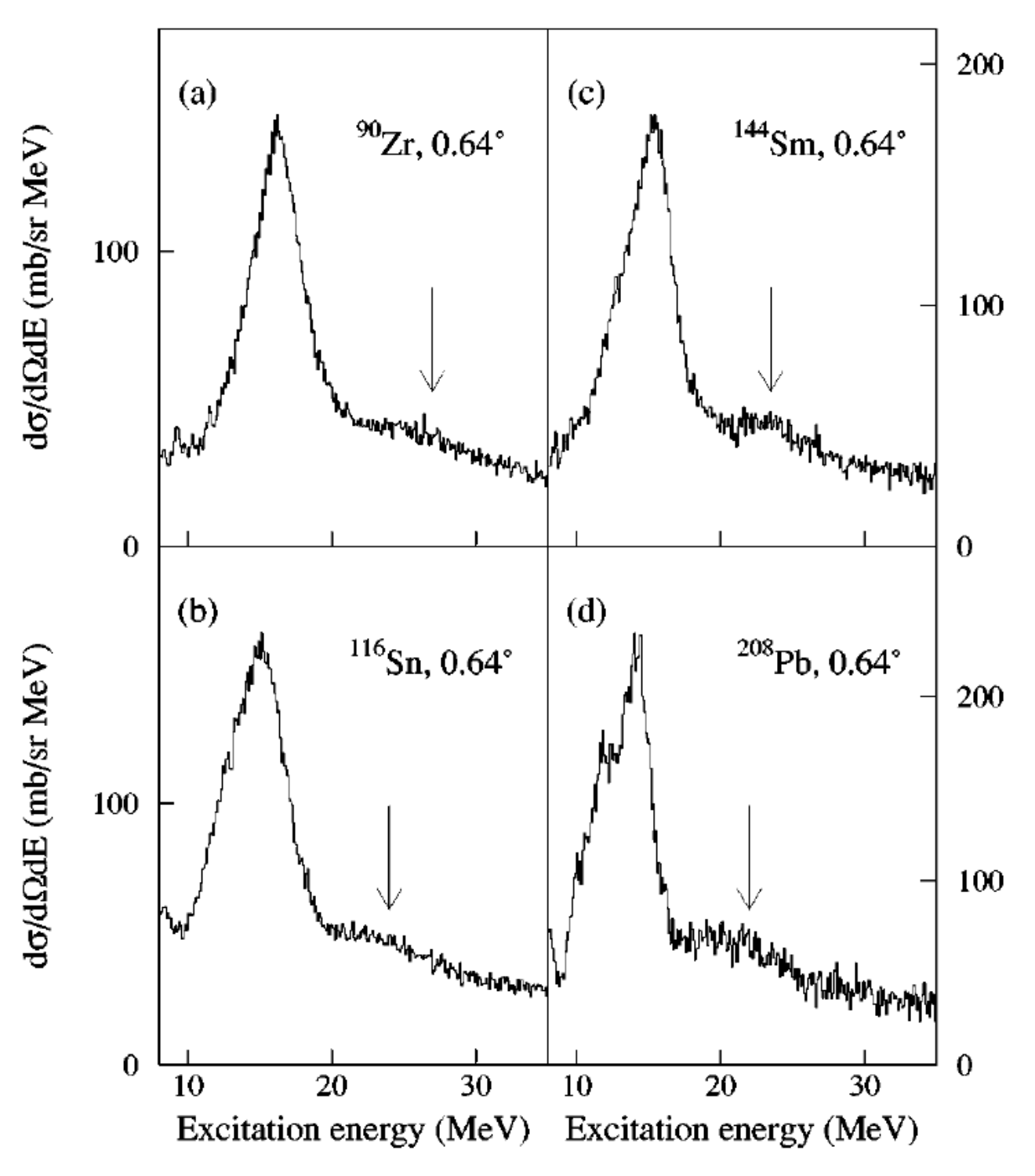}
\caption{``Backgroud-subtracted'' excitation energy spectra for $^{90}$Zr, $^{116}$Sn, $^{144}$Sm, and $^{208}$Pb from the RCNP ($\alpha, \alpha$') work at an averaged scattering angle of $\theta_\mathrm{avg}$ = 0.64$^{\circ}$.  The main ``bump'' consists, primarily, of ISGMR and ISGQR; the secondary ``bump'' is ISGDR+ISHEOR. The arrows indicate the location of the ISGDR as extracted in that work. Figure from Ref. \cite{Uchida_90Zr}}
\label{0degspectra}
\end{figure}

Fig. \ref{subtracted-spectrum} (a) shows the excitation-energy spectrum for $^{144}$Sm at 0$^{\circ}$ obtained from each region. The background spectrum has no discernible structures in the giant resonance region. Clean ``true'' spectra are obtained by subtracting
the instrumental background spectrum from the true+background spectrum, as shown in Fig. \ref{subtracted-spectrum} (b). This spectrum has, practically, no instrumental contributions to the ``background'' which now comprises the true nuclear continuum and, at the highest excitation energies, contributions from three-body channels, such as knock-out reactions, as mentioned earlier in the text. Representative ``background-subtracted'' inelastic scattering spectra for several nuclei from RCNP work are shown in Fig.~\ref{0degspectra}. Incidentally, a similar procedure was used many years ago in giant resonance studies at KVI, Groningen \cite{bran1}.

The backgound-subtracted spectra, so obtained, are used in a multipole-decomposition analysis (MDA) \cite{Bonin1984, Li_2010} to extract the multipole strength distributions. In the MDA process, the experimental cross-sections at each angle are binned into small (typically, $\leq$1 MeV) excitation energy intervals. The laboratory angular distributions for each excitation-energy bin are converted to the center-of-mass frame using the standard Jacobian and relativistic kinematics. For each excitation energy bin, the experimental angular distributions
$\frac{d\sigma^{\rm exp}}{d\Omega}(\theta_{\rm c.m.},E_x)$ are fitted by
means of the least-square method with the linear combination of
the calculated double-differential cross sections associated with different multipoles:
\begin{equation}
\frac{d^{2}\sigma ^{\mathrm{exp}} (\theta_{\mathrm{c.m.}}, E_{x})}{d\Omega dE}= \sum\limits_{L=0}^7 a_{L}(E_{x})\frac{d^{2}\sigma_{L}^{\mathrm{DWBA}}(\theta_{\mathrm{c.m.}}, E_{x}) }{d\Omega dE},  
\label{MDA}
\end{equation}
\noindent
where $a_{L}(E_{x})$ is the EWSR fraction for the $L^{\rm th}$ component, and $\frac{d^{2}\sigma_{L}^{\mathrm{DWBA}} }{d\Omega dE} (\theta_{\mathrm{c.m.}}, E_{x})$ is the cross section corresponding to 100\% EWSR for the $L^{\rm th}$ multipole at excitation energy $E_{x}$, calculated using the distorted-wave Born approximation (DWBA). The fractions of the EWSR, $a_L(E_x)$, for
various multipole components are determined by minimizing $\chi^2$.
This procedure is justified since the angular distributions are well
characterized by the transferred angular momentum $\Delta L$, according to the DWBA calculations for $\alpha$ scattering. For the limited angular range covered in these measurements, summation over $L\leq$7 is more than sufficient to extract the desired strength distributions; indeed, it is not possible to reliably extract the strength distributions for $L\geq$4 
over this limited angular range. The uncertainties in the $a_L(E_x)$ coefficients are estimated by changing the magnitude of the one component $a_{L}(E_{x})$, until refitting by varying the other components resulted in an increase in the $\chi^{2}$ by 1 \cite{YB_24Mg_1999, Itoh_prc2003}. 

The computer codes PTOLEMY \cite{ptolemy1, ptolemy2} and ECIS95 \cite{ecis} were used to perform the DWBA calculations, with the input values in PTOLEMY modified~\cite{Satchler1992} to take into account the correct
relativistic kinematics. The shape of the real part of the potential
and the form factor for PTOLEMY were obtained using the codes
SDOLFIN and DOLFIN \cite{dolfin}. The transition
densities and sum rules for various multipolarities employed in these calculations are obtained from 
Refs.~\cite{Harakeh_book, Satchler1987, Harakeh1981} and the radial moments obtained by numerical integration of the Fermi mass distribution using the parameters $c$ and $a$ from, for example, Ref. \cite{Fricke1995}.

Even though the $\alpha$ particle is isoscalar, the isovector giant dipole resonance (IVGDR) is excited at these beam energies via Coulomb excitation. The cross sections for IVGDR excitation increase with increasing beam energy and can be quite significant, especially for heavy target nuclei \cite{izumoto, shlomo-1}. Because its energy is nearly identical to that of the ISGMR, the IVGDR contribution has to be properly accounted for in the MDA. This is carried out \cite{Darshana2012, Li_2010} by employing IVGDR parameters 
from previously-known photonuclear cross-section data  \cite{Berman_1975} in conjunction with DWBA calculations based on the Goldhaber-Teller model to estimate the IVGDR differential cross sections as a function of scattering angle \cite{Satchler1987}.

To perform the DWBA calculations, one requires appropriate optical model parameters (OMPs). For this purpose, data are obtained for elastic scattering (and inelastic scattering to the low-lying states) over a wide angular range (typically, 0$^{\circ}$--30$^{\circ}$) and the OMPs extracted from fits to the angular distributions of differential cross sections of elastic scattering. 

\begin{figure}[h]
\centering\includegraphics [height=0.25\textheight]{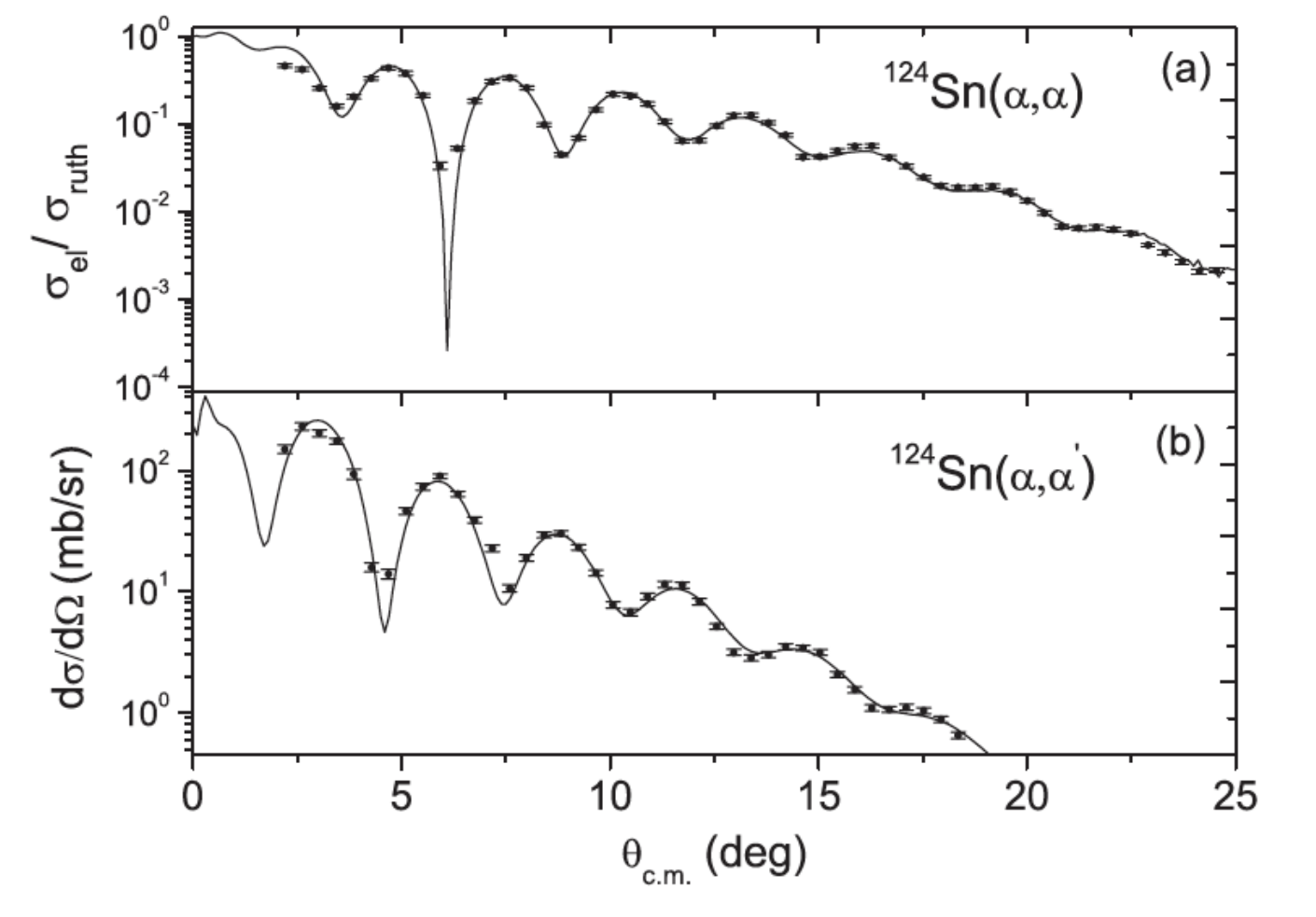}
\caption{(a) Angular distribution of the ratios of the elastic $\alpha$ scattering cross sections to the Rutherford cross sections for $^{124}$Sn at an $\alpha$ energy of 386 MeV. The solid line shows a fit from the optical model form given in the text. (b) Differential cross sections for excitation of the 2$^+$ state in $^{124}$Sn. The solid line shows the calculated cross sections for the state using the OMPs obtained from the fits to the data in (a) and the $B(E2)$ values from Ref. \cite{BE2_24Mg}. Figure from Ref. \cite{Li_2010}.}
\label{OMfits}
\end{figure}

\begin{figure*}[!ht]
\centering\includegraphics [trim= 0.6mm 0.3mm 0.5mm 0.5mm,
angle=360, clip, height=0.50\textheight]{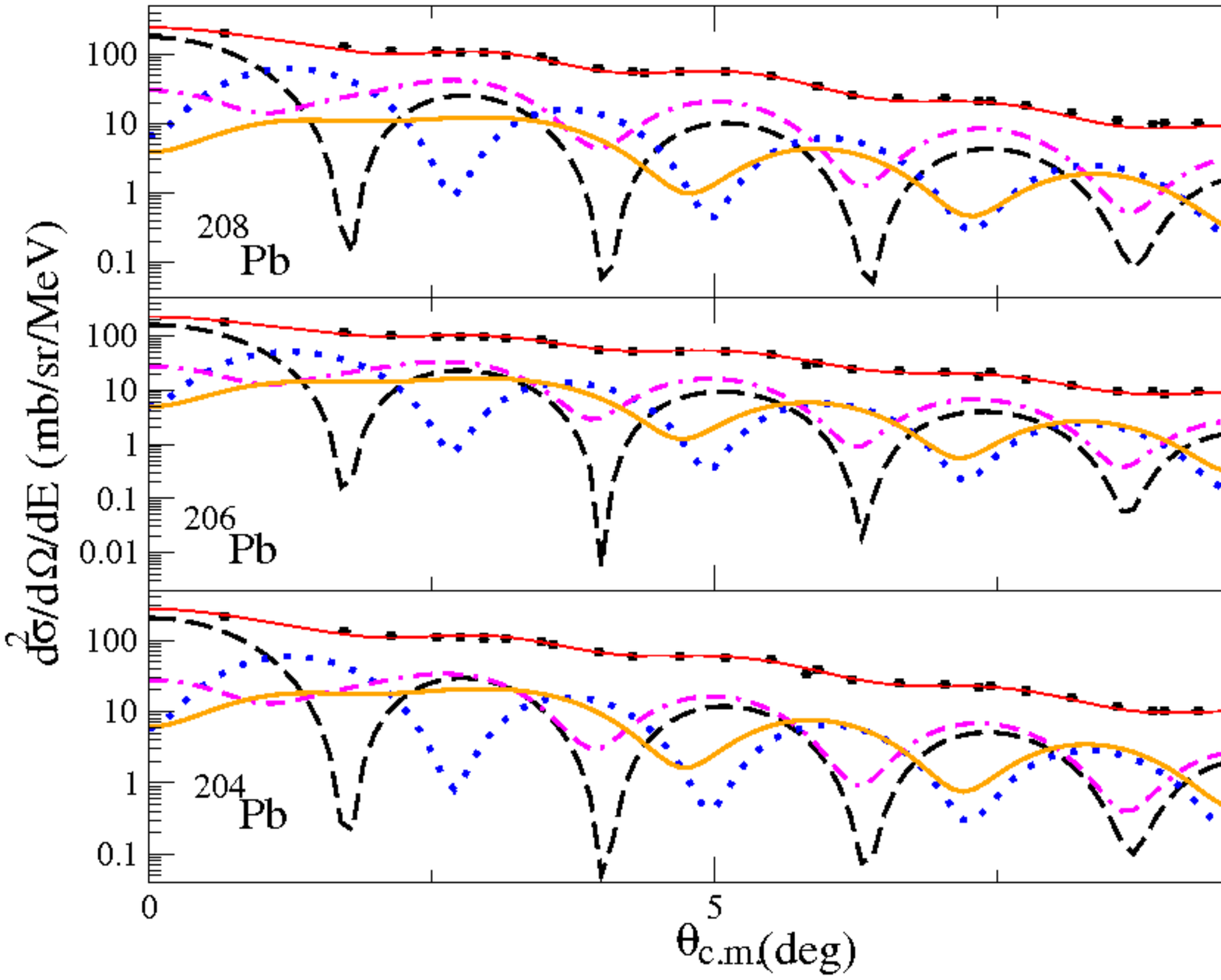}
\caption{(Color online) MDA fits to the experimental angular distributions for inelastic $\alpha$
scattering data for $^{204,206,208}$Pb for a 1-MeV energy bin centered at $E_x$=13.5 MeV. The thin solid line through the data shows the sum of various multipole
components obtained from MDA. The dashed, dotted, dot-dashed, and thick solid curves
indicate contributions from $L$ = 0, 1, 2 and 3, respectively. Figure from Ref.~\cite{Patel_Thesis}.}
\label{MDA_pb}
\end{figure*}

A ``hybrid'' optical-model potential (OMP) proposed by Satchler and Khoa \cite{Satchler_Khoa1997} has been employed in most of  
the RCNP and all TAMU measurements reported here. The real part of the optical potential is generated by single-folding with a density dependent Gaussian $\alpha$-nucleon interaction \cite{Li_2010}, and a Woods-Saxon form is used for the imaginary term. Thus, the total  $\alpha$-nucleus ground-state potential is given 
by:
\begin{equation}
U(r)=-V(r)-\it{i}W/\{1+\exp[(r-R_{I})/a_{I}]\}, 
\label{OpticalPot}
\end{equation}
where $V(r)$ is the real single-folding potential obtained using computer code SDOLFIN \cite{dolfin}  by
folding the ground-state density with the density-dependent
$\alpha $-nucleon interaction:
\begin{equation}
\mathrm{\upsilon_{DDG}}(\mathbf{r},\mathbf{r'},\rho)=-\upsilon[1-\beta \rho(\mathbf{r'})^{2/3}]\mathrm{exp}\left(-\frac{\mathbf{|r-r'|}^2}{t^2}\right). 
\end{equation}

\noindent
Here, $\mathrm{\upsilon_{DDG}}(\mathbf{r},\mathbf{r'},\rho)$ is the density-dependent $\alpha$-nucleon interaction, $\mathbf{|r-r'|}$ is the distance between center-of-mass of the $\alpha$-particle and a target nucleon, $\rho(\mathbf{r'})$ is the ground-state density of the target nucleus at a position $\mathbf{r'}$ of the 
target nucleon, $\beta$ = 1.9 fm$^{2}$, and $t$ = 1.88 fm. In Eq. (\ref{OpticalPot}), $W$ is the depth of the Woods-Saxon type imaginary part of the potential, with the radius $R_{I}$ and diffuseness $a_{I}$.

The imaginary potential parameters ($W$, $R_{I}$, and $a_{I}$), together with the depth of the real part, $V$, are obtained by fitting the elastic-scattering cross sections. The appropriateness of the OMPs so obtained is tested by calculating the cross sections for the low-lying 2$^+$ and/or 3$^-$ states using these parameters and the previously-known transition probabilities for these states \cite{BE2_24Mg, BE3}, and comparing those with the experimental values. Fig. \ref{OMfits} shows the fit to the elastic $\alpha$ scattering data from $^{124}$Sn and the comparison of the experimental and calculated differential cross sections for the first 2$^+$ state in $^{124}$Sn as an illustrative example of this procedure. Note that there is no ``fitting'' involved in Fig. \ref{OMfits}(b) and the calculation is performed for the adopted value for the $B(E2)$ from Ref. \cite{BE2_24Mg}.

We note here that in some cases of RCNP work \cite{Uchida_PLB2003,Uchida_90Zr,Itoh_plb,Itoh_prc2003}, the single-folding model was used, in the same form, for both the real and imaginary parts; the final strength distributions were not appreciably different from those obtained with the aforementioned ``hybrid'' model, however. A similar ``non-hybrid'' model was used also by the TAMU group in analysis of $^{6}$Li-scattering data \cite{DHY2009_MgSi,DHY2009_Sn}.

MDA fits for the energy bin at an excitation energy of 13.5 MeV in $^{204,206,208}$Pb are shown in Fig.~\ref{MDA_pb} for 386-MeV inelastic scattering data from RCNP \cite{Patel_Thesis}; the contributions from the $L$ = 0, 1, 2, and 3 multipoles are also shown.

\begin{figure}[]
\centering\includegraphics [height=0.40\textheight]{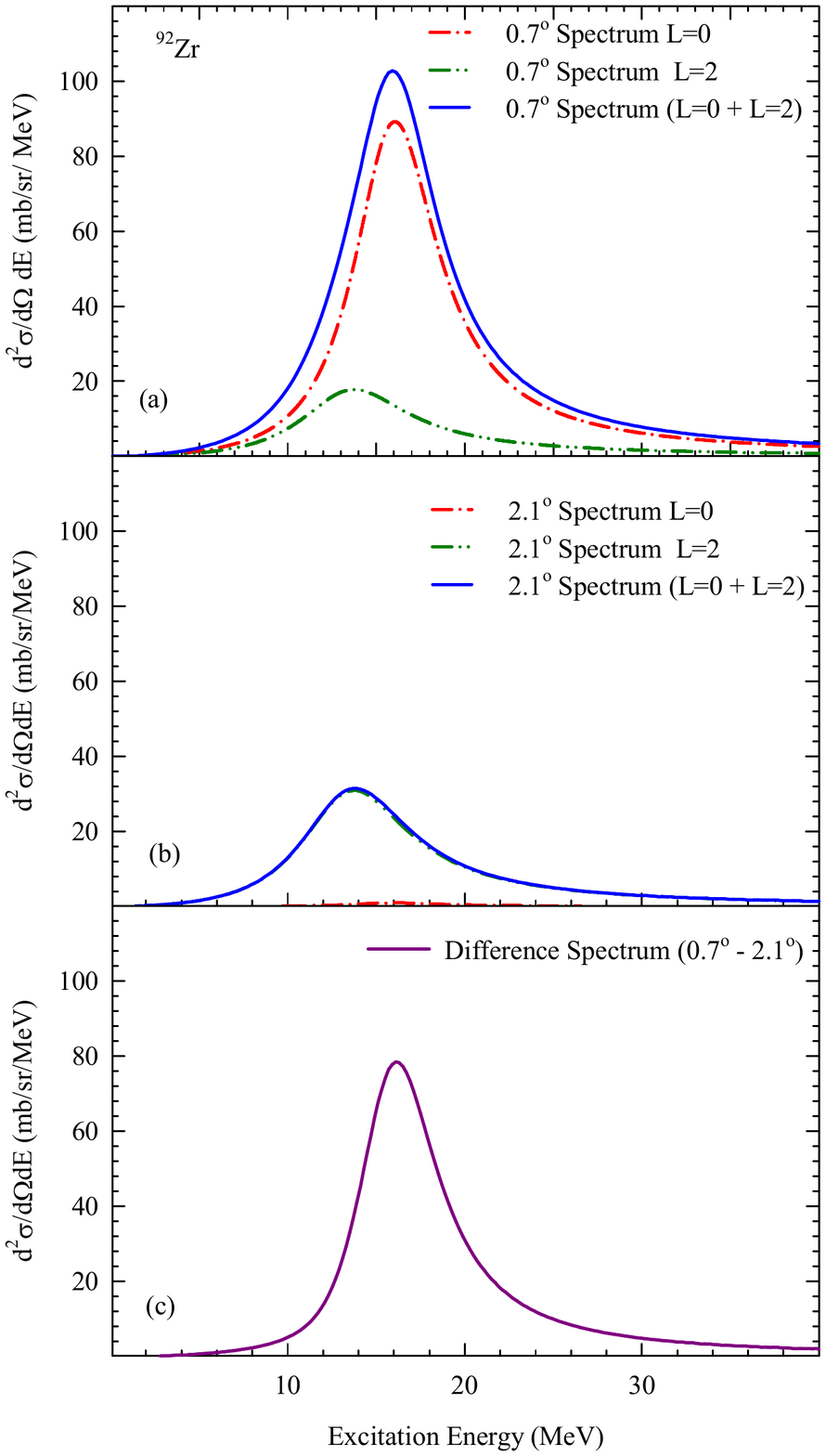}
\caption{
(a) Expected $L$=0, $L$=2 strengths for 100\% EWSR excitation in 386-MeV inelastic $\alpha$ scattering off $^{90}$Zr at 0.7$^{\circ}$. The sum of the two strengths (basically, the inelastic scattering spectrum at this angle) is also shown.
(b) Same as (a) but for 2.1$^{\circ}$. Note that the ISGMR contribution is barely discernible at this angle. (c) The ``difference spectrum'' obtained by subtracting the total ($L$=0 + $L$=2) spectrum in (b) from that in (a). The total spectrum in (c) is essentially the ISGMR.}
\label{subtr}
\end{figure}
The strength distributions for the various multipoles are obtained by multiplying the extracted $a_{L}(E_{x})$'s by the strength corresponding to 100\% EWSR at the given energy $E_x$ \cite{Harakeh_book, Satchler1987, Harakeh1981}:
\begin{equation}
S_0(E_x)=\frac{2\hbar^2A<r^2>}{mE_x}a_0(E_x), 
\end{equation}
in the case of monopole [cf. Eq. (\ref{eq:monopoleEWSR})],
\begin{equation}\label{eq:Sdipole}
S_1(E_x)=\frac{3\hbar^2A}{32\pi mE_x}[11<r^4>-\frac{25}{3}<r^2>^2 
-10\epsilon<r^2>] a_1(E_x), 
\end{equation}
in the case of IS dipole [cf. Eq. (\ref{eq:EWSRdipole_final})\footnote{The extra factor 1/4 mentioned in the footnote of p. 4, is here
included, as well as the factor 3 associated with the sum over the three $M$-components (mentioned at the end of Sec. \ref{general}).}], and
\begin{equation}
S_{L\geq2}(E_x)=\frac{\hbar^2A}{8\pi
mE_x}L(2L+1)^2<r^{2L-2}>a_2(E_x), 
\end{equation}
in the case of the higher IS multipoles.
Here, 
$<r^N>$ 
is the $N^{\rm th}$ moment of the ground-state density, while 
$\epsilon$ has been defined in Eq. (\ref{eq:epsilon}) as 
$\epsilon$=(4/$E_{\rm ISGQR}$+5/$E_{\rm ISGMR}$)$\hbar^2$/3$mA$. 
The centroid energies of the ISGMR and the ISGQR are generally taken as
80~A$^{-1/3}$ MeV and 64~A$^{-1/3}$ MeV, respectively.

\begin{figure}[h]
\centering\includegraphics [height=0.35\textheight]{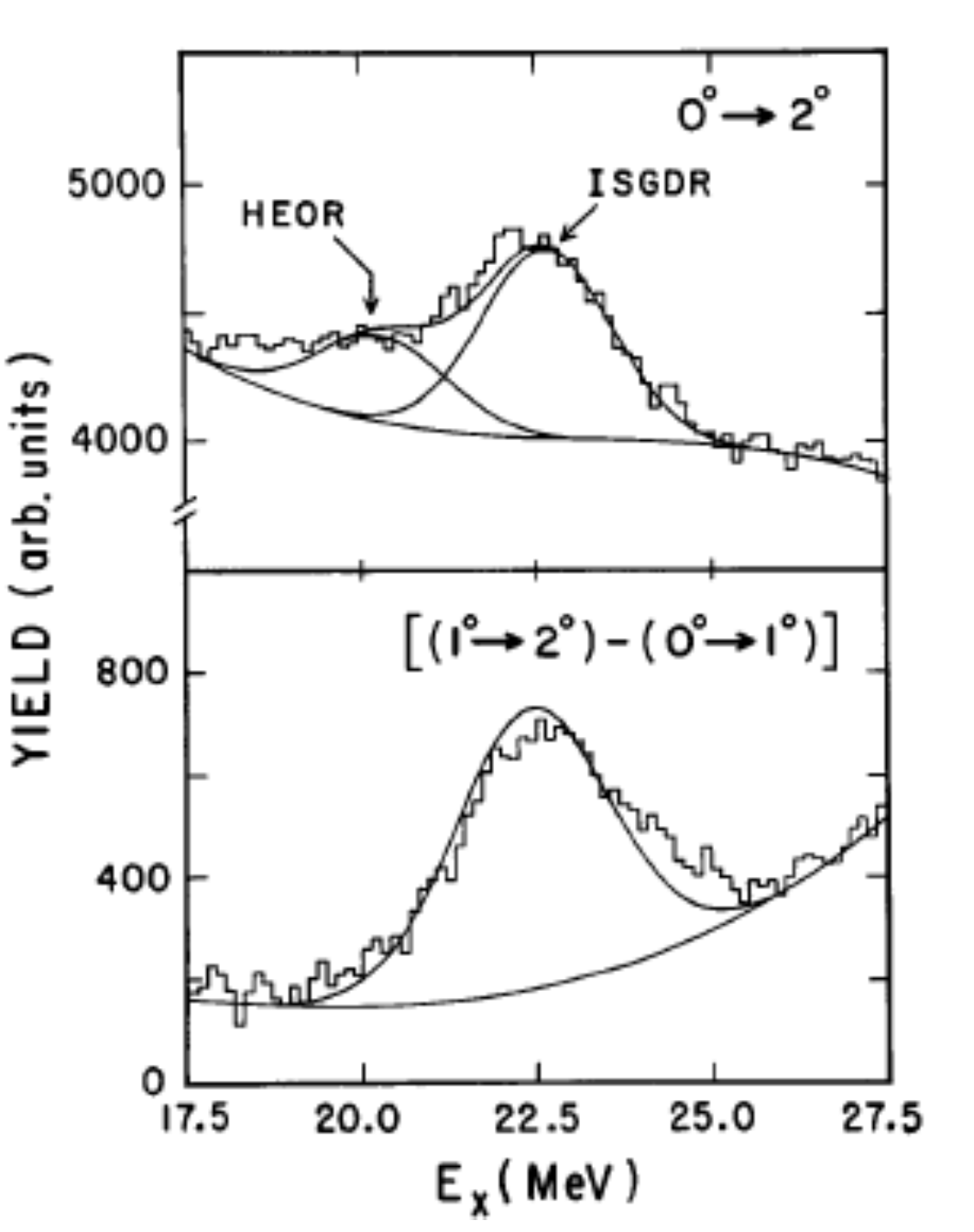}
\caption{(a) Inelastic $\alpha$-scattering spectra for $^{208}$Pb for (0$\pm$2)$^{\circ}$. A two-peak + polynomial-background fit to the data
is shown superimposed with the peaks corresponding to the HEOR and the ISGDR indicated.
(b) The ``difference spectrum''. A fit using peak parameters identical to those in (a) is also shown; note that the fit corresponds to no HEOR strength. Figure from Ref. \cite{benny}.}
\label{isgdr_benny}
\end{figure}

In the absence of data at a sufficient number of angles to perform a proper multipole decomposition analysis, it is possible still to identify the position and width of the ISGMR and ISGDR from a ``difference of spectra'' procedure, if data is available for angles at and near 0$^{\circ}$ (see, for example, 
Ref.~\cite{bran1}). The premise behind this technique is simple: The angular distribution of the ISGMR is maximal at 0$^{\circ}$ and declines sharply to a minimum (see Fig. \ref{angdist}, where this minimum is at about 2$^{\circ}$). The angular distribution of the ``competing'' ISGQR, on the other hand, remains essentially flat over this angular range. Indeed, this is mostly true for all other multipoles, as well as the ``background''. So, the ``difference-spectrum'', obtained from subtracting the inelastic scattering spectrum at the first minimum of the expected ISGMR angular distribution from that at 0$^{\circ}$ (where the ISGMR strength is maximal), essentially represents only the ISGMR strength. This is shown schematically in Fig. \ref{subtr}. The same holds in comparing the angular distributions of the ISGDR and the ISHEOR and this technique was used, for example, in the first clear identification of the ISGDR (see Fig.~\ref{isgdr_benny}) \cite{benny}. Later in this article, this technique will be referred to in discussing some aspects of ISGMR and ISGDR.

%% file: theory.tex
\section{Theoretical models}\label{sec:theory}

\subsection{RPA calculations: status}

As recalled in the Introduction, GRs are collective modes and,
because of their small-amplitude character, they are good examples
of harmonic motion. Consequently,
linear response theory describes their properties rather well in terms of coherent 
superpositions of 1 particle-1 hole (1p-1h) configurations. In nuclear structure, self-consistent 
linear response theory 
is called Random Phase Approximation or RPA. This theory is well described in 
textbooks \cite{Ring-Schuck:1980,Rowe:1970}. 

Although various phenomenological versions 
of RPA exist, only self-consistent RPA based on an effective 
Hamiltonian $H$ or an Energy Density Functional (EDF) is relevant to connect
the ISGMR energy to the nuclear incompressibility. Effective Hamiltonians,
of the type $H_{\rm eff} = T + V_{\rm eff}$, can be used to calculate 
the total energy on the most general Slater determinant $\vert 
\Phi \rangle$, in the form
\begin{equation}
E[\rho] = \langle \Phi \vert H_{\rm eff} \vert \Phi \rangle.
\end{equation}
As our writing implies, the total energy turns out to be (only) 
a functional of the one-body density $\rho$. If we introduce, just for
the sake of clarifying the nomenclature, the energy density $\cal E$, then
\begin{equation}
E[\rho] = \int d^3r\ {\cal E}(\rho).
\end{equation}
Thus, we can also say that we have built a functional of the energy density,
that can depend on the particle density at the same point (local
functional) or at different points (non-local functional): hence, the
name of EDF. Effective Hamiltonians can be seen as EDF generators or,
alternatively, EDFs can be written directly. In principle, one could
write the most general EDF which is consistent with the symmetries of
the system and fit the associated free parameters on observables; in
practice, one has to choose an ansatz (that is, use either a nonrelativistic
or a covariant form, and reduce the terms consistent with symmetries
to a number that is numerically tractable) and a set of observables.
The systematic and statistical errors associated with these choices
are subject of great interest at present \cite{errors}; while the community 
is aiming at finding a universal EDF, there is not a clearly systematic
procedure that leads to it. On these subjects, the literature is huge
but the reader can start from the review paper \cite{Bender}.  
For the following discussions, the main classes of EDFs, that are
Skyrme, Gogny and relativistic functionals, will be relevant (and
in particular RPA built with them). The issue of the pairing part
of an EDF will be also touched. 

\begin{figure}[hbt]
\centering\includegraphics [height=0.35\textheight]{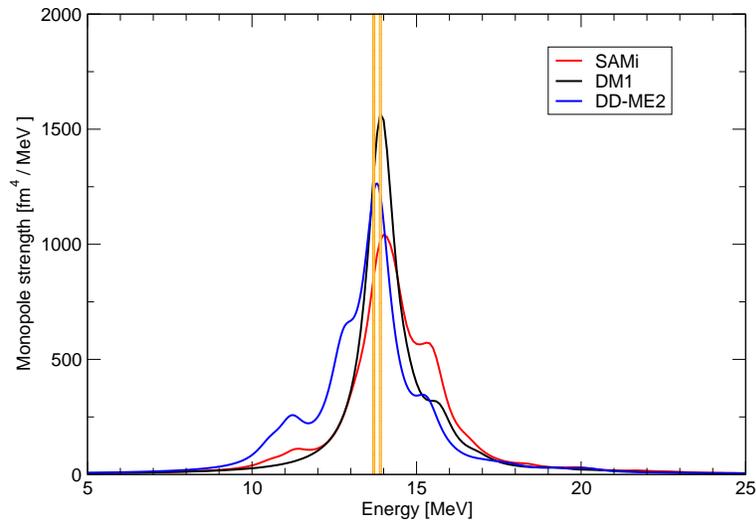}
\caption{RPA calculations of the monopole strength in $^{208}$Pb, 
performed by using the nonrelativistic Skyrme-type functional 
SAMi \cite{RocaMaza2012}, the Gogny-type functional DM1 
\cite{Goriely2009}, and
the relativistic functional DDME2 \cite{Lalazissis2005}.
The vertical lines show the peak energy obtained in the experiments
performed at TAMU (13.9 MeV) and RCNP (13.7 MeV). 
}
\label{gmr_rpa}
\end{figure}

Spherical RPA is now a standard tool. The nonrelativistic Skyrme-RPA implementation on a basis has 
been published and is available in Ref. \cite{Colo:2013}; codes exist, and have been largely exploited, 
in the case of Gogny and relativistic functionals as well. 
An illustrative example of the monopole response in $^{208}$Pb is shown 
in Fig. \ref{gmr_rpa}. 
Standard codes, either using Skyrme, Gogny or relativistic functionals,
do implement full self-consistency at present. 
In Ref. \cite{Sil:2006}, the effect of lacking such full self-consistency
has been carefully checked. It has been seen that dropping the two-body
spin-orbit and Coulomb residual interaction in RPA can produce a shift 
of the ISGMR energy by 100-700 keV (even larger in few cases). 
As we dicuss below, this amounts to an error of up to 10\% in the extraction
of the nuclear incompressibility. Similar effects have been found in the
case of the Gogny force by the authors of Ref. \cite{DeDonno:2014} who have, 
however, considered the case of electromagnetic excitations and not the
ISGMR. 

In the case of exotic nuclei, in which nucleons occupy
more and more weakly bound levels, a proper treatment of the continuum 
is appropriate. More precisely, this is mandatory for low-lying states,
and could be less crucial for high-lying GRs.
For Skyrme forces, 
continuum-RPA in the cooordinate space can be exactly formulated but the practical implementations are
much less than the discrete implementations \cite{HSZ:1996} and they lack full self-consistency. 
By using Gogny forces, it has been seen \cite{DeDonno:2011} that, 
for the case of medium-light neutron-rich nuclei like 
$^{24}$O or $^{52}$Ca, the continuum and discrete RPA results for
high-lying GRs differ by 200-700 keV. The comparison of discrete and
continuum RPA in the case of relativistic functionals has been carried
out in Ref. \cite{Daoutidis:2009} and the results are not too much different than
those in the case of Gogny. A stable nucleus, $^{40}$Ca, has been analysed 
and the difference between the ISGMR energies
calculated either with continuum or discrete RPA is around 200 keV. 

Open shell nuclei require the extension of RPA to Quasiparticle RPA (QRPA). Fully
self-consistent QRPA calculations have been published for spherical nuclei 
using different kinds of functionals \cite{Terasaki:2005,Carlsson:2012,Peru:2005,Paar:2003} 
and, to a lesser extent, also for axially deformed nuclei \cite{Terasaki:2010,
Yoshida:2013,Peru:2008,Arteaga:2008}. In QRPA, one sees the effect of pairing
correlations, which are usually very strong for low-lying states and weak for
GRs, in keeping with the fact that pairing is restricted to a narrow
window around the Fermi energy. Weak pairing effects can nevertheless be of interest
for the precision physics of GRs; in the case of the ISGMR, for instance, one expects that even
shifts of the resonance peak of few hundreds of keV can impact the extraction of
the nuclear incompressibility, as we have already mentioned. In QRPA on top of Hartree-Fock-Bogoliubov (HFB) 
calculations for the ground-state, there are two effects that show up: (1) The p-h 
excitations are replaced by two-quasiparticle excitations that are as a rule
larger; however, this effect is of the order of $\approx$ 2$\Delta$ (where $\Delta$ is
the ground-state pairing gap) close to the Fermi energy but tends to become
negligible when goes far from it; and, (2)  The residual interactions is supplemented by
a pairing contribution which is attractive in the 0$^+$ channel. For this reason, 
it has been found that pairing tends to lower the ISGMR
energies in the Sn isotopes \cite{Li:2008}. More systematic inverstigations 
\cite{Vesely:2012,Avogadro:2013}, 
and their impact on our understanding of the nuclear incompressibility, will
be discussed below. 

In axially deformed nuclei, one expects a splitting of the ISGMR peak. This fact 
is related to the coupling with the ISGQR. We can understand this in simple terms, if 
we consider that only the projection $K$ of the total angular momentum $J$ on 
the symmetry axis is a good quantum number in the case of axially symmetric nuclei.
Therefore, in such nuclei the monopole is coupled with the $K=0$ component of the
ISGQR and this mechanism produces a double peak. In this respect, the splitting
is not simply a function of the deformation parameter $\beta_2$ as in the case
of the IVGDR, but the coupling matrix elements play a role as well. This mechanism
has been studied in the past by using simple models and assumptions. In Ref.
\cite{Abgrall:1980} it has been estimated, based on the cranking model together with
the harmonic and scaling approximations for the vibrational modes, that for
a well deformed nucleus with $\beta_2$ = 0.3 the splitting would be of the
order of $\approx$ 30 A$^{-1/3}$ MeV and, therefore, observable in medium-heavy nuclei
where the width should be smaller. We will discuss below to which extent these
expectations are fulfilled by measurements and microscopic calculations.

Before ending this section, a word should be said about time-dependent HF calculations (TDHF) 
of the ISGMR. In principle, RPA is the small-amplitude limit of TDHF. Therefore,
TDHF would be the appropriate tool to see, at least theoretically, if there is
evidence of large amplitude fluctuations in the monopole case in some cases. Another
advantage of TDHF is that it can be formulated using exact continuum. A longstanding
drawback has always been the fact that in TDHF calculations performed in a finite box, 
it is not straightforward to avoid the reflection of the time-dependent wave function
at the box boundary, which causes a non-physical distortion of the evolution of emitted
wave function components. The most recent attempt to build a continuum TDHF is reported in
Ref. \cite{Pardi:2010}. 

\begin{figure}[hbt]
\centering\includegraphics [height=0.35\textheight]{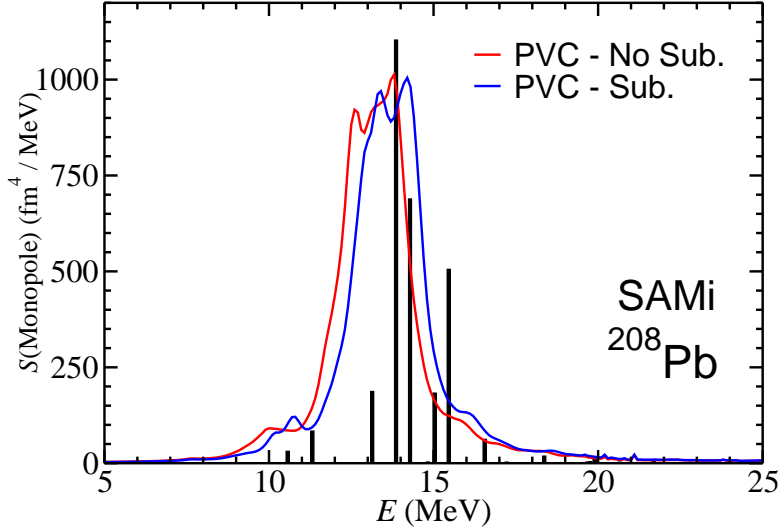}
\caption{Calculation of the ISGMR strength function in $^{208}$Pb 
performed with the Skyrme interaction SAMi \cite{RocaMaza2017}. The
bars correspond to the RPA result while the blue (red) lines
correspond to PVC calculations with (without) subtraction. 
See the text for a discussion.
}
\label{gmr_pvc}
\end{figure}

\subsection{Calculations beyond RPA (2p-2h, particle-vibration coupling)}

RPA calculations take into account the fact that nuclei may not display
a single monopole peak, that is, they include the so-called Landau damping
or fragmentation of the strength coupled to the underlying 1p-1h 
states. They also include the escape width $\Gamma^\uparrow$, which is
associated with particle decay into the continuum; the latter is important in
light nuclei but not in heavier ones. The major source of broadening
for the giant resonance strength is the coupling with more complex
states than the simple 1p-1h (like 2p-2h, 3p-3h ... etc.) that gives
rise to the so-called spreading width $\Gamma^\downarrow$.

Including 2p-2h coupling amounts to solving the so-called second 
RPA (SRPA) equation. This framework was estabilshed long ago, 
and one can find many details, as well as calculations for the monopole
strength that employ a phenomenological Woods-Saxon mean-field plus
Landau-Migdal residual forces, in Ref. \cite{Wambach}. SRPA 
calculations for the ISGMR can be also found in Ref. \cite{Migli1994}. 
In all these cases, approximations with respect to the exact SRPA 
framework have been necessary for computational reasons; the most relevant one is the so-called
``diagonal'' approximation, that is, 2p-2h configurations interact 
with the 1p-1h ones but their mutual interaction is neglected. 
Fully self-consistent SRPA calculations without the diagonal
approximation are extremely demanding from the computational point of
view, and this explains why they have become available only after
a significant temporal gap. The most recent SRPA implementation of 
this type, 
with Skyrme, has been reported in \cite{Gambacurta2015} (see also 
Ref. \cite{Gambacurta2016}).

An analogous and yet different approach is the one based on
the particle-vibration coupling (PVC) idea. The RPA 1p-1h states
are coupled at lowest order with low-lying collective vibrations. 
This approach has been pioneered by the Copenhagen \cite{BohrMott} and
Dubna \cite{Solovev} schools. As in the case of SRPA, one can find
works based on the use of Woods-Saxon mean-field plus Landau-Migdal
residual forces \cite{Kamerdzhiev1994,Kamerdzhiev1997}. Skyrme calculations have 
been performed for the first time in Ref. \cite{Colo1992}. 
Approximations that were mandatory at that time, have been removed
in Ref. \cite{RocaMaza2017}: these are the most recent
self-consistent RPA plus PVC calculations of the ISGMR, together
with the very similar ones reported in Ref. \cite{Lyutorovich2015}.
Extensions to the case with pairing, namely to QRPA plus PVC, are
also available in Ref. \cite{Tselyaev2009}.

All these calculations are quite successful in reproducing most of the spreading width of
the ISGMR, especially in heavy nuclei. 
An example for the case of $^{208}$Pb can be seen in Fig. 
\ref{gmr_pvc}. 
Nevertheless, the precise assessment of the convergence of these 
calculations with respect to the phonon (i.e. vibration) model space 
is still a matter of discussion \cite{Tselyaev2017}. 
Another issue is their relevance in connection 
with the extraction of $K_\infty$. It has been suggested in 
Ref. \cite{Tselyaev2007} that the so-called 
``subtraction method'' should be a way to implement the RPA plus PVC 
and obtain the spreading of the giant resonance strength without
an artificial downward shift of the centroid. This 
optimistic view has been partly questioned in \cite{RocaMaza2017}. 
In the case of the ISGMR, due to cancellations between the different
diagrams that contribute to the shift, this turns out to be not
too large. This can be seen in Fig. \ref{gmr_pvc}. 

Thus, we may have a tentative, positive answer but not a firm, conclusive 
one to two basic questions: The first one
is whether the subtraction method or some other method allows 
to use at the level of RPA plus PVC the same interaction that has been used
at the level of RPA. The second and related question is in which way the
calculations reported in this subsection, that are undeniably 
superior to RPA in keeping with the reproduction of 
the spreading of the experimental
strength function, are relevant to the extraction of $K_\infty$. The ideal
situation would be the one in which the PVC corrections produce a 
realistic width, without moving down the centroid with respect to 
RPA, so that the centroid energy can still be correlated with $K_\infty$.
This situation is realised quite well in the case shown 
in Fig. \ref{gmr_pvc} and discussed in more detailed in Ref. 
\cite{RocaMaza2017}. The centroid energy is 13.7 MeV in RPA 
and the same within PVC with subtraction, while it moves slightly
down to 13.4 MeV without subtraction. It has to be clarified whether
this is general enough to be assumed as a guideline for the future.

\subsection{{\it Ab initio} calculations for light nuclei}

We should remind the reader about the present status of nuclear structure theory. There are several attempts 
to derive properties of finite nuclei starting, as is said, {\it ab initio}. This 
wording has been recently the subject of some debate. If we mean by this a genuine derivation of 
nuclear properties from 
QCD, two paths are currently pursued. One strategy consists in extracting the properties
of nuclear uniform matter (mainly the saturation properties and the equation of state) 
and the properties of few-body systems from lattice calculations with explicit quark degrees of freedom. 
This approach is still striving to reproduce basic properties like binding energies: 
nuclear matter is still underbound, as well as the $^4$He system which displays a binding 
energy of about 5 MeV \cite{Inoue:2013}. Progress in the predictive power of this approach 
is expected in the coming years. Another strategy is based on the use of an {\em effective} 
realization of the QCD Lagrangian, using ideas about chiral symmetry that were
originally proposed by S. Weinberg. Thus, chiral effective field theories (EFT) model nuclei 
in terms of nucleons and not of quarks; nonetheless, their dynamics is governed by a Lagrangian that is constrained by the symmetries of 
QCD and its low-energy behavior, namely one including explicit pionic exchanges; in this case, 
the most advanced implementations of nucleonic lattice simulations lead to good results for 
the ground-state properties of $N=Z$ light nuclei. In principle, these lattice simulations
could be used to look at the radial monopole excitations of those light nuclei; in practice, 
only the Hoyle state of $^{12}$C has been addressed. 

Before the advent of these approaches, the name {\it ab initio} had been used for several decades 
to mean increasingly sophisticated many-body approaches based on point-like nucleons and bare 
two-body and three-body potentials acting them: these can be either phenomenological potentials 
or meson-exchange ones. Among the many-body approaches, we should mention variational methods,
Green's function techniques, the Coupled Clusters method, and in-medium Renormalization Group approaches. 
For quite a long time, the progress along these lines had been slow but recently it has speeded up
so that one can start discussing if and how collective states can be attacked by {\it ab initio} 
approaches.

In Ref. \cite{Bacca:2013} it has been shown that, within an accurate {\it ab initio} framework,
different Hamiltonians including two- and three-body interactions provide very different results
for the transition form factor to the 0$^+$ resonance, while they provide essentially the same
in the case of the elastic form factor. It should be stressed that the results for the transition 
form factor do also all disagree with the experimental findings from $(e,e^\prime)$ scattering.
The resonance under study lies at -8.20 $\pm$ 0.05 MeV, and has a width of 270 $\pm$ 50 keV. 
The obvious question is whether such a state bears any resemblance with the ISGMR in medium-heavy
nuclei. In Ref. \cite{Bacca:2015}, three criteria have been proposed 
to support this resemblance (the peak dominance, the EWSR exhaustion, and
the form of the transition density) and deemed to have been satisfied.
However, the $\alpha$ particle is a very light object and it turns
out to be impossible to relate its incompressibility to that of nuclear
matter.

\subsection{Cluster models}

Especially in light nuclei, some cluster structures are likely to
show up as an effect of the low density. Some authors have argued 
that in these nuclei monopole or isoscalar dipole states may arise, which 
are made up not by the collective vibration of nucleons but by collective
vibrations of nuclear clusters with respect to each other. One candidate
is $^{24}$Mg, where it has been suggested that some peaks
of the ISGMR strength are due either to $^{20}$Ne + $\alpha$, or to $^{12}$C + 
$^{12}$C vibrations, or even to a more exotic vibration of a pentagon
of $\alpha$ clusters (with an additional $\alpha$ lying in the center) 
\cite{Kimura2015}. Similar predictions have been made for some
of the peaks appearing the ISGDR strength of $^{20}$Ne and 
$^{44}$Ti \cite{Kimura2016} (see also \cite{Kimura2017-1,Kimura2017-2}). 

If one assumes a cluster configuration, the IS monopole or dipole 
transitions from the ground state can be estimated analytically 
within a geometrical model based on the masses and the radii of the clusters.
These values of the transition strengths can be confronted with the
single-particle estimates (W.u.) and they may be comparable. 
However, this is simply a very rough indication that cluster vibrations
might be excited. The predictions in the references discussed 
in the previous paragraph have been made by using the Antisymmetrized
Molecular Dynamics (AMD) model. This is a variational model in which
the total wave function is taken as a product of Gaussian wave 
packets, with global antisymmetrization as the fermionic nature of
the particles requires. The molecular dynamics simulations can be
carried out once an effetive interaction is plugged in. In the cases
at hand, some parameterization of the Gogny interaction is
employed. 

While these works present some interest, it should be said that we
still lack any experimental evidence (like $\alpha$-decay
data) that can support the speculations of monopole and dipole
states as cluster vibrations.

%% file: experiment_results_2.tex
\section{Experimental Results}\label{experiment_results}

\subsection{The ISGMR}
\begin{figure*}[!h]
\centering\includegraphics [height=0.48\textheight]{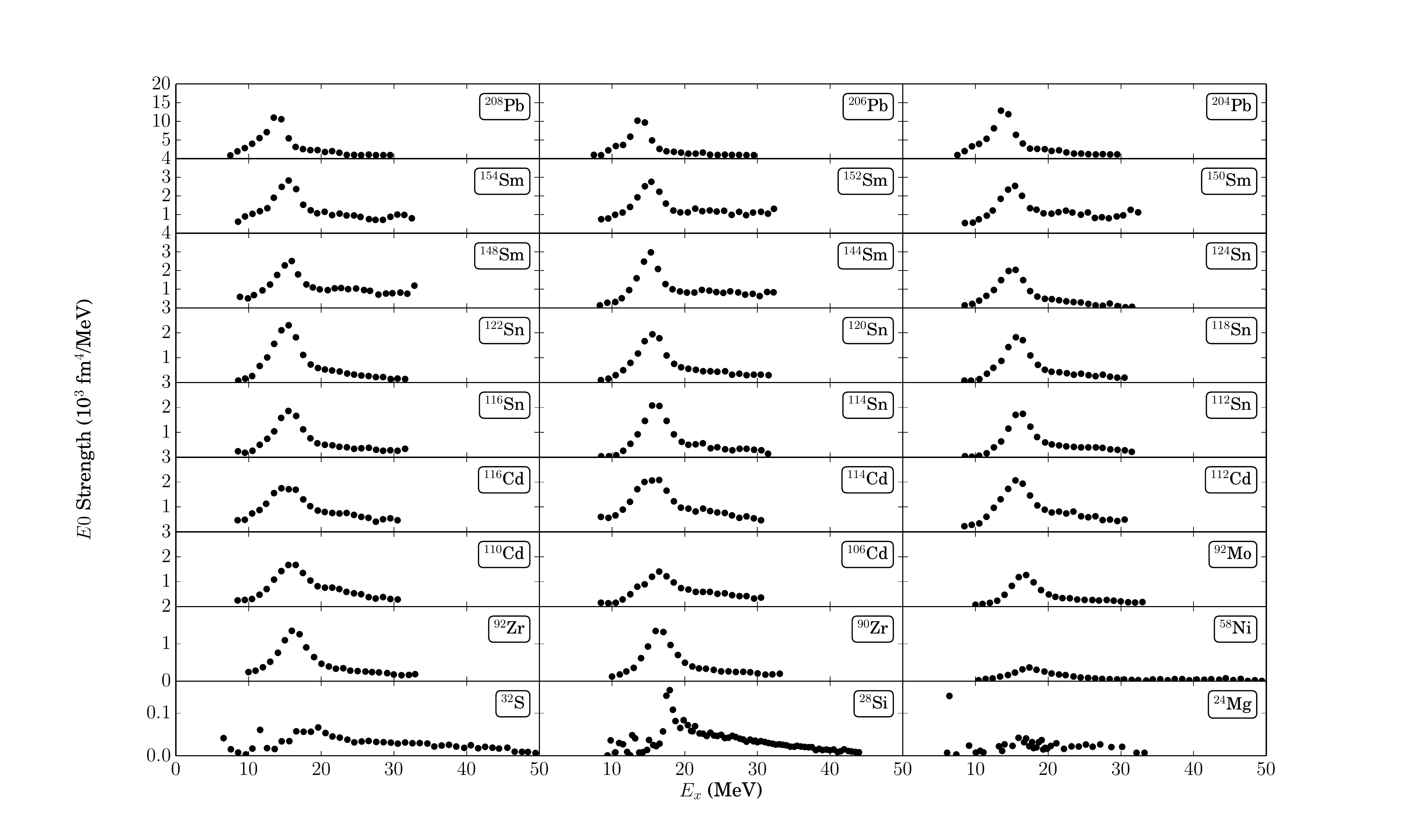}
\caption{ISGMR strength distributions for various nuclei, extracted in the RCNP work. The data are from Refs. \cite{Darshana2013,Itoh_prc2003,Li_PRL2007,Li_2010,Darshana2012,YKGPLB2016,nayak58ni,Itoh_32S,tom-si,yogesh-mg2,yogesh-mg}.}
\label{ISGMR}
\end{figure*}

Experimental strength distributions for ISGMR, based on multipole decomposition analyses, have now been obtained for several nuclei over the range A=24--208. The results from the RCNP work, from small-angle inelastic scattering of 400-MeV $\alpha$-particles are presented in Fig. \ref{ISGMR}. In practically all cases, the ISGMR appears as a single peak exhausting within it nearly 100\% EWSR for $L$=0. (In deformed nuclei, the ISGMR strength distribution has a two-peak structure, as already mentioned.) The associated peak parameters, as well as the various commonly-used moment ratios of the strength distributions, where available, are presented in Table~\ref{ISGMR-table}. The table is comprehensive in that it contains not only the results from RCNP work with $\alpha$ particles, but also their work with the deuteron probe, and the TAMU work with $\alpha$ particles and $^6$Li ions.


A comment about the small, near-constant ISGMR strength at higher excitation energies observed in most of the RCNP work is in order. [A similar effect is observed in the ISGDR strength distributions as well (see below).] This strength is spurious and, in some ways, points to a limitation of the multipole decomposition analysis. While the correct {\em raison d'\^{e}tre} of this extra strength is not quite well understood, this may be attributed, quite reasonably, to contributions to the continuum from three-body channels, such as knockout reactions and quasi-free processes \cite{bran1,nayak58ni}. In the MDA procedure, the continuum underlying the giant resonances is assumed to be composed of contributions from higher multipoles (hence the inclusion of multipoles up to $L$=7, typically, in the MDA). 
The aforementioned three-body processes, which also are forward-peaked in terms of angular distributions, are implicitly included in the MDA as background and may mimic the $L$=0 (and $L$=1) angular distributions, leading to such spurious multipole strengths at higher energies where the associated cross sections are very small.
This conjecture is supported by measurements of proton decay from the ISGDR 
at backward angles wherein no
such spurious strength is observed in spectra in coincidence with the decay
protons~\cite{garg-rev1,hun1,nayak2,hun3}; quasifree knockout results in protons that are forward peaked. This problem 
does not exist in the A \& M work because, as mentioned previously, they subtract all ``background'' from their spectra before carrying out MDA. However, a similar
increase in the ISGMR strength at high excitation energies was reported as well in their results 
in $^{12}$C when they carried out MDA {\em without} first subtracting the continuum from the
excitation-energy spectra~\cite{john}.

\newpage

\begin{table}[t!]

\centering
\caption{Parameters of the ISGMR peaks and moment ratios of the ISGMR strength distributions in stable nuclei as reported by the TAMU and RCNP groups. The probes employed in the measurements are listed for each case. Entries marked with $\star$ indicate that the $\Gamma$ is an RMS width, not that of a fitted peak. Entries marked with $\dagger$ indicate a multimodal strength distribution; in those cases the parameters for only the ``main'' ISGMR peak are included. 
For the TAMU data, the peak parameters correspond to a Gaussian fit, whereas for the RCNP  data, the corresponding parameters are for a Lorentzian fit. 
}
\scalebox{0.6}{
\begin{tabular}{@{}cccccccccc@{}cc}\toprule[0.1ex] 
			Target & Probe && $E_0$ (MeV)&$\Gamma$ (MeV) & EWSR \% & & $m_1/m_0$ (MeV)& $\sqrt{m_1/m_{-1}}$ (MeV) & $\sqrt{m_3/m_1}$ (MeV)& & Ref. \\\cmidrule[0.1ex]{1-2} \cmidrule[0.1ex]{4-6} \cmidrule[0.1ex]{8-10} \cmidrule[0.1ex]{12-12} \\
			$^{16}$O & 240 MeV-$\alpha$ & & - & $8.76 \pm 1.82$ & $48\pm10$ && $21.13\pm0.14$& $19.63\pm0.38$ & $24.89\pm0.59$ & & \cite{lui2001}$^\star$\\		
			$^{24}$Mg & 240 MeV-$^{6}$Li & & - & ${4.98}^{+0.68}_{-0.32}$ & ${106}^{+34}_{-24}$ && ${21.35}^{+0.37}_{-0.26}$& - & - & & \cite{DHY2009_MgSi}$^\star$\\		
			& 240 MeV-$\alpha$ & & - & ${6.5}^{+0.6}_{-0.3}$ & $73\pm 8$ && $21.3\pm 0.4$& - & ${24.0}^{+0.7}_{-0.3}$ & & \cite{DHY2009_redoMg}$^\star$\\
			$^{28}$Si & 240 MeV-$^{6}$Li & & - & ${5.78}^{+1.34}_{-0.34}$ & ${80}^{+35}_{-20}$ && ${20.59}^{+0.78}_{-0.33}$& - & - & & \cite{DHY2009_MgSi}$^\star$\\			
			& 240 MeV-$\alpha$ & & - & $5.9 \pm 0.6$ & $76\pm 7$ && $20.89\pm 0.38$& - & - & & \cite{DHY2007_28Si}$^\star$\\		 
			$^{32}$S & 386 MeV-$\alpha$ & & - &  9.43 & ${108}^{+7}_{-8}$ && ${23.65}^{+0.60}_{-0.66}$ &- & - & & \cite{Itoh_32S}$^\star$\\		 			  
			$^{40}$Ca & 240 MeV-$\alpha$ & & - & $4.88 \pm 0.57$ & $97 \pm 11 $ && $19.18\pm 0. 37$ & $18.3 \pm 0.3$ & $20.6\pm0.4$ & & \cite{DHY2001_40Ca}$^\star$\\	
			$^{48}$Ca & 240 MeV-$\alpha$ & & - & ${6.68}^{+0.31}_{-0.36}$ & ${95}^{+11}_{-15}$ && ${19.88}^{+0.14}_{-0.18}$ & ${19.04}^{+0.11}_{-0.14}$ & ${22.64}^{+0.27}_{-0.33}$ & & \cite{DHY2011_48Ca}$^\star$\\			
			$^{46}$Ti & 240 MeV-$\alpha$ & & $18.44\pm 0.25$ & $9.23 \pm 0.10$ & $62 \pm 11 $ && ${17.66}^{+0.65}_{-0.25}$ & ${18.10}^{+0.50}_{-0.20}$ & ${20.47}^{+1.41}_{-0.49}$ & & \cite{DHY2006_Ti}\\		
			$^{48}$Ti & 240 MeV-$\alpha$ & & $18.73\pm 0.23$ & $8.28 \pm 0.05$ & $84 \pm 11 $ && ${18.80}^{+0.45}_{-0.18}$ & ${18.33}^{+0.36}_{-0.15}$ & ${20.25}^{+0.99}_{-0.28}$ & & \cite{DHY2006_Ti}\\	
			$^{56}$Fe & 240 MeV-$\alpha$ & & ${18.14}^{+0.14}_{-0.15}$ & $7.40\pm0.13$ & ${82}^{+10}_{-8}$ && ${18.35}^{+0.33}_{-0.19}$ & ${17.92}^{+0.26}_{-0.15}$ & ${19.57}^{+0.73}_{-0.16}$ & & \cite{DHY2006_mass60}\\				
			$^{58}$Ni &240 MeV-$\alpha$ & & $18.43\pm0.15$ & $7.41\pm0.13$ & ${82}^{+11}_{-9}$ && ${19.20}^{+0.44}_{-0.19}$ & ${18.70}^{+0.34}_{-0.17}$ & ${20.81}^{+0.90}_{-0.28}$ & & \cite{DHY2006_mass60}\\			
			& 386 MeV-$\alpha$ & & $19.9^{+0.7}_{-0.8}$ & - &  $92^{+4}_{-3}$  & &   - & -& - & & \cite{nayak58ni} \\ 
			$^{60}$Ni & 240 MeV-$\alpha$ & & $17.62\pm0.15$ & $7.55\pm0.13$ & ${67}^{+12}_{-9}$ && ${18.04}^{+0.35}_{-0.23}$ & ${17.55}^{+0.27}_{-0.17}$ & ${19.54}^{+0.78}_{-0.23}$ & & \cite{DHY2006_mass60}\\
			$^{90}$Zr & 240 MeV-$\alpha$ & & 17.1 & 4.4 & 84 && ${17.88}^{+0.13}_{-0.11}$ & ${17.58}^{+0.06}_{-0.04}$ & ${18.86}^{+0.23}_{-0.14}$ & & \cite{Krishi_2015}$^\dagger$\\					
			&386 MeV-$\alpha$ && $16.6\pm0.1$ & $4.9\pm0.2$ &  $101\pm3$ & &   - & - & - & &  \cite{Uchida_90Zr} \\	
			&386 MeV-$\alpha$ && $16.55\pm 0.08$ & $4.2\pm 0.3$ &  $95\pm6$  & &   $18.13 \pm 0.09$ & $17.66\pm 0.07$ & $19.68 \pm 0.13$ & & \cite{YKGPLB2016} \\
			$^{92}$Zr & 240 MeV-$\alpha$ & & 16.6 & 4.4 & 62 && ${18.23}^{+0.15}_{-0.13}$ & ${17.71}^{+0.09}_{-0.07}$ & ${20.09}^{+0.31}_{-0.22}$ & & \cite{Krishi_2015}$^\dagger$\\		
			&386 MeV-$\alpha$ && $16.12\pm 0.04$ & $4.5\pm 0.2$ &  $97\pm3$  & &   $18.05 \pm 0.05$ & $17.52\pm 0.04$ & $19.77 \pm 0.06$ & & \cite{YKGPLB2016}  \\
			$^{94}$Zr &240 MeV-$\alpha$ & & 15.8 & 5.9 & 83 && ${16.16}^{+0.12}_{-0.11}$ & ${15.75}^{+0.27}_{-0.15}$ & ${17.52}^{+0.18}_{-0.14}$ & & \cite{Krishi_2015}$^\dagger$\\		
			$^{92}$Mo & 240 MeV-$\alpha$ & & 16.8 & 4.0 & 42 	&&${19.62}^{+0.29}_{-0.19}$  & - & ${21.68}^{+0.53}_{-0.33}$ & & \cite{DHY_2016_AllMo}$^\dagger$\\						
			&386 MeV-$\alpha$ && $16.79\pm 0.11$ & $4.2\pm 0.4$ &  $84\pm6$  & &   $18.20 \pm 0.13$ & $17.76\pm 0.11$ & $19.64 \pm 0.21$ & & \cite{YKGPLB2016} \\
			$^{94}$Mo & 240 MeV-$\alpha$ & & - & ${5.68}^{+5.53}_{-1.93}$ & ${112}^{+19}_{-12}$ &&  ${17.57}^{+1.14}_{-0.3}$ & ${17.06}^{+0.75}_{-0.19}$ & ${19.62}^{+3.54}_{-1.15}$ & & \cite{Button2016}$^\star$\\
			$^{96}$Mo & 240 MeV-$\alpha$& & 16.4 & 5.7 & 83 && ${16.95}^{+0.12}_{-0.10}$  & - & ${18.18}^{+0.20}_{-0.13}$ & &  \cite{Button2016}$^\dagger$\\
			$^{98}$Mo & 240 MeV-$\alpha$& & 15.7 & 6.5 & 89 && ${16.01}^{+0.19}_{-0.13}$  & - & ${17.29}^{+0.46}_{-0.21}$ & &  \cite{Button2016}$^\dagger$\\
			$^{100}$Mo & 240 MeV-$\alpha$& & 15.8 & 7.1 & 97 && ${16.13}^{+0.11}_{-0.10}$  & - & ${17.35}^{+0.16}_{-0.12}$ & &  \cite{Button2016}$^\dagger$\\								
			$^{106}$Cd&386 MeV-$\alpha$ && $16.50\pm0.19$ & $6.14\pm0.37$ &  -  & &   $16.27\pm 0.09$ & $16.06\pm0.05$ & $16.83\pm0.09$ & & \cite{Darshana2012} \\
			$^{110}$Cd & 240 MeV-$\alpha$& & $15.71\pm0.11$ & ${5.18}^{+0.16}_{-0.17}$ & $86 \pm 10$ && ${15.12}^{+0.30}_{-0.11}$  & ${14.96}^{+0.13}_{-0.12}$ & ${15.58}^{+0.40}_{-0.09}$ & & \cite{DHY2004_Cd}\\			
			&386 MeV-$\alpha$ && $16.09\pm0.15$ & $5.72\pm0.45$ &  -  & &   $15.94\pm0.07$ & $15.72\pm0.05$ & $16.53\pm0.08$  & & \cite{Darshana2012} \\
			$^{112}$Cd &386 MeV-$\alpha$ && $15.72\pm0.10$ & $5.85\pm0.18$ &  -  & &   $15.80\pm0.05$ & $15.59\pm0.05$ & $16.38\pm0.06$  & & \cite{Darshana2012} \\
			$^{114}$Cd&386 MeV-$\alpha$ && $15.59\pm0.20$ & $6.41\pm0.64$ &  -  & &   $15.37\pm0.08$ & $15.37\pm0.08$ & $16.27\pm0.09$ & & \cite{Darshana2012} \\
			$^{116}$Cd & 240 MeV-$\alpha$ & & ${15.17}^{+0.12}_{-0.11}$ & ${5.40}^{+0.16}_{-0.14}$ & $100 \pm 11$ && ${14.50}^{+0.32}_{-0.16}$  & ${14.31}^{+0.20}_{-0.17}$ & ${15.02}^{+0.37}_{-0.12}$ & & \cite{DHY2004_Cd}\\		
			&386 MeV-$\alpha$ && $15.43\pm0.12$ & $6.51\pm0.40$ &  -  & &   $15.44\pm0.06$ & $15.19\pm0.06$ & $16.14\pm0.07$ & & \cite{Darshana2012} \\
			$^{112}$Sn & 240 MeV-$\alpha$ & & $15.67 \pm 0.11$ & ${5.18}^{+0.40}_{-0.04}$ & ${110}^{+15}_{-12}$ && ${15.43}^{+0.11}_{-0.10}$  & ${15.23}^{+0.26}_{-0.14}$ & ${16.05}^{+0.26}_{-0.14}$ & & \cite{DHY2004_Sn}\\			
			&386 MeV-$\alpha$ && $16.1\pm0.1$ & $4.0\pm0.4$ &  $92\pm4$  & &   $16.2\pm0.1$ & $16.1\pm0.1$ & $16.7\pm0.2$ & & \cite{Li_2010} \\
			$^{114}$Sn&386 MeV-$\alpha$ && $15.9\pm0.1$ & $4.1\pm0.4$ &  $104\pm6$  & &   $16.1\pm0.1$ & $15.9\pm0.1$ & $16.5\pm0.2$ & & \cite{Li_2010}  \\
			$^{116}$Sn&196 MeV-d && $15.7\pm0.1$ & $4.6\pm0.7$ &  $73\pm15$  & &   - & - & - & & \cite{Darshana2015} \\			
			& 240 MeV-$^{6}$Li & & $15.58\pm0.18$ & $5.46\pm 0.18$ & ${106}^{+27}_{-11}$ && ${15.39}^{+0.35}_{-0.20}$  & - & - & & \cite{DHY2009_Sn}\\
			& 240 MeV-$\alpha$ & & - & $5.27\pm 0.25$ & $112\pm15$ && $15.85 \pm 0.25$  & - & - & & \cite{dhybg}$^\star$\\	
			& 240 MeV-$\alpha$ & &$15.77\pm0.07$ & - & -&& $-$   & - & - & & \cite{dhyprl}\\	
			&386 MeV-$\alpha$ && $15.4\pm0.1$ & $5.5\pm0.3$ &  $95\pm4$ & &   - & - & - & &  \cite{Uchida_90Zr} \\	
			&386 MeV-$\alpha$ && $15.8\pm0.1$ & $4.1\pm0.3$ &  $99\pm5$  & &   $15.8\pm0.1$ & $15.7\pm0.1$ & $16.3\pm0.2$ & & \cite{Li_2010}  \\	
			$^{118}$Sn&386 MeV-$\alpha$ && $15.6\pm0.1$ & $4.3\pm0.4$ &  $95\pm5$  & &   $15.8\pm0.1$ & $15.6\pm0.1$ & $16.3\pm0.1$ & & \cite{Li_2010}  \\
			$^{120}$Sn&386 MeV-$\alpha$ && $15.4\pm0.2$ & $4.9\pm0.5$ &  $108\pm7$  & &   $15.7\pm0.1$ & $15.5\pm0.1$ & $16.2\pm0.2$ & & \cite{Li_2010}  \\
			$^{112}$Sn &386 MeV-$\alpha$ && $15.0\pm0.2$ & $4.4\pm0.4$ &  $106\pm5$  & &   $15.4\pm0.1$ & $15.2\pm0.1$ & $15.9\pm0.2$ & & \cite{Li_2010}  \\
			$^{124}$Sn & 240 MeV-$\alpha$ & & $15.34\pm0.13$ & ${5.00}^{+0.03}_{-0.53}$ & ${106}^{+20}_{-10}$ && $14.50\pm0.14$  & ${14.33}^{+0.17}_{-0.14}$ & ${14.96}^{+0.10}_{-0.11}$ & & \cite{DHY2004_Sn}\\	
			&386 MeV-$\alpha$ && $14.8\pm0.2$ & $4.5\pm0.5$ &  $105\pm6$  & &   $15.3\pm0.1$ & $15.1\pm0.1$ & $15.8\pm0.1$ & & \cite{Li_2010} \\		
			$^{144}$Sm & 240 MeV-$\alpha$ & & -& $3.40\pm0.2$ & $92\pm 12$ && $15.40\pm0.30$   & - & - & & \cite{dhybg}\\	
			& 240 MeV-$\alpha$ & &$15.16\pm0.11$ & - & -&& $-$   & - & - & & \cite{dhyprl}\\	
			& 386 MeV-$\alpha$ & & ${15.30}^{+0.11}_{-0.12}$& ${3.71}^{+0.12}_{-0.63}$ & ${84}^{+4}_{-25}$ && -   & - & - & & \cite{Itoh_prc2003}\\	
			$^{148}$Sm & 386 MeV-$\alpha$ & & $12.32\pm0.45$& $4.7$ & ${17}^{+3}_{-4}$ && -   & - & - & & \cite{Itoh_prc2003}\\
			&  & & ${15.37}^{+0.14}_{-0.18}$& $3.7$ & ${64}^{+5}_{-24}$ && -   & - & - & & \cite{Itoh_prc2003}\\		
			$^{150}$Sm & 386 MeV-$\alpha$ & & ${12.5}^{+1.7}_{-1.5}$& $4.7$ & $19\pm11$ && -   & - & - & & \cite{Itoh_prc2003}\\
			&  & & $15.48 \pm 0.28$& $3.7$ & ${63}^{+13}_{-28}$ && -   & - & - & & \cite{Itoh_prc2003}\\
			$^{152}$Sm & 386 MeV-$\alpha$ & & ${11.27}^{+0.32}_{-0.54}$& $4.7$ & ${17}^{+2}_{-4}$ && -   & - & - & & \cite{Itoh_prc2003}\\
			&  & & ${15.44}^{+0.12}_{-0.23}$ & $3.7$ & ${73}^{+4}_{-25}$ && -   & - & - & & \cite{Itoh_prc2003}\\	
			$^{154}$Sm & 386 MeV-$\alpha$ & & ${10.83}^{+0.32}_{-0.54}$& $4.7$ & ${17}^{+2}_{-3}$ && -   & - & - & & \cite{Itoh_prc2003}\\
			&  & & ${15.45}^{+0.13}_{-0.16}$ & $3.7$ & ${71}^{+4}_{-23}$ && -   & - & - & & \cite{Itoh_prc2003}\\						 							
			$^{204}$Pb &386 MeV-$\alpha$ && $13.8\pm0.1$ & $3.3\pm0.2$ &  -  & &   - & $13.7\pm0.1$ & - & & \cite{Darshana2013} \\
			$^{206}$Pb &386 MeV-$\alpha$ && $13.8\pm0.1$ & $2.8\pm0.2$ &  - & &   - & $13.6\pm0.1$ & - & &  \cite{Darshana2013} \\		
			$^{208}$Pb& 196 MeV-d && $13.6\pm0.1$ & $3.1\pm0.4$ &  $147\pm18$  & &   - & - & - & & \cite{Darshana2015} \\						& 240 MeV-$\alpha$ & & -& $2.88\pm0.2$ & $99\pm 5$ && $13.96\pm 0.20$   & - & - & & \cite{dhybg}\\			
			& 240 MeV-$\alpha$ & &$13.91\pm0.11$ & - & -&& $-$   & - & - & & \cite{dhyprl}\\	
			&386 MeV-$\alpha$ && $13.4\pm0.2$ & $4.0\pm0.4$ &  $104\pm9$ & &   - & - & - & &  \cite{Uchida_90Zr} \\	
			&386 MeV-$\alpha$ && $13.7\pm0.1$ & $3.3\pm0.2$ &  -  & &   - & $13.5\pm0.1$ & - & &  \cite{Darshana2013} \\				
			
			\bottomrule[0.1ex]
\end{tabular}
}
\label{ISGMR-table}
\end{table}


\begin{figure*}[!h]
\centering\includegraphics [height=0.48\textheight]{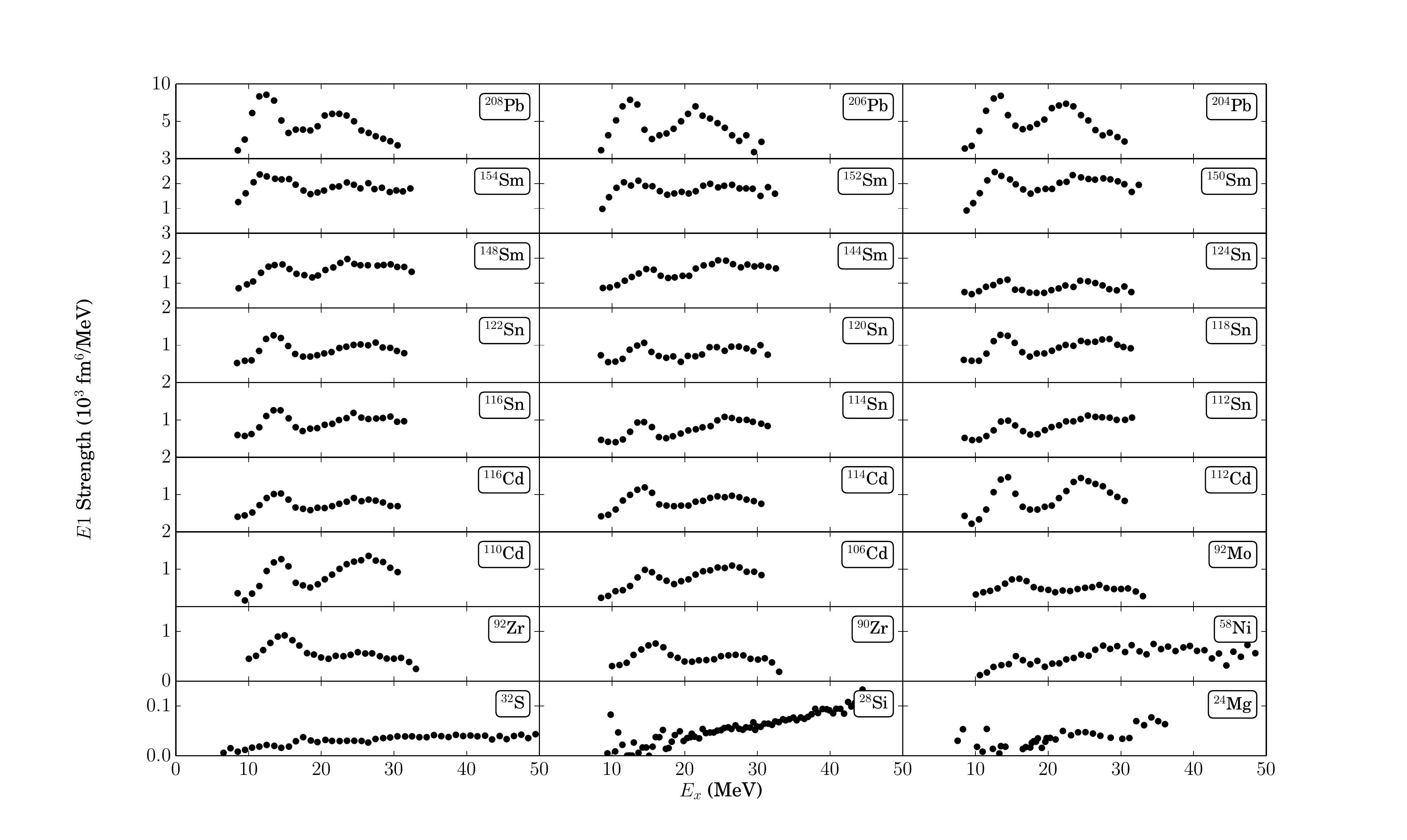}
\caption{ISGDR strength distributions for various nuclei, extracted in the RCNP work. The data are from Refs. \cite{Darshana2013,Itoh_prc2003,Li_PRL2007,Li_2010,Darshana2012,YKGPLB2016,nayak58ni,Itoh_32S,tom-si,yogesh-mg2,yogesh-mg}.}
\label{ISGDR}
\end{figure*}

\begin{table}[!h]
\centering
\caption{Parameters of the two-peak fits (LE component and HE component) to the ISGDR strength distributions in stable nuclei as reported by the TAMU and RCNP groups. The probes employed in the measurements are listed for each case. Entries marked with $\star$ indicate that the $\Gamma$ is an RMS width, not that of a fitted peak. For the TAMU data, the peak parameters correspond to a Gaussian fit, whereas for the RCNP  data, the corresponding parameters are for a Lorentzian fit.}
\scalebox{0.6}{
\begin{tabular}{@{}cccccccccc@{}cc}\toprule[0.1ex]
		 &  && & LE Component &  & & & HE Component & & & \\
		Target & Probe && $E_0$ (MeV)&$\Gamma$ (MeV) & EWSR \% & & $E_0$ (MeV)& $\Gamma$ (MeV) & EWSR \%& & Ref. \\\cmidrule[0.1ex]{1-2} \cmidrule[0.1ex]{4-6} \cmidrule[0.1ex]{8-10} 
		\cmidrule[0.1ex]{12-12} \\
		$^{46}$Ti& 240 MeV-$\alpha$ && $15.94\pm0.31$ & ${6.34}^{+0.63}_{-0.50}$ &  $10\pm4$  & &   $28.30 \pm 0.68$ & ${14.35}^{+0.65}_{-0.79}$ & $67\pm10$ & & \cite{DHY2006_Ti} \\
		$^{48}$Ti& 240 MeV-$\alpha$ && ${15.75}^{+0.31}_{-0.28}$ & ${7.27}^{+0.22}_{-0.24}$ &  $13\pm5$  & &   ${28.82}^{+0.78}_{-0.72}$ & ${12.44}^{+0.56}_{-0.68}$ & $43\pm9$ & & \cite{DHY2006_Ti} \\	
		$^{48}$Ca& 240 MeV-$\alpha$ && ${16.69}^{+0.19}_{-0.13}$ & ${6.24}^{+1.49}_{-0.11}$ &  ${20}^{+12}_{-8}$  & &   ${37.28}^{+0.71}_{-1.98}$ & ${14.95}^{+3.49}_{-0.11}$ & ${160}^{+90}_{-50}$ & & \cite{DHY2011_48Ca} \\					
		$^{56}$Fe& 240 MeV-$\alpha$ && $17.41\pm0.20$ & ${4.03}^{+0.28}_{-0.26}$ &  $5\pm2$  & &   $30.61 \pm 0.29$ & ${20.31}^{+0.39}_{-0.40}$ & $67\pm10$ & & \cite{DHY2006_mass60} \\
		$^{58}$Ni& 240 MeV-$\alpha$ && $17.42\pm0.25$ & ${3.94}^{+0.36}_{-0.34}$ &  $4\pm2$  & &   $34.06 \pm 0.30$ & ${19.52}^{+0.41}_{-0.40}$ & $86\pm12$ & & \cite{DHY2006_mass60} \\	
		& 386 MeV-$\alpha$ && $\sim$16  & - &  -  & &   $30.8^{+1.7}_{-1.1}$ & $-$ & $-$ & & \cite{nayak58ni} \\	
		$^{60}$Ni& 240 MeV-$\alpha$ && $16.01 \pm 0.20$ & ${4.41}^{+0.34}_{-0.22}$ &  $6\pm3$  & &   ${36.11}^{+0.29}_{-0.27}$ & ${27.13}^{+0.43}_{-0.42}$ & $120\pm 16$ & & \cite{DHY2006_mass60} \\		
		$^{90}$Zr& 240 MeV-$\alpha$ && $17.5 \pm 0.2$ & $5.4 \pm 0.7$ &  $9.2 \pm 2.1$  & &   $ 27.4 \pm 0.5 $ & $ 10.1 \pm 2.0$ & $ 49 \pm 6 $ & & \cite{Krishi_2015} \\		
		&386 MeV-$\alpha$ && $17.8 \pm 0.5$ & $3.7 \pm 1.2$ &  $7.9\pm2.9$ & &   $26.9 \pm 0.7$ & $12.0\pm 1.5$ & $67\pm8$ & & \cite{Uchida_90Zr} \\
		$^{92}$Zr& 240 MeV-$\alpha$ && $14.7 \pm 0.3$ & $5.4 \pm 0.7$ &  $5.8 \pm 1.2$  & &   $ 30.0 \pm 0.7 $ & $ 12.9 \pm 2.0 $ & $ 51 \pm 7 $ & & \cite{Krishi_2015} \\
		$^{94}$Zr& 240 MeV-$\alpha$ && $15.7 \pm 0.2$ & $9.0 \pm 1.0$ &  $28 \pm 4$  & &   $ 27.0 \pm 0.5 $ & $ 9.9 \pm 2.0 $ & $ 64 \pm 7 $ & & \cite{Krishi_2015} \\																
		$^{92}$Mo& 240 MeV-$\alpha$ && $17.5 \pm 0.4$ & $5.4\pm0.7$ &  $5.8 \pm 1.1$  & &   $27.6 \pm 0.5$ & $10.2 \pm 2.0$ & $59 \pm 7$ & & \cite{DHY_2016_AllMo} \\	
		$^{94}$Mo& 240 MeV-$\alpha$ && ${15.07}^{+0.22}_{-0.19}$ & ${3.19}^{+0.36}_{-0.22}$ &  $12\pm 2$  & &   ${26.50}^{+0.44}_{-0.42}$ & ${5.99}^{+0.45}_{-0.49}$ & $45 \pm 5$ & & \cite{Button2016} \\			
		$^{96}$Mo& 240 MeV-$\alpha$ && $15.9 \pm 0.3$ & $10.1 \pm 1.1$ &  $17 \pm 2$  & &   $30.0 \pm 0.7 $ & $13.1 \pm 2.9$ & $62 \pm 8 $ & & \cite{DHY_2016_AllMo} \\		
		$^{98}$Mo& 240 MeV-$\alpha$ && $16.0 \pm 0.3$ & $10.9 \pm 1.1$ &  $26 \pm 3$  & &   $27.4 \pm 0.7 $ & $10.8 \pm 3.0$ & $49\pm 8 $ & & \cite{DHY_2016_AllMo} \\			
		$^{100}$Mo& 240 MeV-$\alpha$ && $13.0 \pm 0.3$ & $11.6 \pm 1.2$ &  $18 \pm 3$  & &   $30.1 \pm 0.7 $ & $12.5 \pm 3.5$ & $47 \pm 10 $ & & \cite{DHY_2016_AllMo} \\			
		$^{106}$Cd&386 MeV-$\alpha$ && $14.7\pm0.2$ & $4.2\pm1.2$ &  -  & &   $26.2 \pm 0.4$ & $14.6 \pm 1.9$ & - & & \cite{Patel_Thesis,Patel_Book} \\	
		$^{110}$Cd&240 MeV-$\alpha$ && ${14.47}^{+0.44}_{-0.47}$ & $8.70\pm0.87$ &  $42\pm 11$ & &   ${23.30}^{+0.55}_{-0.48}$ & ${7.32}^{+1.09}_{-0.78}$ & $28\pm 11$ & & \cite{DHY2004_Cd} \\ 
		&386 MeV-$\alpha$ && $14.2\pm0.2$ & $3.6\pm0.6$ &  -  & &   $26.4 \pm 0.3$ & $10.3 \pm 1.3$ & - & & \cite{Patel_Thesis,Patel_Book} \\		
		$^{112}$Cd &386 MeV-$\alpha$ && $14.0\pm0.3$ & $2.9\pm1.0$ &  -  & &   $25.3 \pm 0.7$ & $7.9 \pm 2.4$ & - & & \cite{Patel_Thesis,Patel_Book} \\
		$^{114}$Cd & 386 MeV-$\alpha$ && $13.7 \pm 0.2$ & $5.3 \pm 1.0$ &  -  & &   $25.9 \pm 0.7$ & $13.6 \pm 3.2$ & - & & \cite{Patel_Thesis,Patel_Book} \\
		$^{116}$Cd&240 MeV-$\alpha$ && ${13.94}^{+0.26}_{-0.30}$ & ${8.31}^{+0.59}_{-0.46}$ &  $46 \pm 11$ & &  $23.58 \pm 0.42$ & ${9.22}^{+0.92}_{-0.72}$ & $44 \pm 11$ & & \cite{DHY2004_Cd} \\
		& 386 MeV-$\alpha$ && $13.7 \pm 0.2$ & $4.2 \pm 0.7$ &  -  & &   $25.8 \pm 0.5$ & $12.0 \pm 2.2$ & - & & \cite{Patel_Thesis,Patel_Book} \\
		$^{112}$Sn&240 MeV-$\alpha$ && ${14.92}^{+0.15}_{-0.14}$ & ${8.82}^{+0.26}_{-0.29}$ &  $32\pm 4$ & &   ${26.28}^{+0.32}_{-0.23}$ & ${10.82}^{+0.39}_{-0.36}$ & $70 \pm 10$ & & \cite{DHY2004_Sn} \\			
		&386 MeV-$\alpha$ && $15.4\pm0.1$ & $4.9\pm0.5$ &  -  & &   $26.2 \pm 0.8$ & $16.3 \pm 4.0$ & $102\pm3$ & & \cite{Li_2010} \\	
		$^{114}$Sn&386 MeV-$\alpha$ && $15.0\pm0.1$ & $5.6\pm0.5$ &  -  & &   $26.1 \pm 0.8$ & $13.0 \pm 3.4$ & $123\pm3$ & & \cite{Li_2010} \\	
		$^{116}$Sn&240 MeV-$^{6}$Li && $15.32 \pm 0.20$ & ${5.56}^{+0.20}_{-0.19}$ &  $66\pm10$  & &   $21.73\pm0.20$ & ${2.80}^{+0.26}_{-0.28}$ & ${52}^{+20}_{-14}$ & & \cite{DHY2009_Sn} \\	
		&240 MeV-$\alpha$ && $14.38\pm0.25$ & $5.84\pm0.30$ &  $25 \pm 15$  & &   $25.50\pm0.60$ & $12.0\pm0.6$ &  $61\pm15$ & & \cite{dhybg} \\			
		&386 MeV-$\alpha$ && $15.6 \pm 0.5$ & $2.3 \pm 1.0$ &  $4.9\pm2.2$ & &   $25.4 \pm 0.5$ & $15.7\pm 2.3$ & $68\pm9$ & & \cite{Uchida_90Zr} \\
		&386 MeV-$\alpha$ && $14.9\pm0.1$ & $5.9\pm0.5$ &  -  & &   $25.9 \pm 0.6$ & $13.1 \pm 4.2$ & $102\pm3$ & & \cite{Li_2010} \\	
		$^{118}$Sn&386 MeV-$\alpha$ && $14.8\pm0.1$ & $6.1\pm0.3$ &  -  & &   $26.0 \pm 0.3$ & $13.1 \pm 2.0$ & $120\pm3$ & & \cite{Li_2010} \\		
		$^{120}$Sn&386 MeV-$\alpha$ && $14.7\pm0.1$ & $5.9\pm0.3$ &  -  & &   $26.0 \pm 0.4$ & $13.1 \pm 1.9$ & $150\pm3$ & & \cite{Li_2010} \\										
		$^{122}$Sn&386 MeV-$\alpha$ && $14.4\pm0.1$ & $6.7\pm0.3$ &  -  & &   $26.3 \pm 0.2$ & $12.4 \pm 1.1$ & $147\pm3$ & & \cite{Li_2010} \\		
		$^{124}$Sn&240 MeV-$\alpha$ && $13.31 \pm 0.15$ & ${6.60}^{+0.15}_{-0.13}$ &  $40\pm 4$  & &   ${25.06}^{+0.22}_{-0.21}$ & ${13.87}^{+0.24}_{-0.28}$ & ${93}^{+12}_{-13}$ & & \cite{DHY2004_Sn} \\	
		&386 MeV-$\alpha$ && $14.3\pm0.1$ & $6.6\pm0.3$ &  -  & &   $25.7 \pm 0.5$ & $10.2 \pm 1.6$ & $129\pm6$ & & \cite{Li_2010} \\			
		$^{144}$Sm&240 MeV-$\alpha$ && $14.00\pm0.30$ & $8.0\pm0.60$ &  $32\pm15$  & &   $24.51 \pm 0.40$ & $7.21 \pm 0.40$ & $64 \pm 12$ & & \cite{dhybg} \\											
		&386 MeV-$\alpha$ && $13.04\pm0.34$ & $4.8\pm0.8$ &  $23\pm1$  & &   $25.4 \pm 0.6$ & $19.9 \pm 1.9$ & $109 \pm 2$ & & \cite{Itoh_prc2003}$^\star$ \\	
		$^{148}$Sm&386 MeV-$\alpha$ && $12.95\pm0.45$ & $5.6\pm0.9$ &  $25\pm1$  & &   $25.2 \pm 1.1$ & $19.4 \pm 2.8$ & $103 \pm 3$ & & \cite{Itoh_prc2003}$^\star$ \\						
		$^{150}$Sm&386 MeV-$\alpha$ && $12.91\pm0.61$ & $5.6\pm1.3$ &  $33\pm2$  & &   $25.1 \pm 1.4$ & $20.7 \pm 4.5$ & $122 \pm 5$ & & \cite{Itoh_prc2003}$^\star$ \\						
		$^{152}$Sm&386 MeV-$\alpha$ && $12.77\pm0.37$ & $7.2\pm0.9$ &  $29\pm1$  & &   $25.1 \pm 1.0$ & $21.6 \pm 3.4$ & $103 \pm 5$ & & \cite{Itoh_prc2003}$^\star$ \\
		$^{154}$Sm&386 MeV-$\alpha$ && $12.75\pm0.33$ & $8.2\pm1.0$ &  $32\pm1$  & &   $25.1 \pm 1.0$ & $22.6 \pm 4.2$ & $102 \pm 3$ & & \cite{Itoh_prc2003}$^\star$ \\	
		$^{204}$Pb&386 MeV-$\alpha$ && $12.8\pm0.3$ & $3.6\pm1.3$ &  -  & &   $22.6\pm 0.8$ & - & - & & \cite{Patel_Thesis,Patel_Book} \\
		$^{206}$Pb&386 MeV-$\alpha$ && $12.2 \pm 0.3$ & $4.0 \pm 1.4$ &  -  & &   $22.7 \pm 1.1$ & - & - & & \cite{Patel_Thesis,Patel_Book} \\
		$^{208}$Pb&240 MeV-$\alpha$ && $13.26\pm0.30$ & $5.68 \pm 0.50$ &  $24\pm15$  & &   $22.20\pm 0.30$ & $9.39 \pm 0.35$ & $88\pm15$ & & \cite{dhybg}\\
		&386 MeV-$\alpha$ && $13.0 \pm 0.1$ & $1.1 \pm 0.4$ &  $7.0\pm0.4$ & &   $22.7 \pm 0.2$ & $11.9\pm 0.4$ & $111\pm6$ & & \cite{Uchida_90Zr} \\
		&386 MeV-$\alpha$ && $12.3 \pm 0.3$ & $4.2 \pm 1.3$ &  -  & &   $22.5 \pm 0.9$ & - & - & & \cite{Patel_Thesis,Patel_Book} \\						
		\bottomrule[0.1ex]
\label{ISGDR-table}
	\end{tabular}
}
\end{table}


\subsection{The ISGDR}
Experimental ISGDR strength distributions extracted in the RCNP work are presented in Fig \ref{ISGDR}. 
The properties of the ISGDR peaks, in nuclei where a distinct peak structure is observed, are summarized in Table~\ref{ISGDR-table}. This table is also comprehensive in that it includes all available results from both RCNP and TAMU.

Two points need to be made in this connection: i) No ``peak'' structure is observed for ISGDR in the light nuclei, and generally one sees a monotonously increasing ISGDR strength at excitation energies above ~20 MeV. This is believed to be because of the aforementioned ``spurious strength distributions'' observed also for the ISGMR; this effect is further exacerbated by the fact that the peak energy of the ISGDR itself is rather high ($>$25 MeV) in the light nuclei.
ii) In all heavy nuclei (A$\geq$58), the ISGDR strength distribution has two distinct peaks. 
This ``low-energy'' isoscalar $L$=1 strength (LE) has engendered considerable
interest and argument, as it was hinted in our Introduction. It is present in nearly all of the recent theoretical calculations in some form or the other, and at similar energies, although with varying strength. It has been shown \cite{co00,dv00,jorge2000,shlomo2002} that the centroid of this component
of the $L$=1 strength is independent of the nuclear incompressibility (and, hence, is certainly of ``non-bulk'' nature). While the exact nature of this component is not fully
understood yet, suggestions have been made that this component might
represent the ``toroidal''~\cite{dv00,balb} or the ``vortex'' modes~\cite{dubna1,dubna2}. 
It is impossible to distinguish between the competing possibilities based on currently-available
data~\cite{garg-rev1}; also, it is not at all clear why these exotic modes would be excited with such large cross sections in ($\alpha,\alpha'$) work. There is general agreement, however, that only the
high-energy (HE) component of this bi-modal distribution needs to be
considered in obtaining a value of $K_A$ from the energy of the ISGDR.
Nearly all the expected $L$=1 EWSR strength is observed under this HE peak in all cases.

Incidentally, the ISGMR and ISGDR data in $^{208}$Pb give a consistent value for 
the finite-nucleus incompressibility $K_A$, and hence 
$K_\infty$ \cite{Uchida_PLB2003,Garg_Comex1} (cf. Sec. \ref{sec:kk}). 


\subsection{``Softness'' of Off-shell Nuclei}\label{sec:fluffy}
\begin{figure}[h]
\centering\includegraphics [height=0.30\textheight]{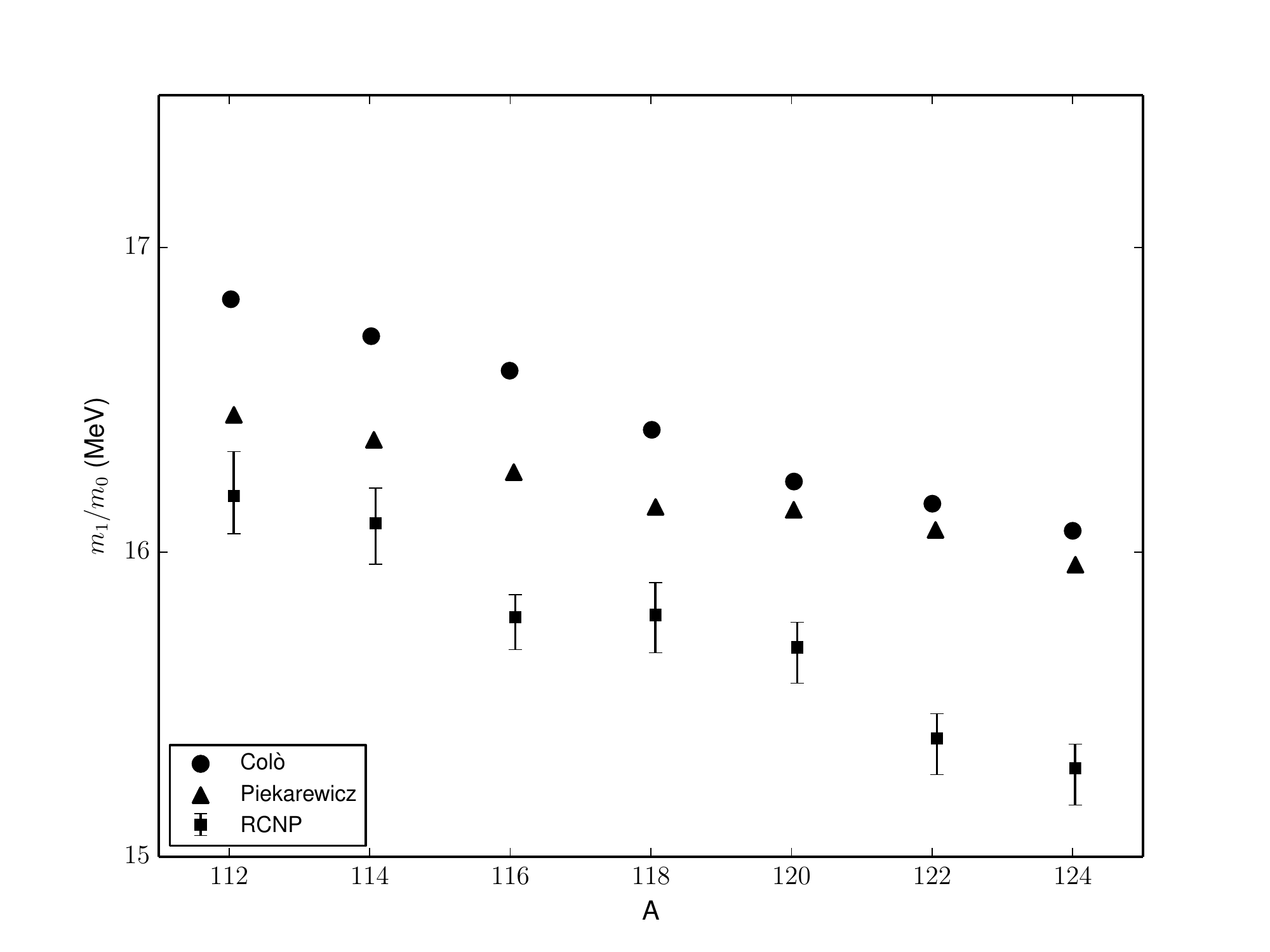}
\caption{Systematics of the moment ratios $m_1/m_0$ for the ISGMR strength distributions in the Sn isotopes. The experimental results (filled squares) are compared with results from nonrelativistic RPA calculations (without pairing) by Col\`o et al. \cite{Colo2004} (filled circles), and relativistic calculations of Piekarewicz \cite{fluffy2} (triangles). Figure adopted from Ref. \cite{Li_2010}.}
\label{fluffy}
\end{figure}

Most nuclei are neither magic nor spherical. Still, in the 1980s and 1990s, but also in the first years after the turn of the new millennium, theoretical attention was focused on $^{208}$Pb and on $^{90}$Zr, the canonical ``doubly closed shell nuclei".
In measurements of ISGMR strengths in the Sn isotopes (A=112--124) \cite{Li_PRL2007,Li_2010}, however, it was observed that the centroids of the strengths of the ISGMR were consistently lower (by as much as 1 MeV) than those predicted by theoretical calculations, both relativistic and non-relativistic, that correctly reproduced the ISGMR energies for $^{208}$Pb and $^{90}$Zr (see Fig.~\ref{fluffy}). This led to the now infamous question: ``Why are Tins So Soft?'' \cite{fluffy1,fluffy2,fluffy6}. [We understand that the author of Ref.~\cite{fluffy2} had originally used the word ``fluffy'', but that was changed by the editors to ``soft'' \cite{fluffy0}; as it happens, ``fluffy'' has found wide informal acceptance among the researchers in the field.] This effect was confirmed in measurements of the ISGMR strength distributions in the Cd isotopes (A=106--116) as well \cite{Darshana2012}.

This question remains unanswered so far, despite several attempts by different theoretical groups. A lot of effort has since been devoted to exploring the effects of ``superfluidity'' on ISGMR, and nuclear incompressibility, in nuclei  \cite{fluffy3,Khan2009,Khan2009b,fluffy5,fluffy6,Vesely:2012,Avogadro:2013}.
Superfluidity is certainly relevant, and depending on the model
can explain a part of the ``fluffiness'' but does not seem
to solve the issue completely.

In Ref. \cite{fluffy3}, QRPA calculations on top
of HFB have been performed for the Sn isotopes and it has been found
that pairing shifts the centroid energy of the ISGMR downwards by 
several hundreds of keV. Still, the value of $K_\infty$ that is
extracted from the Sn data is lower than that extracted from the
Pb data by about 10\%. Pairing introduces a significant source
of uncertainty. Usually, only the ground-state average pairing gap
is used to fix the effective pairing force in nuclei and in such a
way an unambiguous determination of the pairing functional is
simply not possible. Different pairing effective interactions
like volume, surface or mixed pairing forces, can equally well
fit the ground-state average gap and still produce different 
results for other observables like properties of low-lying states
etc. This has been the case in the calculations of Ref. \cite{fluffy3},
in which different pairing forces produce different values for
the ISGMR energy; similar effects are visible in the HF plus BCS
calculations of \cite{fluffy5} which are also extended to Cd and
Pb isotopes.  

In Refs. \cite{fluffy6}, much more systematic investigations 
have been performed, having as object the ISGMR in a large number
of semimagic nuclei along the isotope chart, and including not
only the measured Zr, Sn and Pb isotopes but also neutron-rich
systems. This has allowed, as a by-product, the fit of Eq. 
(\ref{eq:liquid_drop}) within the model, as we have recalled above. 

Along a similar line, also the authors of Ref. \cite{Avogadro:2013} 
have considered a relatively large sample of nuclei and isotopic
chains. They have paid special attention to the self-consistency 
in the pairing channel, and included pairing forces that depend on
the neutron-proton imbalance.

\begin{figure}[hbt]
\centering\includegraphics [height=0.30\textheight]{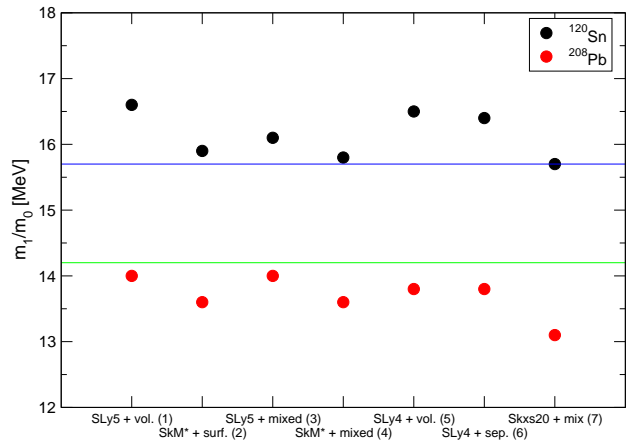}
\caption{ISGMR centroid energies in $^{208}$Pb and $^{120}$Sn 
predicted by different Skyrme models plus different pairing forces.
``Vol.'', ``Surf.'', ``Mixed'' and ``Sep.'' means, respectively,
volume, surface, mixed and separable pairing forces. The 
results are taken from the references mentioned in the text: 
(1) and (2) from \cite{fluffy3}, (3) and (4) from 
\cite{fluffy5}, (5) and (6) from \cite{fluffy6}, (7) from 
\cite{Avogadro:2013}. The horizontal lines correspond to the
experimental data.
}
\label{gmr_sn_pb}
\end{figure}

The conclusion of Refs. \cite{fluffy3,fluffy5,fluffy6} is that it
is difficult to reconcile $^{208}$Pb and Sn data, namely to
reproduce both with a single model having a given value of
$K_\infty$. This point is illustrated in Fig. \ref{gmr_sn_pb}. 
We include for simplicity only $^{120}$Sn and compare it with 
$^{208}$Pb. The centroid energies of the ISGMR in these two nuclei
are shown with black and red circles, respectively. 
We display results coming from different QRPA calculations 
performed in the references we have mentioned, and compare them
with the experimental data (horizontal lines). It is clear that
models that reproduce the ISGMR in $^{208}$Pb tend to predict
an energy in $^{120}$Sn that is too high and, conversely, 
those who perform well in $^{120}$Sn underestimate the ISGMR
energy in $^{208}$Pb. As stated clearly in \cite{fluffy6}, 
the energy difference between the ISGMR in the two nuclei
is systematically overestimated. Compared to $^{208}$Pb, Sn
seem to indicate a value of $K_\infty$ that is 10\% smaller or
even a bit more.

Piekarewicz and Centelles \cite{cente09} created a hybrid model
specifically to reproduce the ISGMR energies in the Sn isotopes; the model has a significantly softer incompressibility coefficient for neutron-rich matter. The predictions of this hybrid model fall within 0.1 MeV of the experimental data for the full Sn isotopic chain if one takes into account the uncertainties in the data. However, although the improvement in the case of the Sn isotopes is significant and unquestionable, an important problem remains: the hybrid model underestimates the GMR centroid energy in $^{208}$Pb by almost 1 MeV.

Further, in calculations using the T5 Skyrme interaction within the quasiparticle time blocking approximation (QTBA) approach, Tselyaev et al. \cite{fluffy8} obtained the ISGMR strength distributions in all the Sn isotopes in good agreement with the experimental data, including the resonance widths. However, T5 has the associated $K_{\infty}$ value of only 202 MeV, which is significantly lower than that extracted earlier from the ISGMR's in $^{208}$Pb and $^{90}$Zr. While the agreement with the experimental data is impressive (and, indeed, reproduces the A-dependence rather well), it does leave the question of ``softness'' of the Sn nuclei unanswered.

Incidentally, calculations in the RMF approach with the DD-ME2 interaction \cite{dario1}, also reproduce the centroids of the ISGMR in the Sn isotopes rather well \cite{dario2}. Concern has been expressed, however, that 
these calculations display significant fragmentation of the ISGMR strength, i.e. the peak and centroid
energies are not quite the same.

The conclusion from all these studies is unequivocal: if from the double-magic nuclei one turns  to Sm, Sn and Cd, data seem to indicate a lower value of $K_\infty$, which clearly points to the present limit of our models and our understanding. As more and more data becomes available on off-shell nuclei, it might be possible to explore this question in more detail theoretically.

\begin{figure}[!h]
\centering\includegraphics [height=0.27\textheight]{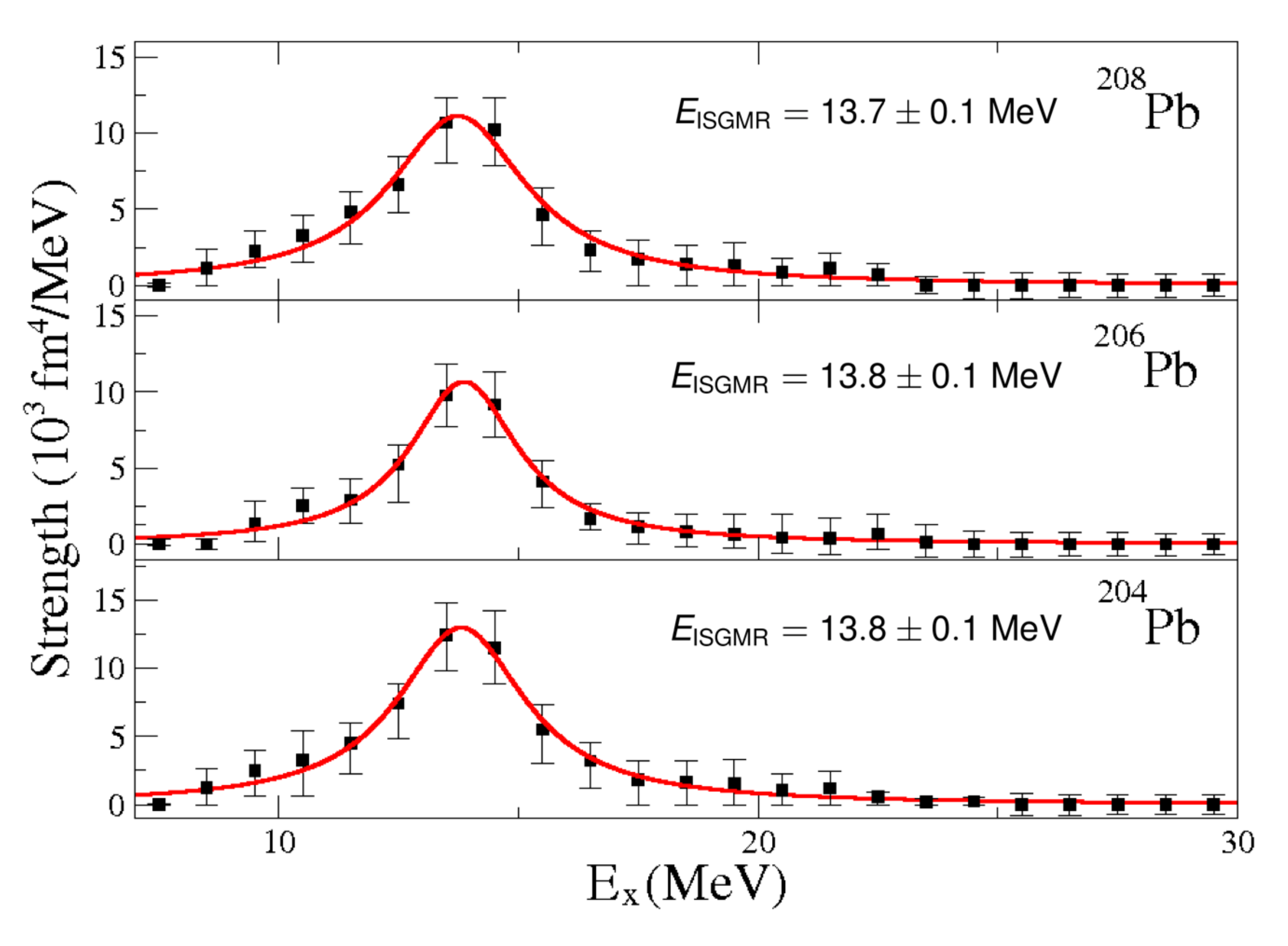}
\caption{The ISGMR strength distributions observed in $^{204,206,208}$Pb. Solid lines represent Lorentzian fits to the data. The peak position is indicated in each case. Figure adopted from Ref. \cite{Patel_Thesis}.}
\label{mem}
\end{figure}

A very intriguing proposal to resolve this question was put forward by E. Khan \cite{Khan2009,Khan2009b} that the mutual enhancement magicity (MEM) effect may play a role in nuclear incompressibility as well. MEM refers to a strong underbinding observed in the Hartree-Fock-Bogoliubov (HFB)  mass formulas for all doubly magic nuclei and their immediate neighbors formed by adding or removing not more than one nucleon \cite{mem1,mem2}. It was argued that $^{208}$Pb, being a doubly-magic nucleus, is ``stiffer'' than the open-shell nuclei and the incompressibility obtained from doubly-magic nuclei would invariably lead to an overestimation of the ISGMR energies in the open-shell nuclei \cite{Khan2009,Khan2009b}. While it was not at all clear as to why the MEM effect in nuclear masses would translate to nuclear incompressibility as well, Khan predicted \cite{Khan2009b} from calculations in the constrained Hartree-Fock-Bogoliubov (CHFB) framework that the ISGMR centroid energy in $^{208}$Pb would be higher than the corresponding values in the $^{204,206}$Pb isotopes by $\approx$600 keV. The predicted excitation-energy difference was large enough to be examined experimentally, considering the current experimental uncertainties of
ISGMR measurements, and these measurements were carried out at RCNP \cite{Darshana2013,Patel_Book}. However, the ISGMR strength
distributions of the $^{204,206,208}$Pb isotopes, measured in the same experiment, were found to be practically identical (see Fig.~\ref{mem}), ruling out any consequences of the MEM effect in nuclear incompressibility, and leaving the question of ``softness'' of the open-shell nuclei an important ``open problem'' in nuclear structure theory \cite{fluffy6}.

%% file: incompressibility_2.tex
\section{Definition of the nuclear incompressibility and its extraction from data}
\label{sec:kk}

In this Section, we define in precise terms the incompressibility of
uniform symmetric matter, $K_\infty$, and deduce an analogous and plausible 
definition for the same quantity associated to a finite nucleus, $K_A$.
We relate these latter to the ISGMR and ISGDR energy.
Some of these concepts have been already introduced in previous review papers 
\cite{Blaizot1980,SKC:2006,Colo:2008}. 

\subsection{Definition of the infinite matter and finite nucleus
incompressibility}\label{sec:ka}

When dealing with bulk materials, one usually define their {\em compressibility} as
\begin{equation}
\chi \equiv -\frac{1}{V}\left( \frac{\partial P}{\partial V} \right)^{-1},
\end{equation}
where $P$ and $V$ are, respectively, pressure and volume. The inverse of this
quantity, namely the incompressibility, $\chi^{-1}$, can be straighforwardly introduced. For instance, water and steel have values of $\chi^{-1}$ of 
2.2 $\cdot$ 10$^9$ and 1.6 $\cdot$ 10$^{11}$ Pa, respectively. 

If we consider a system at zero temperature, with a fixed number of particles $A$, where
the density $\rho=A/V$ is the basic variable, the incompressibility can also be written as
\begin{equation}
\chi^{-1} = \rho^3 \frac{d^2}{d\rho^2}\left( \frac{E}{A} \right).
\end{equation}
Nuclear matter is an ideal object but also a fairly good approximation 
of matter inside
the atomic nuclei. Matter inside neutron stars is also rather uniform 
over a broad density range. Hence, the interest in the incompressibility
in the nuclear case. We consider the case of symmetric matter, 
where the numbers of protons and neutrons are the same.

Symmetric nuclear matter has a state of minimum $E/A$ at the so-called nuclear saturation 
density, $\rho_0$ = 0.166 fm$^{-3}$. Around this minimum, one can write
\begin{equation}
\frac{E}{A}(\rho) = \frac{E}{A}(\rho_0) + \frac{1}{2}K_\infty\left( \frac{\rho-\rho_0}{\rho_0} 
\right)^2 + \ldots, 
\end{equation}
where the factors come from the fact that this expansion has been conceived and
written in terms of the Fermi momentum, $k_F$, and where the compression modulus or nuclear 
incompressibility $K_\infty$ shows up as 
\begin{equation}\label{eq:incompr}
K_\infty = 9\rho_0^2 \frac{d^2}{d\rho^2} \left( \frac{E}{A} \right)_{\rho=\rho_0}.
\end{equation}
This quantity is related to the incompressibility
$\chi^{-1}$ that has been previosuly introduced by 
\begin{equation}
K_\infty = \frac{9}{\rho_0}\chi^{-1}.
\end{equation}
Both quantities are a signature of the curvature of $E/A$ around its minimum or, in other
words, of the stiffness of symmetric nuclear matter. $K_\infty$ has dimension of energy and
can be expressed in MeV. We shall discuss in this paper values for
this quantity of the order of 250 MeV; this means, if we translate back 
in terms of $\chi^{-1}$, that nuclear matter is 
more incompressible than steel by almost 22 orders of
magnitude.

The value of $K_\infty$ has been the subject of early investigations, but all of them amounted
to basically mere speculations before the compression modes of nuclei were 
experimentally
identified: 
Bohr and Mottelson \cite{BohrMott1} mention values of the order of
120-130 MeV, for example. 
The discovery of compressional modes in finite nuclei paved the way
for the investigation of $K_\infty$ discussed in this review. We should, 
therefore, 
introduce the incompressibility of a finite nucleus $A$. 

In finite nuclei, the density is not uniform and so is not a simple number. The simplest
variation that keeps the shape of the system is a variation of only the second moment
of the density i.e. $\langle r^2 \rangle$. If we transform accordingly the second derivative
in Eq. (\ref{eq:incompr}), we obtain the following definition of the finite nucleus incompressibility: 
\begin{equation}\label{eq:KA}
K_A = 4 \langle r^2 \rangle_0^2 \frac{d^2}{d\langle r^2 \rangle^2}
\left( \frac{E}{A}
\right)_{\langle r^2\rangle = \langle r^2 \rangle_0}. 
\end{equation}

In general, for small variations of the nuclear density induced by
an external operator $F$, we can write the nuclear Hamiltonian
as 
\begin{equation}
H' = H + \lambda F,
\end{equation}
where $\lambda$ plays the role of a small perturbative parameter. 
If the expectation value of $H$ had a minimum (for instance, within the 
HF or HFB approximation), the expectation value of $H'$ will have a
minimum as well for every value of $\lambda$. Within these assumptions,
it is straightforward to write
\begin{equation}
\frac{d \langle H' \rangle}{d\langle F \rangle} = 
\frac{d \langle H \rangle}{d \langle F \rangle} + \lambda = 0,
\end{equation}
where all expectation values are ground-state expectation values
(i.e. expectation values in the state with minimal energy). The
last equation implies
\begin{equation}\label{eq:constrained1}
\frac{d \langle H \rangle}{d \langle F \rangle} = -\lambda.
\end{equation}

We can now invoke some theorems that are valid at least within
approximations like the nonrelativistic HF plus RPA, or HFB plus
QRPA. These theorems involve the so-called sum rules associated
with an external operator that have been introduced at the start 
of the paper (Sec. \ref{general}).

The energy-weighted strength function $m_1$ can be obtained
from the {\em Thouless theorem}\footnote{The original paper by
Thouless \cite{Thouless:1961} deals with the RPA case whereas the extension
to QRPA can be found in Ref. \cite{Khan:2002}.}, according to which
\begin{equation}\label{eq:EWSR}
m_1 = \sum_n E_n \vert \langle n \vert F \vert \tilde 0 \rangle \vert^2 
= \langle 0 \vert \left[ F, \left[ H, F \right] \right] \vert 0 
\rangle.
\end{equation}
In this equation, $\vert \tilde 0 \rangle$ is the (Q)RPA ground-state
as above but $\vert 0 \rangle$ is the HF ground-state: this feature
makes the expectation value of the double commutator easy to
be evaluated \cite{Harakeh_book}. Another theorem, still due to Thouless, 
is called dielectric theorem\footnote{Also in
this case, the original paper by Thouless \cite{Thouless:1960} deals
with HF plus RPA and the extension to HFB plus QRPA is much more
recent and can be found in Ref. \cite{Capelli:2009}.}, 
and concerns the inverse-energy weighted sum rule $m_{-1}$. It states that
\begin{equation}\label{eq:constrained2}
m_{-1} = \frac{1}{2}\frac{d\langle F \rangle}{d\lambda} = \frac{1}{2}\frac{d^2\langle H \rangle}
{d\lambda^2}.
\end{equation}

We now denote $\langle H \rangle$ by $E$ and we join the result of Eq. 
(\ref{eq:constrained1}) with the dielectric theorem (\ref{eq:constrained2}), so that we arrive at
\begin{equation}
\frac{d^2E}{d\langle F \rangle^2}=-\frac{d\lambda}{d\langle F\rangle}=
-\frac{1}{2m_{-1}}.
\end{equation}
We now restrict ourselves to the monopole operator. The EWSR provided by (\ref{eq:EWSR}) reads
\begin{equation}
m_1 = \frac{2\hbar^2}{m}\langle r^2 \rangle_0.
\end{equation}
Therefore, the finite nucleus incompressibility (\ref{eq:KA}) becomes
\begin{equation}\label{eq:ka}
K_A = 4 \langle r^2 \rangle_0^2 \frac{d^2}{d\langle r^2 \rangle^2}
\left( \frac{E}{A}
\right)_{\langle r^2\rangle = \langle r^2 \rangle_0} =
\frac{m \langle r^2 \rangle^2_0}{A\hbar^2}\frac{m_1}{m_{-1}}.
\end{equation}
The ISGMR energy can be identified with the so-called constrained energy $E_c \equiv \sqrt{\frac{m_1}{m_{-1}}}$.
From the latter formula we then deduce
\begin{equation}\label{eq:e_vs_ka}
E_{\rm ISGMR} = \sqrt{\frac{\hbar^2 A K_A}{m\langle r^2 \rangle_0}},
\end{equation}
which is precisely Eq. (3.48) of \cite{Blaizot1980}.
If we assume that the matter radius is known with sufficient
accuracy (for instance, because elastic scattering data are obtained
at the same time when the monopole is excited and measured), then
this latter relation shows how the ISGMR energy can provide
the finite nucleus incompressibility. 

Another possibility consists in identifying the ISGMR energy with 
$E_s \equiv \sqrt{\frac{m_3}{m_{1}}}$, that is, the so-called scaled energy. A
formula similar to Eq. (\ref{eq:e_vs_ka}) can be derived \cite{Treiner1981}, and
in general this leads to a larger value of $K_A$ since 
$E_s$ is larger than $E_c$. Using the Thomas-Fermi theory in the limit of very large
$A$, the authors of Ref. \cite{jennings:1980} have deduced that $K_A^s$ extracted from 
$E_s$ should be larger by a factor 10/7 than $K_A^c$ extracted from $E_c$. 
Anyway, a relationship between the ISGMR energy
and the square root of $K_A$ can be inferred in either case, and the precise value of
the factor is not relevant to our discussion below.
 
The same formulas can be also found in \cite{Stringari1982}, where a simple yet
realistic form for the energy functional is assumed and the hydrodynamic model is employed.
In this paper, also the expressions of $K_A$ both from constrained and scaled ISGDR energies are
reported [cf. Eqs. (12) and (13).] The correlation between ISGDR energies and the 
square root of $K_A$ is found in these expressions as well. This is the key point for the following.

Now, we move to the extrapolation
to the infinite matter case. 

\subsection{Relationship between $K_A$ and $K_\infty$}
\label{sec:ka_kinf}

One could assume the existence of a relationship between the finite nucleus
incompressibility $K_A$ and the nuclear matter incompressibility $K_\infty$, 
in terms of an expansion of the same kind as the liquid-drop formula, the so-called 
leptodermous expansion:
\begin{equation}\label{eq:liquid_drop}
K_A = K_\infty + K_{surf}A^{-1/3} + K_{\tau} \left( \frac{N-Z}{A} \right)^2 
+ K_{Coul} Z^2A^{-1/3}.
\end{equation}
Blaizot, in Sec. 6.2 of \cite{Blaizot1980}, has started from a simple 
liquid-drop expression for $E/A$ and has extracted formally the terms 
defined by this, through the use of Eq. (\ref{eq:ka}). One key ingredient in
this derivation are the formulas related to the existence of the 
saturation density $\rho_0$. From the discussion by Blaizot, it is also
useful to retain the simple but important concept that the
coefficients that appear in Eq. (\ref{eq:liquid_drop}) cannot be
interpreted as being second derivatives of the corresponding
coefficients in the mass formula. Some other works also deal with the derivation of Eq. (\ref{eq:liquid_drop}) 
(cf. e.g. \cite{Satpathy1999}).

Eq. (\ref{eq:liquid_drop}) can be used (or not used!) in different ways. Three 
strategies are relevant for our discussion.
\begin{enumerate}
\item One could trust Eq. (\ref{eq:liquid_drop}) literally, and try
to determine the parameters appearing therein from experimental data.
\item One could use Eq. (\ref{eq:liquid_drop}) in a very mild way, 
and infer from it only a linear correlation between $K_A$ and 
$K_\infty$. 
Together with Eq. (\ref{eq:e_vs_ka}) and the considerations made below that formula, this linear 
correlations would imply, among other things, that
\begin{equation}
\frac{\delta K_\infty}{K_\infty} \approx 
2 \frac{\delta E_{\rm ISGMR}}{E_{\rm ISGMR}};
\end{equation}
we will use this rule of thumb throughout the paper.
\item One could try different ways to correlate $K_\infty$ to
data, as the previous two strategies may be not free of
difficulties.
\end{enumerate}
The strategy \# 3 is mainly used by the authors of Refs. 
\cite{KMV:2012,KM:2013}. 
Most of the work of the last two decades, and consequently most of
our discussion here, is based on strategy \# 2. As we discuss in this subsection, 
strategy \# 1 has many drawbacks. A general consensus has emerged that
a far better procedure is to trust microscopic theories (mainly those
based on EDFs as outlined in Sec. \ref{sec:theory}) that can calculate 
on equal footing both $K_A$ (or, equivalently, the ISGMR energy) and 
$K_\infty$. If a given theory can reproduce the experimental 
values of $K_A$, the associated value of $K_\infty$ can be 
considered the correct one. Unfortunately, we will conclude at the end
that there is not at present a single model that can reproduce
all measured monopole data. By collecting the efforts that theorists have
made so far, we will set limits more than pin down a unique value. 
This partial and not total success of strategy \# 2 has motivated
some attempt to go back to strategy \# 1 as a
qualitative guideline, or within localized regions of the isotope chart.

We start with discussing strategy \# 1. After the first systematic 
measurements of the monopole reasonances, in the 1980s, there 
was hope that they could lead to extrapolating the value 
of $K_\infty$ \cite{Blaizot1980}. This hope faded during 
the 1990s; in fact, trying to use quantitatively the 
liquid-drop expansion (\ref{eq:liquid_drop}) in order to 
extract the values of the different terms, using the
experimental values of $K_A$ available at that time, was proven 
to be unreliable \cite{Pearson1991,SY:1993}. Data were scarce
and, even more importantly, parameters had correlations. In practice, the authors 
of \cite{SY:1993} have shown that values between 
200 MeV and 350 MeV are all possible in following this procedure. This is why, at the turn 
of the century, consensus has emerged that only theory can provide a bridge
between the measurements 
in finite nuclei and the value of $K_\infty$.
Recently, other theoretical papers that we shall discuss below 
\cite{Vesely:2012,Vinas2015} have strenghtened the rationale behind this belief.

A further problem in the use of Eq. (\ref{eq:liquid_drop}) is
the proper definition of the energy to be inserted in $K_A$ so that
the liquid-drop expansion is meaningful. In Sec. \ref{sec:ka} 
we have seen that using the constrained or the scaled energy
does not lead to the same value of $K_A$. 
The former (latter) choice is associated with using the sum rule 
$m_{-1}$ ($m_3$), that is, with measurements
at low (high) energy that may suffer from larger errors than
those around the main peak. Microscopic models based on EDFs are
free from these drawbacks; one compares the ISGMR and ISGDR energies
with experiment directly, and in this way one can get 
rid of the ambiguities related to the use of either the constrained or
the scaled energy.

Despite all the serious warnings, a recent attempt to revisit the
liquid drop expansion to extract $K_\infty$ and the other
parameters that appear in Eq. (\ref{eq:liquid_drop}) 
can be found in Ref. \cite{Stone2014}. The final result
provides larger $K_\infty$ than all the attempts based on
other strategies, although there is not absolute incompatibility. Ref. \cite{Stone2014} 
also constitutes an excellent overview of our present understanding
of the various terms of Eq. (\ref{eq:liquid_drop}) although the pairing term is not considered.
The results obtained in Ref. \cite{Stone2014} are also
reported in Table \ref{table:Kcoeff}.

The best known term, albeit the least important, is of course 
the Coulomb term that reads:
\begin{equation}
K_{Coul} = \frac{3}{5}\frac{e^2}{r_0} \left[ 
1 - \frac{27\rho_0^2}{K_\infty}\frac{d^3}{d\rho^3}
\left( \frac{E}{A}
\right)_{\rho=\rho_0} \right].
\end{equation}
The second term in the parenthesis introduces an uncertainty 
on this quantity, which is nonetheless estimated to be less
than about 20\% in Ref. \cite{Sagawa2007}, where the value 
$K_{Coul} = 5.2 \pm 0.7$ MeV is quoted, extracted from a host
of EDFs. Other estimates do not significantly vary from this
range. 

In the case of $K_{surf}$, a certain tendency to be related to 
$-K_\infty$ has been pointed out in Ref. \cite{Patra2002}, where 
extended Thomas-Fermi calculations have been carried out using 
EDFs in semi-infinite nuclear matter. However, we cannot make any firm
statement about the value of $K_{surf}$. The authors 
of Ref. \cite{Stone2014} claim that the (negative) ratio $K_{surf}/K_\infty$ 
is much larger in absolute value, for example. 
Certainly, as we shall argue below, our incomplete knowledge 
of the density dependence of the effective nuclear interaction or
of the EDF, plays a significant role here.

A similar statement can be made for the symmetry term, $K_\tau$. 
However, more detailed investigations have concerned this term
recently, as we discuss below. 

Even more debated is the curvature term $K_{curv}A^{-2/3}$ which is another
possible extension of the liquid-drop expansion. Last but not least, we 
should mention the possibility that Eq. (\ref{eq:liquid_drop}) is
complemented with a pairing term $K_{pair}$. This possibility has
been explored in Ref. \cite{Khan2010} by employing some effective 
pairing interactions that produce realistic values of the pairing gaps
in uniform matter. The conclusion is that the pairing term has only
a small effect, of the order of 2\% or slightly more. We will use
this fact below.

All this discussion should make the reader aware of our present understanding 
of the various terms of the liquid-drop model expansion 
of $K_A$. We end this subsection by stressing that, 
{\em within a given model}, it is possible to calculate $K_A$ in
as many nuclei as desired and then fit the results with a formula
of the type (\ref{eq:liquid_drop}). This is a useful test, albeit
biased by the model assumptions.

Extrapolations of $K_A$ for large $A$ are extremely demanding. Already 
in Ref. \cite{Bohigas1976}, it was shown that one needs to go to
extremely large nuclei to recognise the behaviour of $K_A$ in terms of
the liquid drop model formula. More recently, in the case of the
binding energy formula and with modern EDFs, it has been confirmed 
in \cite{leptodermous_PG} that one needs to go to nuclei with 
$\approx$ 10$^6$ nucleons, discard the Coulomb force, deal with
shell corrections, and still be left with significant uncertainties.

The authors of \cite{Vesely:2012} have not attempted such an extrapolation
but simply fit their QRPA results with Eq. (\ref{eq:liquid_drop}),
by including a surface symmetry term but not curvature or pairing terms. 
Such an attempt should be taken as a parametrization of the theoretical 
results and/or a check of general properties and model features. We report
in Table \ref{table:Kcoeff} the results obtained by fitting one 
of the models, namely UNEDF0 complemented with zero-range pairing. 

One should note that, 
in Ref. \cite{Vesely:2012}, values of $m_1/m_0$ have been 
used, while this is not supported by the theoretical arguments 
discussed in Sec. \ref{sec:ka}.
This may explain why the values for $K_\infty$, like the one in the
Table, are different than the nominal values associated with the EDF.

Within a similar philosophy, but in a different way, the authors of 
Ref. \cite{Vinas2015} have used the NL3 functional and 
calculated the monopole energies in 750 nuclei within the extended Thomas-Fermi approach. Then, they 
have extracted the coefficients of Eq. (\ref{eq:liquid_drop}) by means of a fit. The values obtained
therefrom are incompatible with the nominal values of NL3 obtained in infinite or semi-infinite matter.
This is due to the approximated form of Eq. (\ref{eq:liquid_drop}). 
The conclusions of Refs. \cite{Vesely:2012,Vinas2015} points clearly to the drawbacks that are inherent
in the use of the leptodermous expansion of Eq. (\ref{eq:liquid_drop}).

\begin{table}[h]
\centering
\begin{tabular}{cccc}
\hline
& Ref. \cite{Stone2014} & Ref. \cite{Vesely:2012} \\
\hline
$K_\infty$ & 250 -- 315 MeV & 257 $\pm$ 4 MeV \\
$K_{surf}/K_\infty$ & -2.4 -- -1.6 & -1.6 $\pm$ 0.06 \\
$K_\tau$ & -810 -- -370 MeV & -550 $\pm$ 30 MeV \\
$K_{\tau,s}$ & -1020 -- 160 MeV & 740 $\pm$ 100 MeV \\
$K_{Coul}$ & -5.2 $\pm$ 0.7 MeV & -5.1 $\pm$ 0.4 MeV \\
\hline
\end{tabular}
\caption{Very different attempts to extract the values of the coefficients 
of the terms appearing in Eq. (\ref{eq:liquid_drop}). See the text for a
complete discussion.
}\label{table:Kcoeff}
\end{table}

\subsection{Extraction from $^{208}$Pb and $^{90}$Zr data}

A first, certainly too naive, approach assumes that the 
exact EDF has all the correct terms and reproduces systematically the ISGMR energies
of nuclei. Restricting first to heavy, magic and spherical nuclei, 
among which $^{208}$Pb is the obvious benchmark, one could claim
that the EDF that reproduces the ISGMR energy in $^{208}$Pb has
the correct incompressibility. Different functionals are, however, 
built by using different kinds of ansatz. In particular, they
are characterised by different kinds of density dependence. This
creates a model dependence in the extraction of $K_\infty$, even
restricting to a single nucleus.

This has been clearly illustrated for the first time in Ref. \cite{Colo2004}.
In that work, Skyrme forces with different density dependence have
been determined by using the same fitting protocol and reaching
comparable quality. If one changes the density dependence, one
can reproduce the ISGMR energy in $^{208}$Pb either with forces
having $K_\infty$ around 230 MeV or around 250 MeV. This range spans, to
a good extent, also the incompressibility values associated with
Gogny and RMF functionals. Going back to Fig. \ref{gmr_rpa}, one can 
note the similar results obtained with three functionals that have,
respectively, $K_\infty$ = 225 MeV (D1M), $K_\infty$ = 230 MeV (SAMi), 
and $K_\infty$ = 251 MeV (DD-ME2). We can add a specific word
concerning relativistic functionals. Point-coupling Lagriangians 
introduced in Ref. \cite{Niksic2008} reproduce the ISGMR 
in $^{208}$Pb with the same quality as DD-ME2 and with a lower
value of $K_\infty$, that is, 230 MeV. The conclusion, which
was already drawn a few years ago by several authors, is that
choosing a nonrelativistic or a covariant formulation is not
at all the key element that contributes to our remaining
uncertainty on $K_\infty$. 

Usually, the calculations that reproduce the ISGMR in $^{208}$Pb 
work in the case of $^{90}$Zr as well. Looking (mainly) at those
nuclei, the conclusion that $K_\infty$ should be in the range 
240 $\pm$ 20 MeV has been reached since about one decade 
\cite{SKC:2006}. We now turn our attention to non-magic nuclei.

\subsection{Extraction from larger sets of nuclei: superfluid systems and more}

Most nuclei are neither magic nor spherical. In the 1980s and 1990s, but also
in the first years after the turn of the new millennium, still attention
was focused on $^{208}$Pb. Systematic measurements on 
the Sn isotopes raised the concern whether superfluidity affects
the ISGMR and, in turn, if superfluid systems have a different
incompressibility. Several works have since been devoted to the ISGMR
in superfluid nuclei \cite{fluffy3,Khan2009,Khan2009b,fluffy5,fluffy6,Avogadro:2013}.

The issue has been already discussed in Sec. \ref{sec:fluffy}, where
we had concluded that is difficult to reconcile $^{208}$Pb and Sn data, namely to
reproduce both with a single model having a given value of
$K_\infty$ (cf. Fig. \ref{gmr_sn_pb}). 

Along a similar line, the authors of Ref. \cite{Avogadro:2013} 
have considered a relatively large sample of nuclei and isotopic
chains. They have paid special attention to the self-consistency 
in the pairing channel, and included pairing forces that depend on
the neutron-proton imbalance. Our aforementioned conclusion 
is also echoed by the authors of Ref. \cite{Avogadro:2013}, 
and made more general: if from the double-magic nuclei one turns 
to Sm, Sn and Cd, data seem to indicate a lower value of $K_\infty$, which
clearly points to the present limitations of our models and our understanding.

\subsection{Alternative methods}

The impasse produced by the question ``Why is Sn (with other open-shell
nuclei) so fluffy ?'' has motivated some attempts to shift the current
attitude towards the problem of the nuclear incompressibility.

As we have already mentioned, the discrepancy between the incompressibility 
from different nuclei point to our incomplete understanding of EDFs,
or to the fact that they are characterised by different balance
of volume, surface and asymmetry incompressibility. The surface term
is certainly one of the first which must be put under scrutiny. 

The importance of the nuclear surface is the starting point of the
reasoning done in Refs. \cite{KMV:2012,KM:2013}; therein, the authors 
have pointed out that in medium-heavy nuclei
the average density is not the saturation density but rather
something around $\approx$ 0.1 fm$^{-3}$, which is defined as crossing
density $\rho_c$. Many nuclear
properties like the symmetry energy or the pairing gap are known in fact
to be sensitive to this average density. 
At such a density models like different EDFs tend to give similar predictions for different
observables. Consequently, the authors of \cite{KMV:2012,KM:2013} have 
defined a density-dependent incompressibility that reads
\begin{equation}\label{eq:rhoincompr}
K(\rho) = 9\rho_0^2 \frac{d^2}{d\rho^2}\left( \frac{E}{A} \right)_{\rho},
\end{equation}
and shown that this quantity has the smallest disperson, when
microscopically calculated, around the crossing density $\rho_c$. The spread
in the microscopic predictions is, then, associated with the derivative
of the density-dependent incompressibility around $\rho_c$, or with
\begin{equation}
M_c \equiv \left. 3\rho_c \frac{d}{d\rho} K(\rho) \right\vert_{\rho_c}.
\end{equation}
This claim is supported by correlation plots that show better $E_{\rm ISGMR}-M_c$ 
correlations than $E_{\rm ISGMR}-K_\infty$ correlations, although in a
limited number of nuclei. It remains to be ascertained to which extent
the limited number of parameters in the EDFs affects these conclusions.

%% file: deformation_plus_other_3.tex
\section{Effects of Deformation on ISGMR and ISGDR}
\label{deformation}

\begin{figure}[h]
\vspace*{-0.2cm}
\centering\includegraphics [height=0.50\textheight]{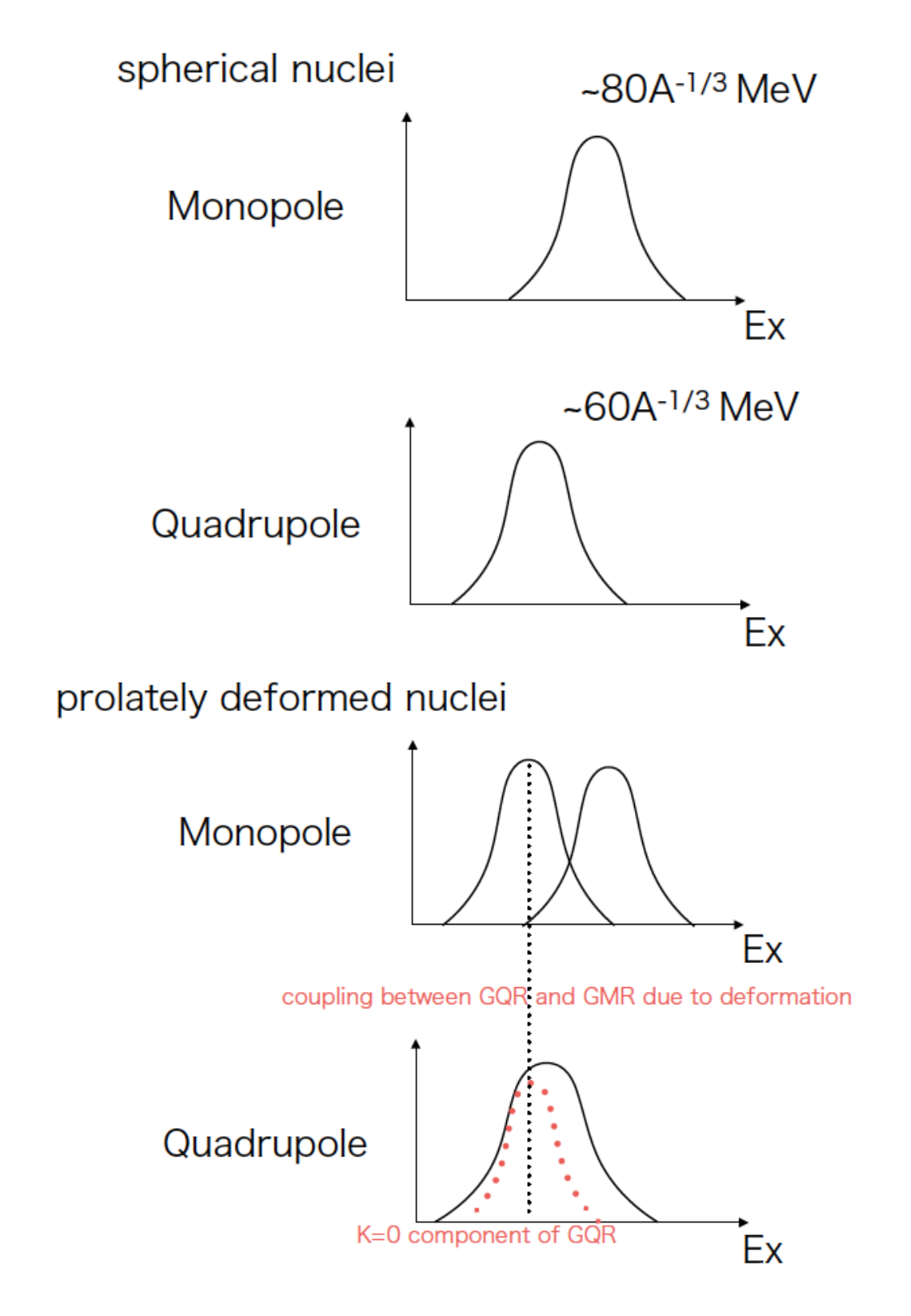}
\caption{Schematic representation of the effect of ground state deformation on the ISGMR and ISGQR. Figure courtesy of K. Yoshida, Kyoto University, Japan.}
\label{splitting}
\end{figure}

The splitting of the IVGDR into two components in deformed nuclei has been known for a very long time; this splitting was attributed to the different frequencies of dipole oscillations along the major and minor axes of a symmetric ellipsoidal nuclear shape \cite{Harakeh_book}. The ISGQR, on the other hand, exhibited only a small broadening due to deformation of the ground state: the $L$=2 resonance splits into three components corresponding to the $K$ quantum numbers $K$=0, 1, and 2. These components are rather closely-spaced leading to an overall increase in ISGQR width in a deformed nucleus ($^{154}$Sm, for example), as compared with that in a spherical nucleus ($^{144}$Sm)  \cite{kishimoto}. Naively, one would have expected the $L$=0 ISGMR to remain unaffected by the deformation of the ground state; however, one does see a ``splitting'' of the ISGMR strength in deformed nuclei, as first reported by Garg et al. in $^{154}$Sm \cite{umesh_prl2}. This ``splitting'' results from a coupling of the $K$=0 component of the ISGQR with the ISGMR. The monopole and quadrupole vibrations in the deformed nuclei no longer have a unique $J^\pi$, each containing a mixture of $L$=2 and $L$=0 instead. Thus, there are two $K$=0 states, the lower predominantly $L$=2, but containing significant $L$=0 strength; the upper predominantly $L$=0 but with a small amount of $L$=2 strength \cite{umesh_prl2}. This is represented schematically in Fig.~\ref{splitting}.

\begin{figure}[h]
\centering\includegraphics [height=0.4\textheight]{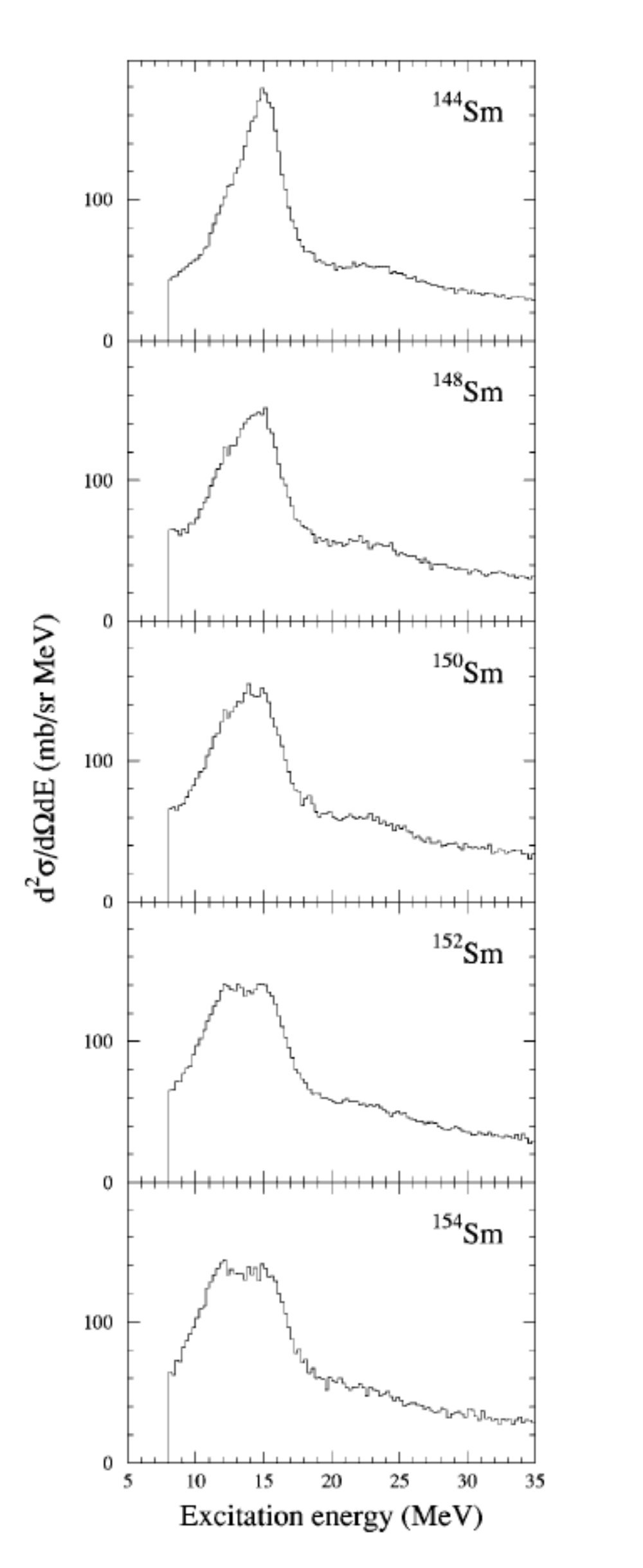}
\caption{Inelastic $\alpha$ scattering spectra at $\theta_{\rm av}$=0.7$^{\circ}$ and $E_\alpha$=386 MeV for $^{144-154}$Sm. Figure from Ref.~\cite{Itoh_prc2003}.}
\label{isgmr_sm}
\end{figure}

In the Sm isotopes, which range from the ``spherical" $^{144}$Sm (deformation parameter $\beta_2$ = 0.09) to the well-deformed $^{154}$Sm ($\beta_2$ = 0.34), the evolution of the ISGMR strength as a function of increasing deformation is observed rather succinctly \cite{Itoh_prc2003}: a single peak in $^{144}$Sm transmutes into two clearly discernible components in case of $^{154}$Sm. Indeed, this transmutation is evident even in the ``0$^{\circ}$'' inelastic scattering spectra (see Fig.~\ref{isgmr_sm}). A clear two-component structure in the ISGMR strength distribution was reported in the A \& M work as well \cite{dhybg,dhy-sm2}.

\begin{figure}[h]
\centering\includegraphics [height=0.25\textheight]{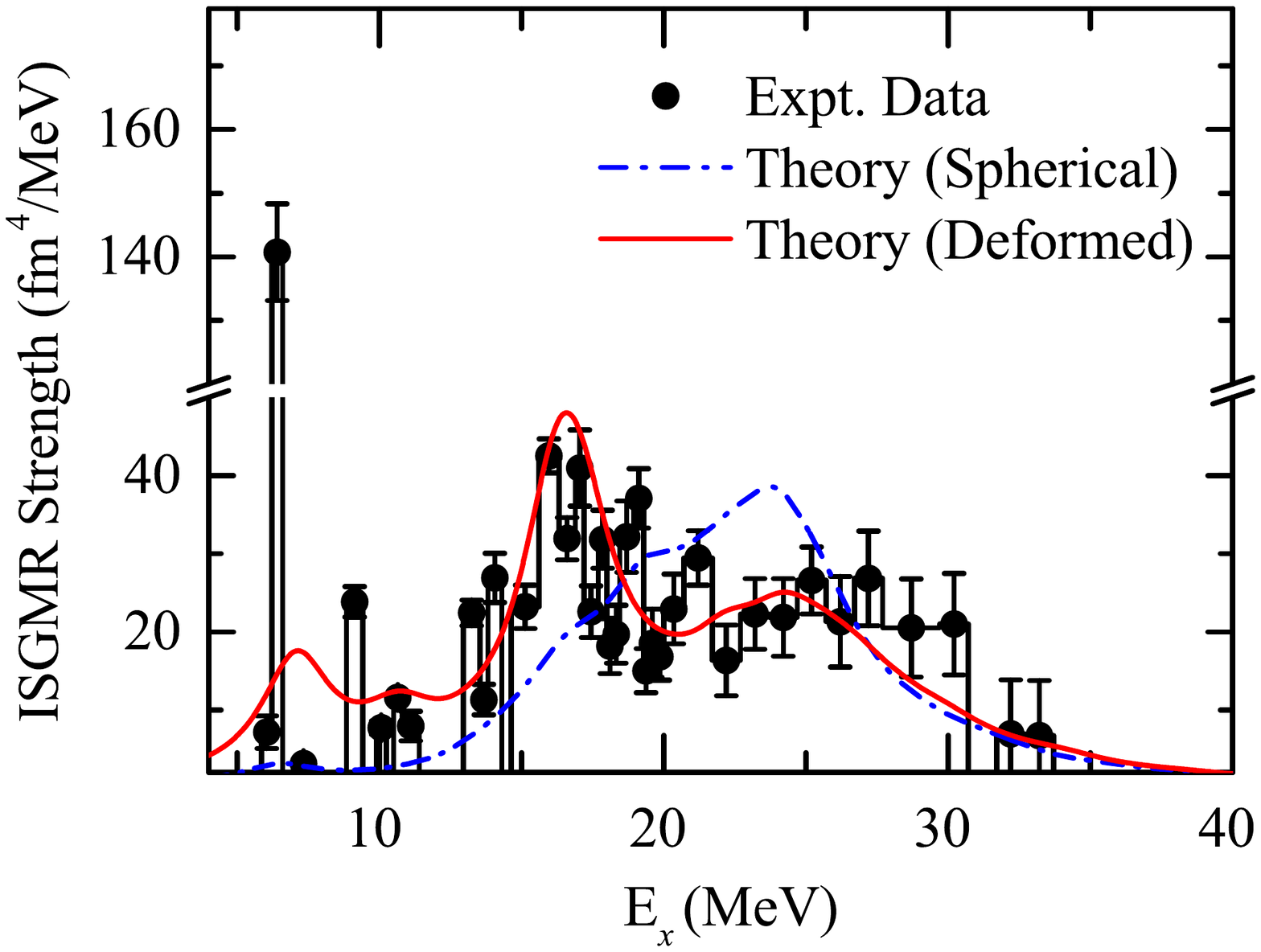}
\caption{ISGMR strength distribution in $^{24}$Mg (solid circles). The dash-dotted (blue)  and solid (red) lines  show microscopic calculations for spherical and prolate ground-state deformation, respectively. Figure from Ref. \cite{yogesh-mg}.}
\label{mg24}
\end{figure}

The effect of deformation on ISGDR is similar in that there is coupling between the $K$=1 (as well as the $K$=0) components of the ISGDR ($L$=1) and the ISHEOR ($L$=3). However, because of the aforementioned LE component of the ISGDR even in the spherical nuclei, this coupling is not as clearly evident as in case of the ISGMR. However, two effects are discerned in going from $^{144}$Sm to $^{154}$Sm, both consistent with the coupling between the $K$=1 components of the two resonances \cite{Itoh_plb,Itoh_prc2003}: i) the relative strength of the LE component of the ISGDR increases smoothly with nuclear deformation, whereas the strength of the HE component remains constant; ii) the width of the LE component also increases with increased deformation. A direct comparison of the ISGDR strength in the deformed nucleus, $^{154}$Sm, is complicated, of course, by the uncertain nature of the LE component.

The most interesting result on the effect of deformation on the ISGMR strength was presented recently in the nucleus $^{24}$Mg \cite{yogesh-mg2,yogesh-mg}. Generally, the ISGMR strength (indeed, all multipole strengths) in the lighter-mass nuclei (A$<$58) is fragmented over a wide excitation energy range and does not form a nice ``peak'' as in the higher-A nuclei (see, for example, Refs. \cite{lui2001,Itoh_32S}). With that, any effects of deformation would be expected to be very difficult to discern. However, in recent RCNP work on this deformed nucleus, a two-peak structure was observed in the ISGMR strength distribution, indicative of the ``splitting'' of the ISGMR. The observed strength distribution is in good agreement with microscopic calculations for a prolate-deformed ground state for $^{24}$Mg, carried out in a deformed Hartree-Fock-Bogoliubov (HFB) approach and the quasiparticle random-phase approximation (QRPA) with a Skyrme and Gogny energy-density functional \cite{Yoshida2010,Peru2008}, and is in contrast with that expected if a spherical ground state is assumed for this nucleus (see Fig.~\ref{mg24}). Another set of calculations, with the SkM*, SVbas, and SkP$^{\delta}$ interactions, further confirms that the $E$0 peak at $E_x\sim$16 MeV is caused by the deformation-induced coupling of the ISGMR with the $K$=0 part of the ISGQR \cite{kvasil}.
This is the first time that such a splitting had been observed in a light-mass nucleus, indeed in any nucleus other than the well-deformed Sm nuclei (as discussed above) and $^{238}$U \cite{kvi_prl}. A similar effect has since been observed in the oblate-deformed nucleus $^{28}$Si as well \cite{peach}. Recently, Kvasil and 
collaborators \cite{Kvasil2015,Kvasil2016} have carried out a detailed and systematic theoretical investigation of the effects of deformation on ISGMR within the QRPA approach, using two different Skyrme forces. Consistent with the experimental results and previous theoretical work, they find that the ISGMR broadens and attains a two-peak structure due to the coupling with the quadrupole giant resonance. 

\section{Are There Other Structure Effects on ISGMR?}
Since giant resonances and nuclear incompressibility are collective (bulk) phenomena, one expects only a smooth variation of the properties of the ISGMR with mass (e.g. the $A^{1/3}$ dependence of the energy) and one does not expect 
very strong 
variations related to the internal structure of the nuclei (the shell model orbitals being populated, for example). 

\begin{figure}[!h]
\centering\includegraphics [height=0.28\textheight]{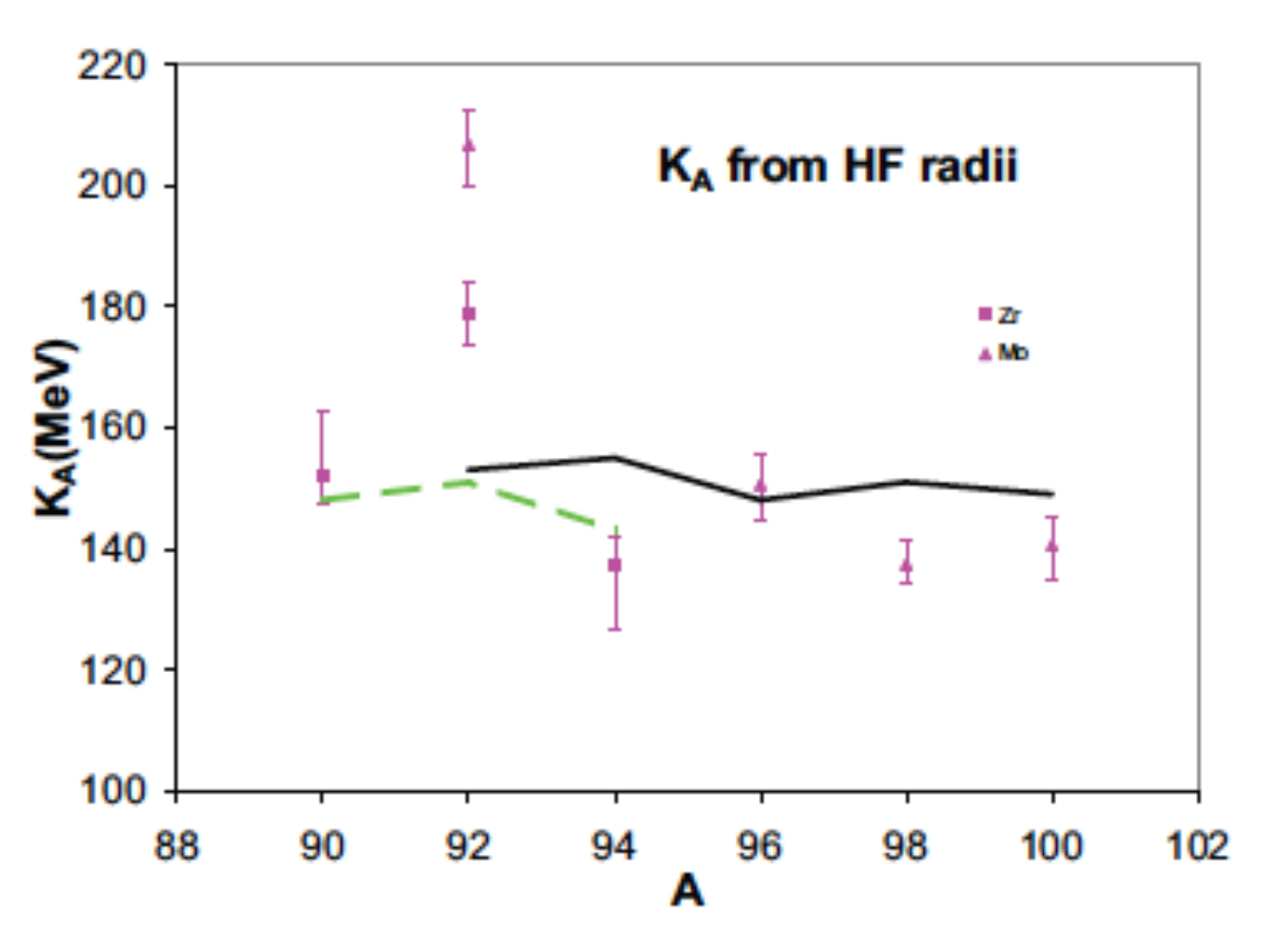}
\caption{The scaling model $K_A$ values obtained from the measured scaling energies $\sqrt{m_3/m_1}$ are shown for the Zr isotopes (squares) and for the Mo isotopes (triangles) plotted versus A. The error bars reflect the uncertainties in $\sqrt{m_3/m_1}$. Also shown are lines connecting the HF-based RPA values of $K_A$ calculated within HF-RPA using the KDE0v1 interaction for the Zr (dashed line) and Mo (black line) isotopes. Figure from Ref. \cite{dhy_a90_2}.}
\label{tamu_a90}
\end{figure}

\begin{figure}[!h]
\centering\includegraphics [height=0.45\textheight]{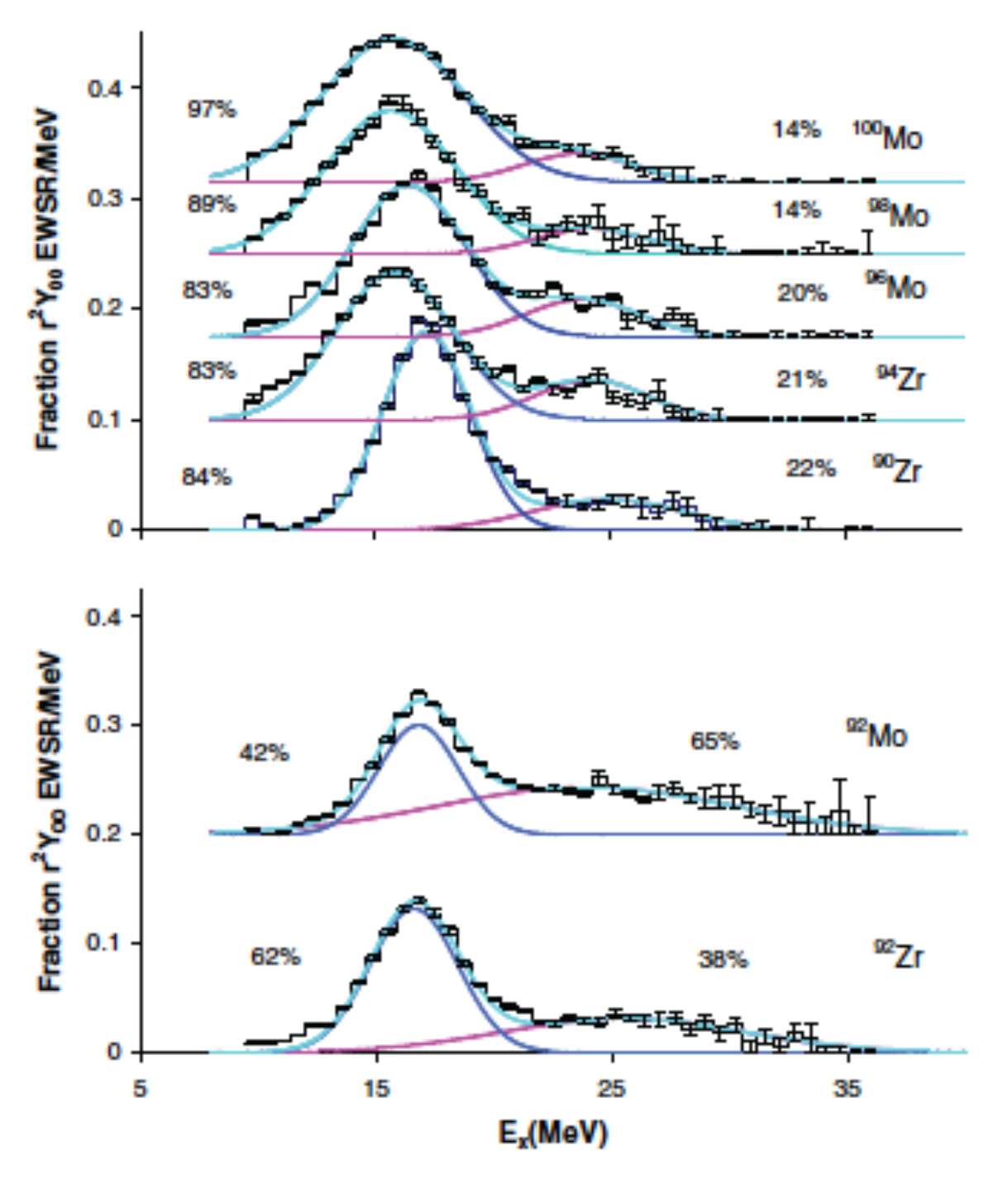}
\caption{The black histograms show the fraction of the $r^2Y_{00}$ sum rule obtained for Mo and Zr isotopes plotted as a function of excitation energy. Superimposed are Gaussian fits to the two components of the distributions as well as the sum of the fits. On the left side are the strengths of the lower energy peak while on the right side the strengths of the higher energy peaks are listed, all given as a percentage of the $r^2Y_{00}$ sum rule. Figure from Ref. \cite{dhy_a90_2}.}
\label{tamu_a90_2}
\end{figure}

This picture of purely collective behavior was put to question by recent results on the Zr and Mo isotopes reported by the TAMU group \cite{dhy_a90_2,DHY_2016_AllMo,Krishi_2015,Button2016}. They observed a dramatic variation in the extracted ISGMR strength distributions in these nuclei. In particular, the A=92 nuclei, $^{92}$Zr and $^{92}$Mo, emerged quite disparate from the others: the ISGMR
energies ($E_{\rm ISGMR}$) for $^{92}$Zr and $^{92}$Mo were observed to be, respectively, 1.22 and 2.80 MeV higher than that of $^{90}$Zr. Consequently, the $K_A$ values determined for $^{92}$Zr and $^{92}$Mo were, respectively, $\sim$27 MeV and $\sim$56 MeV higher than the $K_A$ for $^{90}$Zr (see Fig.~\ref{tamu_a90}). This was the consequence of significant ISGMR strength at higher excitation energies, constituting an additional ``peak'' at $E_x\sim$25 MeV. This second peak was excited with much higher strength in the A=92 cases when compared with that in the other measured Zr and Mo nuclei (see Fig.~\ref{tamu_a90_2}). These results implied significant nuclear structure contribution to the nuclear incompressibility in this mass region. Such nuclear structure effects have not been observed in any of the investigations of ISGMR going back to its first identification in the late 1970's \cite{muhsin1,dhy1} and, indeed, were contrary to the standard hydrodynamical picture associated with collective oscillations \cite{stringari}.

\begin{figure}[!h]
\centering\includegraphics [height=0.27\textheight]{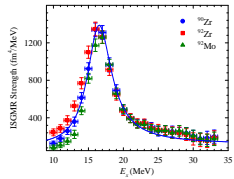}
\caption{ISGMR strength distributions for $^{90}$Zr (filled circles), $^{92}$Zr (filled squares), and $^{92}$Mo (filled triangles). The solid line represents the Lorentzian fit for $^{90}$Zr. Figure from Ref. \cite{YKGPLB2016}.}
\label{und_a90}
\end{figure}

An experiment was subsequently performed at RCNP to further investigate and elucidate this unusual, and unexpected, effect. In ``background-free'' inelastic $\alpha$-scattering experiments on $^{90,92}$Zr, and $^{92}$Mo, no evidence was found for this anomalous effect in the A=92 nuclei \cite{YKGPLB2016,gupta-private}. This was clear already in the 0$^{\circ}$ spectra (where the ISGMR is excited maximally) and in the ``difference spectra'' (which contain primarily the ISGMR strength; see the discussion in the section on Experimental Techniques): these spectra are virtually identical for the three nuclei. This was further corroborated by extracting the ISGMR strengths using MDA; the extracted strengths were also observed to be nearly identical and, in particular, no discernible differences were observed in the the strength near $E_x\sim$25 MeV for the three nuclei (see Fig.~\ref{und_a90}).

These RCNP results appear to establish clearly, and strongly, that determination of nuclear incompressibility in nuclei is not influenced in any appreciable manner 
by the underlying nuclear structure.

The obvious question is why the RCNP results are so different from those obtained by the TAMU group. The answer, most likely, lies in the way the ``background'' in the inelastic scattering spectra is accounted for in the two approaches (see the discussion in Sec. \ref{experiment_techniques}). In the RCNP work, all instrumental background is eliminated because of the superior optical properties of the Grand Raiden Spectrometer (see, e.g., Fig.~\ref{bgsubtraction}), leaving the physical continuum as part of the excitation-energy spectra. In the TAMU work, an empirical background is subtracted which leads to subtraction of the physical continuum as well. It is quite possible, and perhaps likely, that this background subtraction approach is responsible for the differing strengths observed for various nuclei in their work. Since there is no arbitrariness involved in the background-subtraction procedure employed in the RCNP work, it has been argued that their results may be deemed more reliable \cite{YKGPLB2016}.

%% file: ktau_2.tex
\section{$K_\tau$ from ISGMR in Sn and Cd Isotopes}
Measuring the ISGMR in a series of isotopes can provide an ``experimental'' value for the asymmetry term of nuclear incompressibility, $K_{\tau}$ [see Eq.~(\ref{eq:liquid_drop})]. As pointed out by Patel et al. \cite{Darshana2012},
the finite-nucleus asymmetry term $K_{\tau}$, although closely related, should not be confused with the corresponding term in infinite nuclear matter -- a quantity also denoted by $K_{\tau}$ at times, but written here as
$K_{\tau}^{\infty}$. Indeed, $K_{\tau}^{\infty}$ should never be regarded as the $A\!\rightarrow\!\infty$ limit of the finite-nucleus asymmetry $K_{\tau}$ \cite{Blaizot1980}. Yet the fact that $K_{\tau}$ is both experimentally accessible and strongly correlated with $K_{\tau}^{\infty}$ is vital in placing stringent constraints on the
density dependence of the symmetry energy. Recall that $K_{\tau}^{\infty}$ is simply related to a few fundamental parameters of the equation of state \cite{cente09}:
\begin{equation}
 K_{\tau}^{\infty}= K_{\rm sym}-6L-\frac{Q_{0}}{K_{\infty}}L\,,
\end{equation}
where $Q_{0}$  the ``skewness'' parameter of symmetric nuclear matter and $L$ and $K_{\rm sym}$,
respectively, are the slope and curvature of the symmetry energy. It is the strong sensitivity of $K_{\tau}^{\infty}$ to the density dependence of the symmetry energy that makes the present study of critical importance in constraining the EoS of neutron-rich matter.

In Eq.~(\ref{eq:liquid_drop}), 
we assume $c\sim -1$ \cite{Patra2002} and, as we have discussed in Sec. 
\ref{sec:ka_kinf}, $K_{Coul}$ is essentially a model-independent term \cite{Sagawa2007}. 
In this way, over a series of isotopes, the only term that varies in any significant manner is the one corresponding to the neutron-proton asymmetry, $(N - Z)/A$ and a value for the associated coefficient, $K_{\tau}$ may be obtained from a fit to the isotopic data. An analysis of this kind was first carried out by Sharma et al. in the late 1980s~\cite{mms}, but was later much improved by  Li et al. over the even-A $^{112-124}$Sn isotopes \cite{Li_PRL2007,Li_2010}. 

Rearranging Eq.~(\ref{eq:liquid_drop}), one gets:
\begin{eqnarray}\label{eq:ktau}
K_A - K_{Coul}\frac{Z^2}{A^{4/3}} \approx K_\infty(1+cA^{-1/3})+K_\tau\left( \frac{N-Z}{A} 
\right)^2, 
\end{eqnarray}
\noindent
which, {\em for all practical purposes}, is a simple quadratic equation in the asymmetry term, $(N - Z)/A$ (the $A^{-1/3}$ term would vary only very little over the range of isotopes under consideration, leading to reasonably treating the first term on the right hand side as a constant). 
From a quadratic fit to this equation, with $K_A$'s derived from their measurements of ISGMR in the Sn isotopes at RCNP, Li et al. obtained an ``experimental'' value of $K_{\tau}$ = -550$\pm$100 MeV. In subsequent measurements on five Cd isotopes (A=106, 110, 112, 114, 116), the same group obtained a value $K_{\tau}$ = -555$\pm$75 MeV, in excellent agreement with the value obtained from the Sn isotopes \cite{Darshana2012}; in both cases, the quoted uncertainties include the effects of the uncertainty in the $K_{Coul}$ term, for which a value of 5.2$\pm$0.7 MeV was used from Ref.~\cite{Sagawa2007}. This value of $K_{\tau}$ is consistent with  $K_\tau=-370 \pm$120 MeV obtained from the analysis of the isotopic transport ratios in medium-energy heavy-ion reactions \cite{Chen2009}, $K_\tau=-500^{+120}_{-100}$ MeV obtained from constraints placed by neutron-skin data from anti-protonic atoms across the mass table \cite{Centelles2009}, and $K_\tau=-500 \pm$50 MeV obtained from theoretical calculations using different Skyrme interactions and relativistic mean-field (RMF) Lagrangians \cite{Sagawa2007}.

\begin{figure}[!h]
\centering\includegraphics [height=0.27\textheight]{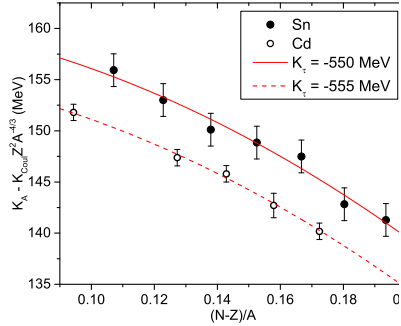}
\caption{The difference $K_A-K_{Coul}Z^{2}A^{-4/3}$ from ISGMR in the Sn and Cd isotopes plotted as a function of the asymmetry parameter, 
$(N-Z)/A$. The solid and dashed lines represent quadratic fits to the respective data. Data from Refs. \cite{Li_2010,Darshana2012}.}
\label{ktau}
\end{figure}

Fig.~\ref{ktau} shows the fits to the Sn and Cd data from Refs. \cite{Li_2010,Darshana2012} taken together.

A caveat to the discussion above: the $K_{\tau}$ obtained from the measurements on the Sn and Cd isotopes is only an ``average" value, and the data cannot disentangle the volume symmetry term from higher-order effects like the surface symmetry. It is possible, then, to execute similar fits including higher-order terms and obtain very different values for $K_{\tau}$, as has been shown by Pearson et al. \cite{Pearson2010}: in fact, introducing the surface symmetry amounts to replacing
\begin{equation}
K_{\tau} \left( \frac{N-Z}{A} \right)^2 \ \ \ \rightarrow
\left( K_{\tau,vol} + K_{\tau,surf} A^{-1/3} \right)
\left( \frac{N-Z}{A} \right)^2.
\end{equation}

%% file: decay_2.tex
\section{Decay measurements}\label{decay}

Measurements of the charged-particle and neutron decays of the giant resonances can provide important information on the microscopic nature of the resonances, in particular on the particle-hole states involved. Because the giant resonances, in general, are located well above the particle separation thresholds, particle emission is the dominant decay process that takes place and can occur either from the initial 1p-1h state, leaving a single-hole state in the A$-$1 nucleus (direct component), or from the states with partially or completely equilibrated configurations, resulting in an evaporation-like spectrum of the emitted particles; the latter is termed the statistical component. The observation of particle emission spectra observed in coincidence with the excitation of giant resonances can provide useful information on the evolution of the decaying configuration and on the microscopic structure of the giant resonance \cite{Harakeh_book,hun4}.

Furthermore, the coincidence technique with particle decay gives a useful means to suppress the backgrounds in inelastic scattering spectra and to reliably isolate resonances strengths. Indeed, such measurements have been found, in many instances, to be very effective in eliminating both the instrumental background and the nonresonant continuum, thus overcoming the problems connected with background subtraction discussed earlier, and allowing for a better determination of the giant-resonance gross properties \cite{Harakeh_book,hun4}.

Several decay measurements were carried out in the 1980's focused on the  particle decay of ISGMR and ISGQR. While charged-particle decay occurs preferentially in the light nuclei (A$\lessapprox$60), neutron emission is the dominant decay mode for the medium-to-heavy mass nuclei because of Coulomb barrier effects. Also, in most cases, the observed decay, typically, had a large statistical component, making the separation of the direct decay component an experimentally challenging task. Still, an excess of neutron emission over statistical model calculations was identified for the ISGMR in some cases 
(see, for example, the measurements reported in Refs. \cite{bran1,Bracco1989} and
the corresponding theoretical calculations \cite{Colo1992}).

The situation turns out to be somewhat different (in the positive sense) for the ISGDR. Continuum RPA (CRPA) calculations predicted
significant decay widths from the ISGDR in $^{208}$Pb to some proton-hole states in $^{207}$Tl, and a similar behavior in other medium- and heavy-mass nuclei \cite{urin1,urin2}. Charged-particle decay measurements are, in general, easier than their neutron-decay counterparts and the predicted decay branching ratios were sufficiently large to make such measurements quite feasible within the beam times typically granted for an experiment at international accelerator facilities.

\begin{figure}[!h]
\centering\includegraphics [height=0.32\textheight]{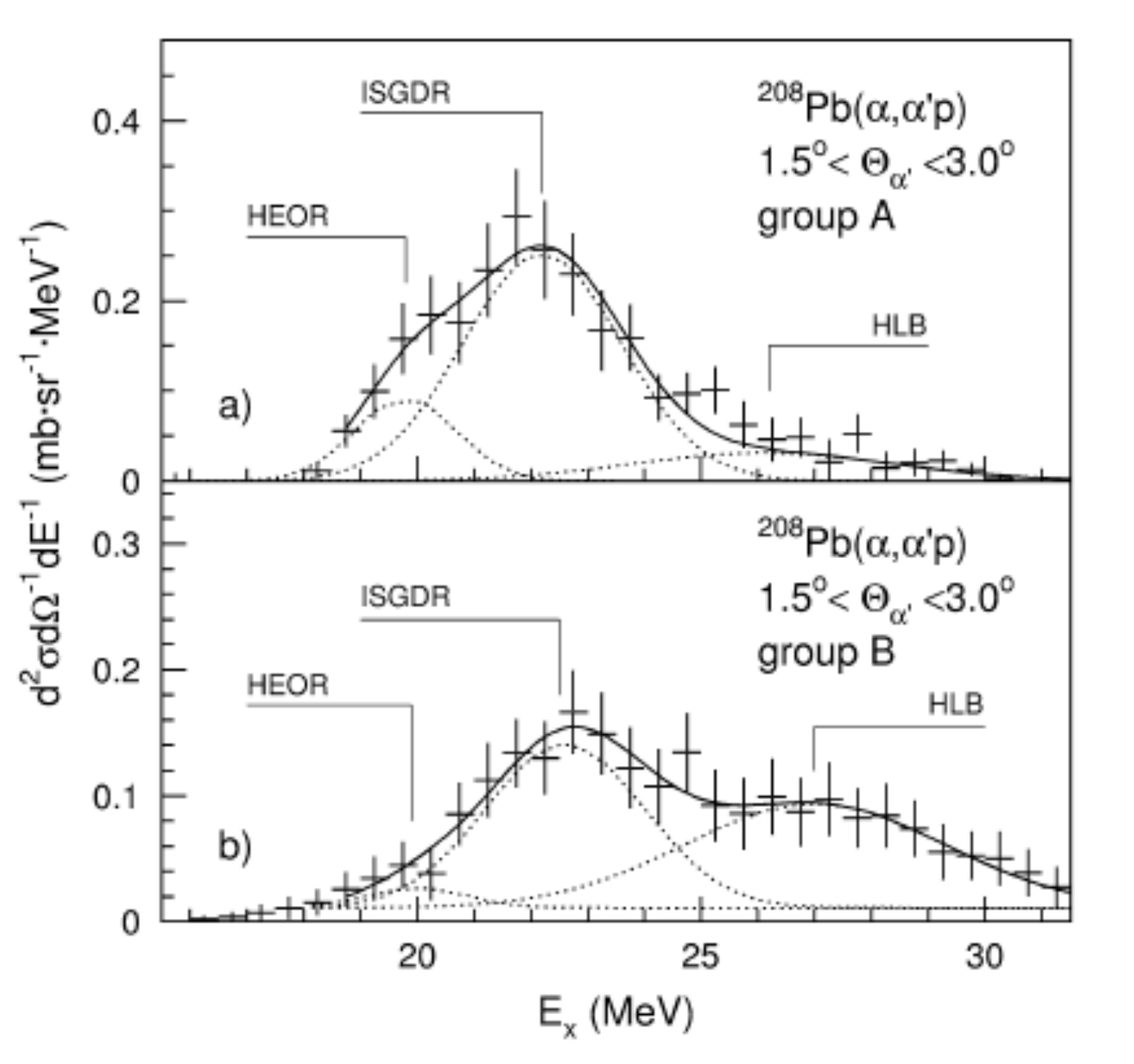}
\caption{Double-differential cross sections for $^{208}$Pb($\alpha,\alpha'p$), as a function of the excitation energy in $^{208}$Pb, gated on the indicated scattering angles and measured in coincidence with direct-decay protons of (a) group A and (b) group B. The fits of the HEOR, ISGDR and the higher-lying bump (HLB) are also shown. Group A and Group B refer to two enhanced structures in the final-state spectrum of the low-lying proton-hole states in $^{207}$Tl generated by gating on the ISGDR region, and on scattering angles of $\theta_{c.m.}$. = 1.5$^\circ$--3.0$^\circ$. Figure from Ref. \cite{hun1}.}
\label{pb_decay}
\end{figure}

Two sets of proton-decay measurements have been performed, at KVI, Groningen, on $^{58}$Ni, $^{90}$Zr, $^{116}$Sn, and $^{208}$Pb, using 200-MeV $\alpha$-particle beams provided by the AGOR superconducting cyclotron facility \cite{hun1,hun3,hun4}, and at RCNP, on $^{58}$Ni  and $^{208}$Pb, using 400-MeV $\alpha$ particles \cite{nayak58ni,nayak2}. In all cases, significant decay was observed to specific particle-hole states in the daughter nuclei, in qualitative agreement with the predictions of the theory; indeed, the agreement with theory was termed excellent for $^{208}$Pb \cite{hun1,hun3,nayak2}. An example of the quality of the decay spectra from the KVI measurement on $^{208}$Pb is provided in Fig.~\ref{pb_decay}.

\begin{figure}[!h]
\centering\includegraphics [height=0.45\textheight]{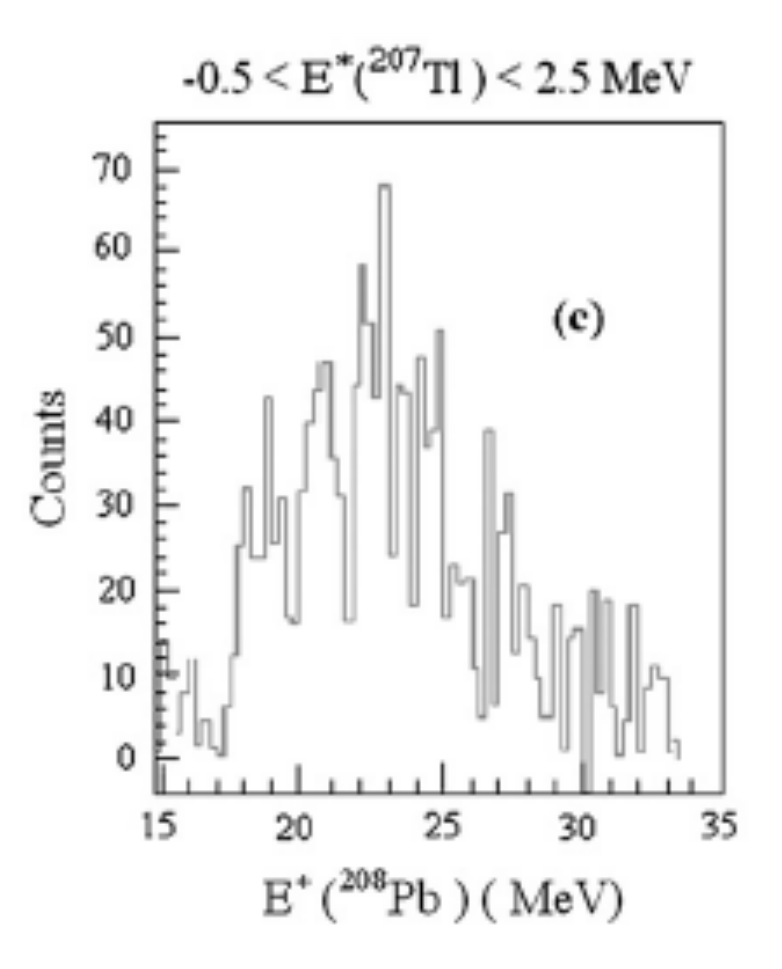}
\vspace*{-1cm}
\caption{Projection of 2-dimensional coincidence data onto the excitation-energy axis in $^{208}$Pb for the indicated excitation-energy range in $^{207}$Tl. The spectrum shows primarily the true HE-ISGDR strength distribution. Figure from Ref. \cite{nayak2}.}
\label{pure_isgdr}
\end{figure}

A very intriguing aspect of the KVI measurement on $^{208}$Pb was the observation of the structure identified as the higher-lying bump (HLB) in Fig.~\ref{pb_decay}. DWBA calculations favored an $L$=2 character for this structure, observed over the region of $E_x$=25--31 MeV. It was suggested that this ``bump'' corresponded to the excitation of the overtone of the ISGQR, the response to the second-order quadrupole  operator. The excitation energy and width of this bump was reported as: $E_x$ = 26.9$\pm$0.7 MeV and $\Gamma$ = 6.0$\pm$1.3 MeV  \cite{hun1}. If the quadrupole character for this mode were confirmed, it would correspond to the first observation of the third compressional mode after the ISGMR and ISGDR.

Unfortunately, this mode was not unambiguously identified in subsequent RCNP $p$-decay measurements on the same nucleus \cite{nayak2}. While there exists some strength in the RCNP decay spectra at the location of the HLB reported in the KVI work, it is too small to allow arriving at any definitive conclusion about its multipole character. This was most likely a consequence of the fact that the excitation cross section for ISGDR is significantly larger when compared with the $L$=2 excitation cross section in the angular range 1.0$^\circ$--1.5$^\circ$ covered in the RCNP measurement \cite{nayak2}. This structure has also not been observed, so far, in any other nuclei on which singles or coincidence measurements have been made.

The observation of HLB is, nonetheless, a testament to the power of coincidence measurements--this strength would be ``buried'' in the continuum, and hence unobservable, in the singles measurements. Another useful result of these measurements was the clear evidence that the $L$=1 strength observed at high excitation energies in the ($\alpha,\alpha'$) spectra did not belong to the ISGDR; as seen in Fig.~\ref{pure_isgdr}, that ``spurious'' strength is largely absent in the $^{208}$Pb($\alpha,\alpha'p$) coincidence spectra \cite{nayak2}.

%% file: unstable_nuclei_2.tex
\section{Measurements in Nuclei Far from Stability}\label{unstable_nuclei}

As radioactive ion beams of higher intensities become available, the investigation of the compression-mode resonances in nuclei far from stability becomes extremely interesting in order to understand and delineate the effect of large neutron-proton asymmetries on the nuclear incompressibility. For one, the effect of the asymmetry term in going from incompressibility of individual nuclei to incompressibility of infinite nuclear matter is quite important but not well understood. There also is the intriguing possibility of the observation of the Òsoft GMR,Ó akin to the soft giant dipole resonance (the so-called pygmy dipole resonance) observed in the halo nuclei. Thus, one would be looking at two nuclear incompressibilities: one for the Òcore,Ó the other between the core and the ``halo'' or the ``skin''. 

Calculations by Sagawa et al. \cite{Hamamoto:1997,Sagawa1998} had indicated a threshold effect in the monopole response resulting in considerable ISGMR strength at low energies in nuclei far from the stability line. Similar effects have been  predicted in calculations by Khan et al. \cite{Khan2011} and J. Piekarewicz \cite{Piekarewicz:2017} as well.  Although the conclusions of Ref.~\cite{Khan2011} were later questioned in Refs.~\cite{Sagawa2014,jorge2015}, the nature of ISGMR strength in nuclei far from the stability line remains of tremendous current interest. 

These measurements have to be performed in the inverse kinematics mode with the concomitant problem of very low velocities of the recoiling target nuclei at forward-angles essential for identifying the ISGMR with multipole-decomposition analysis. As stated before, the best experimental probes for the investigation of the ISGMR are deuterons and $\alpha$ particles. For the excitation energy range corresponding to the ISGMR, the expected energies of the recoiling particles is in the range of $\sim$100 keV to $\sim$2 MeV. In a standard thin target and particle telescope set-up, this energy would necessitate use of very thin targets ($\sim$100 $\mu$g/cm$^2$) and detectors that have practically no dead-layers or entrance foils. Considering the intensities available for radioactive ion beams, this makes these measurements practically impossible.

The only reasonable option at present appears to be the use of an active-target timing projection chamber (AT-TPC). In an active target system, the detector gas employed in the TPC also acts as the target. Such a system, in principle, can have an angular coverage close to 4$\pi$, a low-energy threshold, and large effective target thickness, alleviating all the problems mentioned above associated with the inverse kinematics measurements with radioactive ion beams \cite{mittig-ahn}. 

The first such experiment, meant primarily to establish the technique, was performed at the GANIL facility in France, using the AT-TPC system MAYA \cite{maya}. A $^{56}$Ni beam at energy of 50 MeV/nucleon was incident on MAYA filled with deuterium gas at a pressure of 1050 mbar which is equivalent to a pure deuterium target of 1.6 mg/cm$^2$-thickness \cite{ni56}. Even with an effective data taking time of only 15 hours and an average beam intensity of 5$\times$10$^4$ pps, it was possible to observe the ``bump'' corresponding to the ISGMR+ISGQR in the spectrum of recoil deuterons (see Fig.~\ref{ni56fig} below). Moreover, the ISGMR and ISGQR components were distinguished on the basis of MDA, leading to excitation energy values for these resonances consistent with the known ISGMR and ISGQR energies for the nearby stable nucleus $^{58}$Ni (see Table \ref{ISGMR}).
 
\begin{figure}[!h]
\centering\includegraphics [height=0.30\textheight]{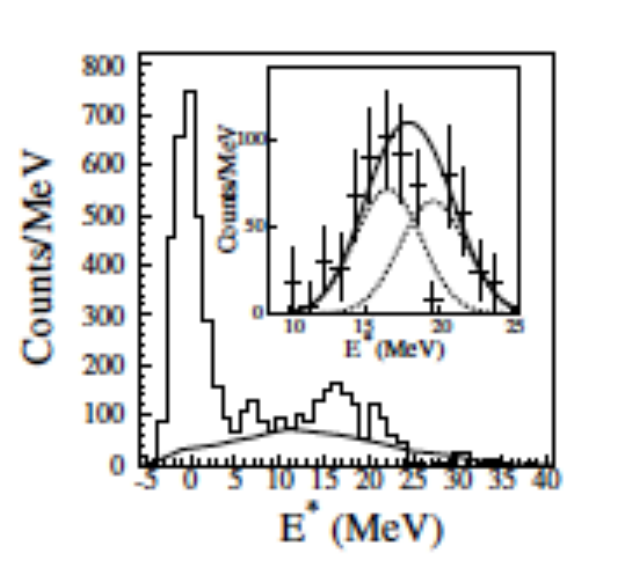}
\vspace*{-0.5cm}
\caption{$^{56}$Ni excitation energy spectrum deduced from the deuteron kinematics and corrected for geometrical efficiency. The background that was subtracted is shown by the solid line. The inset shows the background subtracted inelastic data fitted with Gaussian distributions located at 16.5 and 19.5 MeV for the ISGQR and the ISGMR, respectively. Figure adopted from Ref. \cite{ni56}.}
\label{ni56fig}
\end{figure}

This experiment, as noted, employed deuterium gas as the target, even though inelastic scattering of $\alpha$ particles had been established for a long time as the preferred method to excite the ISGMR and there was a large data set validating the MDA in extracting the ISGMR strength distributions based on DWBA calculations. Also, break-up of deuteron adds significantly to the ``background'' in the final spectra: since the detector had to be optimized for detection of very low deuteron energies, it was not possible to separate protons from deuterons based on range versus charge measurements. Indeed, the background shown in Fig.~\ref{ni56fig} arises primarily from deuteron break-up and was estimated from direct kinematic measurements for $^{58}$Ni at 50 MeV/nucleon performed previously \cite{dxp}. 
The choice of deuterium gas in MAYA had to do with a major practical consideration in that gaseous detectors spark at high voltages when filled with pure Helium gas. A further difficulty arises because of the need for a ``mask'' to absorb the electrons resulting from the high ionization of the gas by the incoming beam \cite{ni68_2}. These aspects are discussed further later in this section.

\begin{figure}[!h]
\centering\includegraphics [height=0.23\textheight]{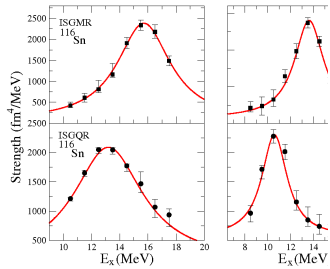}
\caption{ISGMR and ISGQR strength distributions for $^{116}$Sn and $^{208}$Pb obtained in the RCNP $(d,d')$ work. The solid red lines are Lorentzian fits to the data. Figure from Ref. \cite{Darshana2013}.}
\label{ddprime}
\end{figure}

Although, as an isoscalar particle, the deuteron may be thought of as a probe ideally suited for investigation of the ISGMR, it has not been used much for the purpose, save for some very early experiments carried out in France (see, for example, Refs. \cite{ddprime1,ddprime2}). Specifically, Willis et al. \cite{ddprime2} had performed $(d,d')$ measurements on several nuclei using a beam of 54 MeV/nucleon energy and extracted strength distributions of various giant resonances based on a peak-fit analysis, a method deemed less reliable than the MDA technique currently in use. As deuteron appeared to be the probe of choice for measurements with radioactive ion beams, it was important to validate in known cases the results obtained in $(d,d')$ work via direct comparison with results obtained from inelastic $\alpha$ scattering.
This was carried out by Patel et al. \cite{Darshana2013,Patel_Thesis,Patel_Book} at RCNP. In this first investigation of giant resonances with a deuteron probe at a beam energy amenable to cross sections for excitation of ISGMR required for radioactive ion beam experiments, they employed a 196-MeV deuteron beam to obtain the now standard at RCNP ``background free'' inelastic scattering spectra for $^{116}$Sn and $^{208}$Pb. They extracted ISGMR and ISGQR strength functions using the MDA technique (see Fig.~\ref{ddprime}) and demonstrated that the ISGMR strength may be extracted reliably with the deuteron probe \cite{Darshana2013}. The properties of ISGMR and ISGQR extracted in this work agree very well with the previous values from inelastic $\alpha$ scattering (see Tables 2 and 3 in Ref. \cite{Darshana2013}), establishing, in the process, that small-angle deuteron inelastic scattering can serve as a reliable tool for investigation of ISGMR in nuclei far from stability, using the rare isotope beams now becoming available at facilities worldwide.

As  mentioned earlier, there are some practical issues that initially dissuaded experimentalist from using $(\alpha,\alpha')$ in inverse kinematics measurements with radioactive ion beams. A major one was the sparking of pure Helium gas at high voltages typically used in the AT-TPC systems. This may be alleviated by quenching the gas with a suitable admixture; however, the concern always was that the admixed gas (CF$_4$ or CO$_2$ were the possibilities initially considered) would give rise to unacceptably high backgrounds, especially because of much higher expected cross sections for scattering off $^{12}$C. Another possible complication arose from the high amplification of the electrons emanating from ionization of the detector gas by the beam particles, with these electrons overwhelming the collecting wires in the TPC. This problem was solved by use if an electrostatic ``mask'', placed just below the beam \cite{mask}; this device suppresses the collection of electrons generated by the beam on the central wires. [Such a ``mask'' is not required in the $(d,d')$ measurements because the electron amplification was significantly smaller.] 

In two measurements, performed also at GANIL, the aforementioned problems associated with using $(\alpha,\alpha')$ for ISGMR using an AT-TPC device were largely overcome \cite{ni68_1,ni68_2,ni56_2}. Helium gas was used at a pressure of 500 mbar in MAYA, with 5\% of CF$_4$ as the quencher. Two different radioactive ion beams, $^{56}$Ni ($\sim$2$\times$10$^4$ pps) and $^{68}$Ni ($\sim$4$\times$10$^4$ pps), both at 50 MeV/nucleon incident energy, were employed, with data taken on $^{68}$Ni also for $(d,d')$ to obtain a direct comparison between the two probes; the statistics in the $(d,d')$ data was much lower than in $(\alpha,\alpha')$ because of the lower cross section for $(d,d')$, as also the experimental conditions of pressure and high voltage employed in the two experiments  \cite{ni68_1,ni68_2}. In all cases, a number of narrow peaks are observed in the inelastic scattering spectra sitting atop, so to speak, a large background.

\begin{figure}[!h]
\centering\includegraphics [height=0.45\textheight]{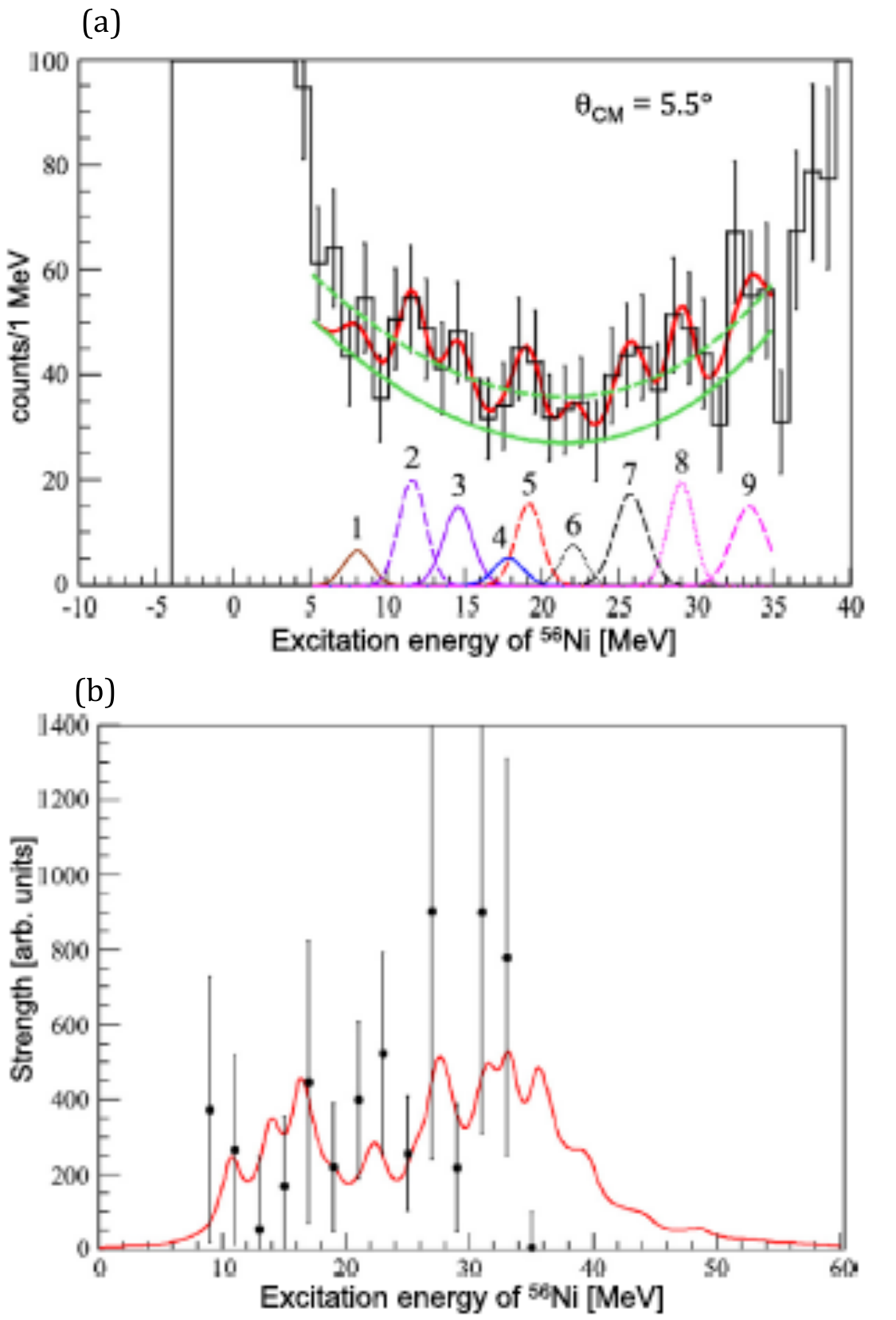}
\caption{(a) Excitation-energy spectrum of $^{56}$Ni for a 1$^{\circ}$ bin centered around 5.5$^{\circ} \theta_{CM}$ angle is shown.  The thin solid line visible at the bottom of the data represents the background fit. The thick solid line is the result of the fit to the data with nine Gaussian peaks and a fitted background. The nine Gaussian peaks are shown separately. (b) Comparison of the experimental ISGDR strength distribution in $^{56}$Ni (solid circles with error bars), obtained from the MDA, is shown together with the prediction of an HF-RPA calculation (solid line). Figure from Ref. \cite{ni56_2}.}
\label{bagchi}
\end{figure}

In $^{68}$Ni, a ``straight'' background was subtracted from the inelastic scattering spectra. Complementary analyses of fitting a number of peaks to the spectra, and performing MDA, led to identification of ISGMR strength over $E_x$=11--23 MeV, with a dominant component at $E_x$=21.1 MeV; this result is consistent with the $(d,d')$ data \cite{ni68_1,ni68_2}. A possible indication of a soft isoscalar monopole resonance, mixed with ISGDR strength, was also found at 12.9$\pm$1.0 MeV, in the fitting method in the $(\alpha,\alpha')$ data. 
In case of $^{56}$Ni \cite{ni56_2}, the background was rather large and its shape was approximated by a polynomial of order 4. A total of 9 peaks were identified in the spectra and used in peak-fitting.  The centroid position of the ISGMR was found to be 19.1$\pm$0.5 MeV which compares well with the value 19.5 MeV obtained in the previous measurement on this nucleus \cite{ni56}. The authors also reported identification of the ISGDR strength over $E_x$=10--35  MeV, albeit with large uncertainties; the observed strength distribution was consistent with the predictions of the HF-based RPA calculations from Auerbach et al. \cite{naftali}. The $(\alpha,\alpha')$ results on $^{56}$Ni are presented in Fig.~\ref{bagchi}.

All these measurements were plagued by low statistics, high background, low energy resolution, and limited excitation energy range, rendering accurate and unambiguous determination of the ISGMR and ISGDR strengths very difficult and leading to large uncertainties. Their success, and true importance, lies, nevertheless, in establishing the suitability of the AT-TPC set-up for measuring the properties of unstable and exotic nuclei available at rare-isotope beam facilities the world over. It is highly likely that the forthcoming new AT-TPCs, currently under development \cite{mittig-ahn,pancin}, will provide significantly higher energy and angular resolutions, making high-quality small-angle inelastic scattering measurements without these problems feasible.

A more recent measurement on ISGMR strength in the doubly closed shell nucleus $^{132}$Sn has been carried out at the RIKEN RIB Facility in Japan \cite{ota-garg}. The newly-developed active-target system, CAT \cite{ota_cat}, was employed with deuterium gas. The primary aim of this measurement is to obtain a more precise value for $K_{\tau}$, in conjunction with the ISGMR data available on the stable Sn isotopes. Results from this measurement are awaited. In the event, because of the relatively high intensity of $^{132}$Sn beams available at RIKEN ($>$ 5 x 10$^4$ pps), it is anticipated that the ISGMR would be observed with better statistics than observed so far in other measurements.

A possible, and novel, way of measuring the compression-mode giant resonances in unstable nuclei using stored beams has recently been introduced \cite{exl_gsi}. In a demonstration experiment, carried out at GSI, Darmsrtadt, a $^{58}$Ni beam of 100 MeV/nucleon energy was incident on an He gas-jet target internal to ESR, the heavy-ion storage ring at GSI. Luminosities of the order of 10$^{25}$--10$^{26}$ cm$^2$ sec$^{-1}$ were achieved in the measurement, and inelastically scattered $\alpha$ recoils were measured at very forward angles ($\theta_{c.m.}\leq1.5^{\circ}$) with ultra-high vacuum compatible detectors. The results indicated a dominant contribution from the ISGMR, exhausting 79$^{+12}_{-11}$\% EWSR at an excitation energy consistent with the previously reported values for ISGMR in this nucleus. This experiment used a stable beam (the luminosity for an unstable beam would have been too low for any meaningful experiment) and the statistics in the final spectra were rather low; still, it points to the possibility of such measurements with the advent of future facilities, such as FAIR, when much higher luminosities might be feasible.

%% file: conclusions_2.tex
\section{Conclusions}\label{conclu}

The story of the compression modes in nuclei is not a new one. Since
the discovery of the ISGMR in the late 1970's, interest has been devoted
to the incompressibility of finite nuclei, to the extrapolation
to the case of infinite nuclear matter, and to the relationship
with the physics of core-collapse supernova and neutron stars.
The status up to the turn of the new millennium has been reviewed
in previous papers and books; so the purpose of the current work
is to report on significant results of the past 15-20 years.

Experimental techniques have significantly improved during this period,
and this has led to new measurements of the ISGMR and ISGDR that
have highlighted or disproven new effects. Exclusive measurements
(particle-decay) on the ISGDR are also among the new achievements.
At the same time, there is a new host of fully self-consistent
calculations; linear response theory has been fully implemented
both with nonrelativistic and covariant functionals, with a special
focus also on the pairing part, which is relevant for superfluid
nuclei. We have discussed these new steps forward in this review,
and stressed the consequences on the extraction of the nuclear
incompressibility $K_\infty$. While $K_\infty$ extracted from the
doubly-magic nuclei like $^{208}$Pb has been reported to be around
240 $\pm$ 20 MeV, the value associated with Sn and other open-shell
nuclei seems to be lower. This issue of the softness, or fluffiness,
of the open-shell nuclei has not been resolved so far.

Effects of deformation and other nuclear structure effects on the
compression modes have also been
analyzed. Whether they can modify our understanding
of $K_\infty$, remains to be seen, however.

The new frontier is certainly the exploration of exotic, neutron-rich
nuclei. In long isotopic chains, as the difference between the neutron
and proton number increases, it is known that protons become more bound
while neutrons occupy higher levels that lie close to the continuum.
The large difference between the Fermi energies of protons and
neutrons may produce a decoupling between the well-bound nucleons
in the ``core'' and the less bound neutrons in the skin or halo.
The question to be answered is whether these two components might
behave like two fluids with different incompressibilities. Measurements with radioactive ion beams have been initiated over the past decade and the results appear to be quite promising. With development of new experimental equipment and techniques, it is very likely that important new results will become available in the next few years.

%% file: acknowledgments_2.tex
\section*{Aknowledgments}

We wish to thank Mr. Kevin Howard for his help with some of the figures and tables.
This work has been supported in part by the U. S. National Science
Foundation (Grant No. PHY-1713857).